\documentclass[aps,superscriptaddress,nofootinbib,tightenlines,showkeys,notitlepage,11pt]{revtex4-2}
\usepackage{graphicx}% Include figure files
\usepackage{dcolumn}% Align table columns on decimal point
\usepackage{amsfonts,amsthm,stmaryrd,mathtools,mathrsfs,bm}%bm  bold math
\usepackage{hyperref}% add hypertext capabilities
\usepackage{xcolor}
\usepackage{color}

\usepackage{caption}
\usepackage[export]{adjustbox}
\usepackage{float}
\makeatletter
\let\newfloat\newfloat@ltx
\makeatother

\usepackage{algorithm}
\usepackage{algorithmicx}
\usepackage{algpseudocode}

\usepackage{listings}
\usepackage{array}
\usepackage{multirow}
\usepackage{dsfont}

\usepackage{scalerel}

\newcommand{\red}{\color{red}}
\newcommand{\blue}{\color{blue}}

\newcommand\nn{\nonumber}
\newcommand{\q}{\quad}

\newcolumntype{L}[1]{>{\raggedright\let\newline\\\arraybackslash\hspace{0pt}}m{#1}}
\newcolumntype{C}[1]{>{\centering\let\newline\\\arraybackslash\hspace{0pt}}m{#1}}
\newcolumntype{R}[1]{>{\raggedleft\let\newline\\\arraybackslash\hspace{0pt}}m{#1}}

\def\threeJ{ \{ { 3j} \} }
\def\fourJ{ \{ { 4j} \} }
\def\sixJ{ \{ { 6j} \} }
\def\fifJ{ \{ { 15j} \} }
\def\cohj{\scalebox{1.3}{\bf c}}
\def\intw{\scalebox{1.25}{$i$}}
\def\d{  {\rm d} }

\begin{document}

\title{Efficient Tensor Network Algorithms for Spin Foam Models}

\author{Seth K. Asante}
\email{seth.asante@uni-jena.de}
\affiliation{Theoretisch-Physikalisches Insit\"{u}t, Friedrich-Schiller-Universit\"{a}t Jena, Max-Wien Platz 1, 07743, Jena, Germany}

\author{Sebastian Steinhaus}
\email{sebastian.steinhaus@uni-jena.de}
\affiliation{Theoretisch-Physikalisches Insit\"{u}t, Friedrich-Schiller-Universit\"{a}t Jena, Max-Wien Platz 1, 07743, Jena, Germany}

\begin{abstract}  ~

Numerical computations and methods have become increasingly crucial in the study of spin foam models across various regimes. This paper adds to this field by introducing new algorithms based on tensor network methods for computing amplitudes, focusing on topological SU(2) BF and Lorentzian EPRL spin foam models. By reorganizing the sums and tensors involved, vertex amplitudes are recast as a sequence of matrix contractions. This reorganization significantly reduces computational complexity and memory usage, allowing for scalable and efficient computations of the amplitudes for larger representation labels on standard consumer hardware---previously infeasible due to the computational demands of high-valent tensors.

We apply these tensor network algorithms to analyze the characteristics of various vertex configurations, including Regge and vector geometries for the SU(2) BF theory, demonstrating consistent scaling behavior and differing oscillation patterns. Our benchmarks reveal substantial improvements in computational time and memory allocations, especially for large representation labels. Additionally, these tensor network methods are applicable to generic 2-complexes with multiple vertices, where we introduce partial-coherent vertex amplitudes to streamline the computations. The implementation of these algorithms is available on GitHub for further exploration and use. 

\end{abstract}


\maketitle
\tableofcontents 

~

\section{Introduction}

Recent advances in spin foam models \cite{Bambi:2023jiz,Engle:2023qsu,Livine:2024hhc} have been driven by the development and application of various numerical techniques \cite{Dona:2022yyn} tailored for different regimes. These techniques facilitate the computation of spin foam amplitudes, which are crucial for understanding the dynamics of discrete quantum geometries as derived from loop quantum gravity (LQG). The quantum numbers assigned to these quantum geometries are described by unitary irreducible representation labels of the underlying gauge group of the model. In the quantum regime, where these representation labels are small, numerical libraries such as  \texttt{sl2cfoam} \cite{Dona:2019dkf} and its latest version, \texttt{sl2cfoam-next} \cite{Gozzini:2021kbt,Dona:2022dxs}, have been developed to compute amplitudes for the Lorentzian Engle-Pereira-Rovelli-Livine (EPRL) model \cite{Engle:2007uq,Engle:2007wy}. These libraries are well optimized for high-performance computing, with significant improvements in the latest version.  These numerical tools have been utilized to study various aspects of spin foam models, including exploration of different physical scenarios \cite{Dona:2020tvv,Dona:2022vyh,Frisoni:2022urv}. Monte Carlo methods have also been used within spin foam models, employing techniques such as deforming integration contour via Lefschetz thimbles \cite{Han:2020npv}. Moreover, different sampling methods over the representation labels \cite{Dona:2023myv,Steinhaus:2024qov} have been implemented to enhance convergence of the spin foam amplitudes.

For explorations in the semi-classical or asymptotic regime, numerical methods such as the complex critical points program \cite{Han:2021kll,Han:2023cen,Han:2024lti} have been developed to identify semi-classical geometries in the limit of large representations. To bridge the gap between the numerical methods for quantum amplitudes and those for asymptotic regimes, a hybrid algorithm was proposed in \cite{Asante:2022lnp}. Additionally, numerical methods \cite{Bahr:2015gxa,Bahr:2016hwc} based on symmetry reductions have been employed to study the renormalization aspects of spin foam models. Effective spin foam models \cite{Asante:2020qpa,Asante:2021zzh,Asante:2020iwm}, which rely on a quantum superposition of discrete area configurations using area Regge calculus, provide efficient numerical methods to investigate the dynamics of super-imposed discrete-area geometries. There are also well-developed computational tools and techniques  \cite{Dittrich:2021kzs,Dittrich:2022yoo,Borissova:2022clg} for studying the continuum limit of these discrete geometries which are crucial for understanding continuum limit of spin foam dynamics.

One reason for the reliance on recent numerical approaches is the intrinsic difficulty of handling spin foam amplitudes analytically. These amplitudes are typically represented by highly oscillatory functions and involve integrations over numerous variables. Numerical methods manage the oscillatory and complex nature of the transition amplitudes by explicitly utilizing efficient recoupling symbols, such as Clebsch-Gordan coefficients and Wigner $\{ nj\}$-symbols \cite{Speziale:2016axj}, along with specific algebraic identities. These recoupling symbols  simplify the combinatorial structure of spin networks and facilitate the expression of transition amplitudes as sums and products of representation labels and intertwiners. The computations are then efficiently  performed through tensor contractions, where recoupling symbols feature in the components of the involved tensors. Spin foam amplitudes evaluated using tensor contractions are more computationally feasible than performing direct sums or the oscillatory integrals. Despite this, these calculations can be computationally expensive, both in terms of memory usage and computation time, with costs escalating as the representation labels become large and the degrees of freedom to sum over increase. The tensors involved in the amplitudes can be of high-valence, adding to the computational resources required for their contractions. For instance, the $\fifJ$-symbol which appears in the construction of a spin foam vertex amplitude, is typically computed as a 5-valent tensor and contracted with other tensors. Storing and contracting such $\fifJ$-symbols becomes computationally expensive as their representation labels gets larger.  %construction and

To address these challenges, tensor network methods and techniques can be employed. Tensor networks \cite{Orus:2013kga,Biamonte:2017dgr}, developed in the context of quantum many-body physics, provide a compact and efficient representation of high-dimensional tensors through the contraction of smaller, low-rank tensors.  Tensor network methods enhance the efficiency of tensor contractions and allow for better scalability when dealing with the large-scale computations, by breaking down high-valence tensors into networks of lower-valence tensors. This approach significantly reduces both the computational complexity and the memory requirements, making it feasible to perform computations with less computer resources. This makes them a promising tool to support numerical computations of spin foam models. Previous works \cite{Dittrich:2011zh,Dittrich:2013voa,Dittrich:2014mxa,Delcamp:2016dqo,Asante:2022dnj} on applications of tensor network methods within spin foam models have focused on renormalization and coarse graining aspects.  Here, we aim to optimize the evaluation of spin foam amplitudes by employing tensor contractions of smaller, low-valence tensors. By organizing the computation into sequences of smaller tensor contractions, we can significantly enhance both the efficiency and scalability of the calculations. This method allows for more effective handling of the highly complex amplitudes in spin foam models.

In this paper, we present tensor network algorithms for an efficient computation of coherent amplitudes within spin foam models. We provide tensor network algorithms for computing both SU(2) BF spin foam  and EPRL spin foam vertex amplitudes. Our methods optimize the computation of a coherent vertex amplitude by reorganizing the sums and products of the recoupling symbols in the definition of the amplitudes, resulting in contractions between matrices which are tensors of low-valence. This optimization not only minimizes memory usage but also accelerates the computational time needed for computing coherent vertex amplitudes. For example, we are now able to compute SU(2) BF equilateral vertex amplitude at a large boundary spin value $j=200$ in few seconds on a consumer hardware, a task that is impractical using previous methods. We make use of tensor network notations to streamline the discussion and to clearly illustrate how the tensor contractions are performed. The tensor network methods are also applicable to deal with spin foam amplitudes associated with a generic 2-complex having multiple vertices. A direct implementation of the tensor network algorithm for the SU(2) BF model is made available on the \href{https://github.com/Seth-Kurankyi/su2bf-TNAlgo}{GitHub repository} \cite{AsanteTN2024}.

The rest of the paper is organized as follows: Section \ref{sec:su2bf} provides an overview of the spin foam models, providing  notations and conventions which are used through out this paper. In Section \ref{sec:spintensors}, we review how spin foam amplitudes are computed as sums of contractions between tensors constructed from recoupling symbols. Section \ref{sec:newvertex} details the tensor network methods, including the organization of the $\sixJ$-symbols and coherent vectors to construct `coherent $\sixJ$-intertwiner matrices'. We provide tensor network algorithms utilizing the contraction of matrices to compute vertex amplitudes. Section \ref{sec:results} presents numerical results and benchmarks demonstrating the efficiency and accuracy of our algorithm applied to various vertex amplitudes. Examples of the vertex configurations used for the results and benchmarks are described in Appendices \ref{sec:geoms} and \ref{sec:Asymptotics}. In Section \ref{sec:ideas}, we discuss ideas for applying tensor network methods to compute spin foam amplitudes for generic 2-complexes with bulk vertices. Lastly, Section \ref{sec:conclude} concludes with a summary and directions for future research.

\section{Spin Foam Amplitudes: Notation and Conventions}\label{sec:su2bf}

Spin foam models are discrete path integrals or state-sum models for {`quantum geometries'} regularized on a 2-dimensional cell (2-complex) $\Delta^{\!*}$ commonly chosen to be dual to a triangulation $\Delta$ of a spacetime manifold. These models are defined by assigning local amplitudes to the building blocks of the discretized 2-complex. The components of a 2-complex are summarized in Table \ref{tab:2-complex}. Its boundary components are made up of links and nodes describing a \emph{spin-network} graph. The local amplitudes assigned to the  2-complex are constructed from the representation theory of the underlying gauge group $\cal G$ of the spin foam model. In this article, we shall focus on the four dimensional BF spin foam model based on SU(2) gauge group, and the Engle-Pereira-Rovelli-Livine-Freidel-Krasnov (EPRL-FK) model based on $\rm SL(2,\mathbb C)$ gauge group. The SU(2) BF model is a topological state-sum model based on a quantization of 4D BF topological action. Refer to \cite{Baez:1999sr,Barrett:2009as,Freidel:2007py} for more details on the relation between BF theory and spin foam models.  

\begin{table}[ht!]
\centering
\captionsetup{justification = justified}
\renewcommand*{\arraystretch}{1.4}
\begin{tabular}{|c|c|c|c|}\hline 
2-complex ($\Delta^{\!*}$) & \, Internal \, & \, Boundary \, & Triangulation ($\Delta$)  \\ \hline \hline
vertex $v$ & $v$ & - & 4-simplex $ \sigma$ \\ \hline
edge $e$ & ${e}_i $ & node $n $ & tetrahedron $ \tau$ \\ \hline
face $f$ & $f_{i}$ & link $ \ell$ & triangle $ t$ \\ \hline
\end{tabular}
\caption{Components of a 2-complex $\Delta^{\!*}$ dual to a 4D triangulation $\Delta$}
\label{tab:2-complex}
\end{table}

Spin foam models for quantum gravity, such as the Barrett-Crane \cite{Barrett:1997gw,Barrett:1999qw,Jercher:2022mky} and EPRL-FK models \cite{Engle:2007uq,Engle:2007wy,Freidel:2007py}, are closely related to BF state-sum models. These models are constructed as constrained versions of the BF models. The quantum geometry data, described by group representation theory labels, are assigned to the spin-network boundary graph. Loop quantum gravity (LQG), a canonical framework for quantum gravity, provides well-defined mathematical structures for the construction of spin networks associated with 3D boundary graphs \cite{Ashtekar:1996eg,Rovelli:1995ac}. Thus, spin foam amplitudes of state-sum models, are considered transition amplitudes for LQG boundary states. In the following section, we shall focus on the SU(2) BF spin foam model to introduce the notations and conventions for defining spin foam amplitudes.

\subsection{ SU(2) BF State-Sum Model}

The spin foam model for the SU(2) BF theory  is defined by decorating a discretized manifold or its dual 2-complex $\Delta^{\!*}$ with functions constructed from SU(2) representation theory.  Each face $f$ of the 2-complex is assigned a spin $j_f$ labeling a SU(2) unitary irreducible representation $D^{j_f}$, while each edge $e$ is assigned an intertwiner $i_e$---an invariant map between tensor products of spin representations.  The combinatorial structure of a 2-complex $\Delta^{\!*}$, which is dual to a four dimensional triangulation $\Delta$ is such that each edge constitutes four faces, while each vertex is comprised of five edges and ten faces.  Each vertex is dual to a 4-simplex, with its edges dual to tetrahedra and its faces are dual to triangles.

The amplitude associated with a 2-complex $\Delta^{\!*}$ is obtained by summing over all possible assignments of bulk spin representation labels and intertwiner labels of products of local amplitudes. The amplitudes generally take the form
\begin{equation}\label{transitionAmp}
A_{\Delta} (\{ j_{\ell} \},\{ \intw_{n} \}) = \sum_{ \{j_{f_i} \} } \sum_{ \{i_{e_i} \} } \,  \prod_{f} A_{f}(j_f) \,  \prod_{e} A_{e}(i_e) \,  \prod_v   A_v (j_f,i_e) 
\end{equation}
where $A_f(j_f) = 2j_f+1$ represents the face amplitude \cite{Bianchi:2010fj} and $ A_e(i_e)$ representing the edge amplitude is chosen  to be dimension of the intertwiner or inverse of the norm of the intertwiner  \cite{Livine:2024hhc}. In SU(2) BF model, the vertex amplitude $A_v$ is given by the $\fifJ$-symbol (of the first kind) \cite{Ooguri:1992eb,Barrett:1997gw}, defined in terms of Wigner $\sixJ$-symbols (\texttt{wigner6j}) as follows:

\renewcommand*{\arraystretch}{1.3}
\begin{eqnarray}\label{15jsymbol} 
\fifJ (j_{ab}; {\red i_e}) &=& \includegraphics[valign=c,scale=0.95]{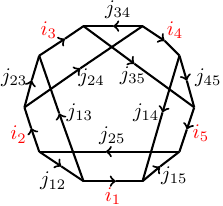} = \begin{Bmatrix}  \red i_1 & j_{13} & \red i_3 & j_{35} &  \red i_5 \\ j_{12} & j_{23} & j_{34} & j_{45} & j_{15}\\  j_{25} &  \red i_2 & j_{24} &  \red i_4 & j_{14}   \end{Bmatrix}    \nn \\
&=&  (-1)^{  \sum_m {\red i_m }+ \sum_{n<m} j_{mn}  }  \sum_{x} (2x+1) \, \begin{Bmatrix}  \red  i_1 & j_{25} & x \\   \red i_2 & j_{13} &  j_{12}   \end{Bmatrix}   \begin{Bmatrix}  \red  i_2 & j_{13} & x \\  \red  i_3 & j_{24} &  j_{23}   \end{Bmatrix}  \begin{Bmatrix}   \red i_3 & j_{24} & x \\   \red i_4 & j_{35} &  j_{34}   \end{Bmatrix} \times \nn \q \\
&& \q \q \q  \q \q \q  \quad \quad \times \begin{Bmatrix}  \red  i_4 & j_{35} & x \\   \red i_5 & j_{14} &  j_{45}   \end{Bmatrix}  \begin{Bmatrix}   \red i_5 & j_{14} & x \\   \red i_1 & j_{25} &  j_{15}   \end{Bmatrix}  
\end{eqnarray}
where the summation variable $x$, referred to as the virtual spin, ranges from  $\max(  |i_1 - j_{25} |,   | i_2 - j_{13}|  , |i_3-j_{24}| , |i_4-j_{24}| , |i_5-j_{24}| ) $  to $\min(i_1+j_{25}, i_2+j_{13},i_3+j_{24},i_4+j_{35},i_5+j_{14} )$. We have followed the conventions and orientations used in \cite{yutsis1962mathematical}  for the definition of the $\fifJ$-symbol of the first kind in Equation \eqref{15jsymbol}.  The intertwiners\footnote{The intertwiner labels are highlighted in red color in Equation \eqref{15jsymbol}.} assigned to the edges are denoted by $i_a$, and the spins $j_{ab}$ are associated with the faces dual to the  triangles. 

The spin foam amplitudes \eqref{transitionAmp} can be represented in a  coherent basis by introducing coherent states for the boundary links \cite{Livine:2007vk,Freidel:2007py}. Each intertwiner of a boundary node is represented in the coherent state basis by  group averaging the tensor product of the four coherent states associated with the node. A coherent state associated with a boundary link is characterized by a spin $j_{ab}$ and a unit normal vector ${\bf n}_{ab} \in S^2$ on the 2-sphere.  Considering a single vertex $v$ with five boundary edges, the vertex amplitude in the coherent basis is given by the integral expression

\begin{equation}\label{su2cohvertex}
A_v( j  ,  {\bf n} ) =  (-1)^\chi \int_{\rm SU(2)} \left( \prod_{a=1}^4 \d G_a \right)\,  \prod_{1\leq a<b \leq 5} \langle   j_{ab},  {\bf n}_{ba}   \triangleleft  {\cal J}\,|\, G_{a}^\dagger \, G_b \,|\, j_{ab}, {\bf n}_{ba}  \rangle
\end{equation}
where $G_a$ is a SU(2) group element in the fundamental representation, ${\blue \cal J} : \mathbb C^2 \rightarrow \mathbb C^2$ is an anti-linear map defined by  $(z_0, z_1) \mapsto (-\bar z_1, \bar z_0)$,  inducing a real structure on $S^2$, and $\langle \,,\rangle$ denote the invariant inner product on the spin $j_{ab}$ representation. One of the group integrals associated with the five edges in the definition of the coherent vertex amplitude \eqref{su2cohvertex} is redundant and is therefore gauge fixed to identity. The sign factor $(-1)^\chi$ is determined by the graphical calculus and orientations relating the spin network diagram.

The coherent states can also be introduced for every bulk face to enable a full coherent representation of the spin foam amplitude. In the integral representation of the amplitude  \eqref{transitionAmp}, each vertex amplitude is given by the expression \eqref{su2cohvertex} through an insertion of a resolution of identity of each bulk spin $j$ in terms of coherent states, given by 
\begin{equation}
\mathds 1_j = (2j+1) \int_{\rm SU(2)/U(1)} \d {\bf n}  \,\, |j,{\bf n} \rangle \langle j,{\bf n}  | \,\, ,
\end{equation}
As a result, the spin foam amplitude for a generic 2-complex is expressed as products of coherent vertex amplitudes \eqref{su2cohvertex}. This procedure, however, introduces numerous integration variables for the normal vectors associated with the bulk spins in addition to the group integrals.  The high dimensionality of the integration variables involved makes explicit evaluation of the coherent spin foam amplitudes challenging.  Even for the a single vertex, where all edges are boundary (hence no integration of the normal vectors), there are  a total of $12$ integration variables for the four SU(2) group integrations. The high dimensionality, combined with the highly oscillatory nature of the integrand, renders explicit numerical integrations inefficient due to slow convergence. Conversely, the integral representation is suitable for a stationary phase approximation.  

Typically, the group integrals \eqref{su2cohvertex} can be performed analytically. Through the Peter--Weyl theorem, functions of SU(2) can be decomposed into linear combinations of functions of the spin network basis labeled by spins and intertwiners \cite{Ashtekar:1996eg}. This decomposition generally allows the group integrations to be performed, resulting in sums of intertwiner variables\footnote{The group integrations associated with the bulk edges are replaced by sums over bulk intertwiner variables in the spin network representation.}, as already expressed in the amplitude \eqref{transitionAmp}. The coherent vertex amplitude in the spin network basis, therefore, results in a sum over intertwiner labels given by

\begin{eqnarray}\label{eqn:cohvertex}
A_v( j  ,  {\bf n} ) = (-1)^\chi \sum_{i_1, \cdots ,i_5} \, \, \fifJ ( j_{12}, \ldots ,j_{45}; i_1,\ldots,i_5)    \prod_{e=1}^5  d_{i_e} {\cohj}_{i_e}(j,{\bf n} )   
\end{eqnarray}
where ${\blue d_{i}} := 2i+1$ is the dimension factor of the intertwiner label $i$. The coherent $\fourJ$-symbol\footnote{A graphical notation of the coherent-$\fourJ$ symbol (see reference \cite{Dona:2017dvf}) comes with an orientation for each spin. Flipping an orientation of spin introduces a phase factor for the spin. The anti-linear map $\cal J$ acting on the normal vectors in the definition of the coherent vertex amplitude \eqref{su2cohvertex} leads to the choice of the orientation of the $\fifJ$-symbol and coherent-$\fourJ$ symbols employed in Equation \eqref{eqn:cohvertex}. }   $\blue {\cohj}_{i_e}$ associated with the boundary edge $e$ is explicitly given in terms of the Wigner $\{\rm 3j\}$-symbols (\texttt{wigner3j}) by 

\begin{eqnarray}\label{coh4vectors}
\hspace{-15pt} {\cohj}_{\red \bm i} (j,{\bf n})  = \hspace{-10pt}\sum_{m_1, \cdots, m_4} (-1)^{{\red \bm i}+m_1+m_2} \begin{pmatrix} j_1 & j_2 & \red \bm i \\ m_1 & m_2 & -m_1-m_2  \end{pmatrix}  \begin{pmatrix} \red \bm i & j_3 & j_4 \\ m_1 +m_2 & m_3 & m_4  \end{pmatrix}  \prod_{k=1}^4 D^{j_k}_{j_k , m_k}({\bf n}_k)  .
\end{eqnarray}
Here, \( D^{j}_{j , m}({\bf n}) = \langle j,m|{\bf n} | j,j\rangle \) are the coefficients of the Wigner-$D$ matrix (\texttt{wignerDjm}) in the highest weight spin basis for the SU(2) group element representation of the normal vector $\bf n$. The computation of the coherent symbols \eqref{coh4vectors} has been optimized in our current implementation, as available on the \href{https://github.com/Seth-Kurankyi/su2bf-TNAlgo}{GitHub repository} \cite{AsanteTN2024}. This implementation shows significant improvements over previous methods employed in \cite{Asante:2022lnp}. The optimization is achieved by caching the \texttt{wignerDjm} functions since they are independent of the intertwiner index $i$ in Equation \eqref{coh4vectors}. This approach allows for an efficient storage, and scalability of these coherent vectors.

In summary, using coherent states as boundary data gives the  coherent representation of the SU(2) BF spin foam amplitude for a 2-complex expressed in the intertwiner basis as 

\begin{equation}\label{Coh_amplitude}
\hspace{-12pt} A_{\Delta} ( \{ j_{\ell} \}, \{ {\bf n}_\ell \}) = (-1)^{\chi} \sum_{ \{j_{f_i} \} } \sum_{ \{i_{e} \} } \,  \prod_{f}  \,d_{j_{f}}  \prod_{e} \,\, d_{i_{e}} \,  \prod_v  \fifJ (j_f,i_e) \prod_{{n}} {\cohj}_{i_n} (j,{\bf n}) ,
\end{equation}
where $n$ represent the boundary edges (nodes) and $\chi$ depends only on the boundary spin representation labels.

\section{Spin Foam Amplitudes as Tensor Network Contractions}\label{sec:spintensors}

In both the spin-network basis and the coherent representation of the spin foam amplitudes, each vertex amplitude is given by the $\fifJ$-symbol, which is a function of ten spin labels $j_f$ associated with its faces and five intertwiner labels $i_e$ associated with its edges. By treating the spin representation labels as parameters, each $\fifJ$-symbol represent a 5-valent tensor (referred to as {\blue $\fifJ$-tensor} for short) with the five intertwiners as its indices. The $\fifJ$-tensor is depicted as:

\begin{equation}\label{fifJtensor} 
\fifJ^{({ J}_{ab})}_{\red i_1 i_2 i_3 i_4 i_5} \equiv \begin{picture}(22,25)
\put(14,2){\includegraphics[scale=0.38,valign=c]{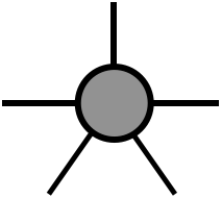} }
\put(30,28){$\red  i_1$}  \put(57,1){$\red  i_2$ \q .}  \put(47,-19){$\red  i_3$}  \put(15,-19){$\red  i_4$}  \put(2,1){$\red  i_5$}
\end{picture}  \vspace{10pt}  \q \q  \vspace{0.6em}
\end{equation}
All associated spins are summarized into the parameter $\blue J_{ab}$. The sum over the bulk and boundary intertwiners in the coherent amplitude \eqref{Coh_amplitude} results in a contraction between the vertex amplitudes and products of the dimension factors and coherent boundary intertwiners. For a generic 2-complex, the dimension and phase factors can always be absorbed in either the vertex amplitudes or the boundary coherent $\fourJ$-symbols. Consequently, each bulk intertwiner label is contracted between a pair of vertex amplitudes, while each boundary intertwiner is contracted between a vertex amplitude and a coherent $\fourJ$-symbol.

The product of the dimension factor $d_i$ and the coherent $\fourJ$-symbol associated with each boundary edge/node in the amplitude \eqref{eqn:cohvertex} can be represented as a 1-valent tensor (or a vector) indexed by an intertwiner label, and depicted as 
\begin{equation}\label{cohIntwT}
 d_{\red i_1}  \cohj_{\red i_1}(j,{\bf n}_1)  \equiv \, \includegraphics[scale=0.4,valign=c]{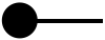}  \, {\red i_1} . 
\end{equation}  
This combination is referred to as the {\blue coherent $\fourJ$-vector}.  Again, all the spin labels and unit normal vectors associated with the coherent-$\fourJ$ vector are treated as parameters.  Thus, the sum over the intertwiner labels in the SU(2) coherent vertex amplitude \eqref{eqn:cohvertex} gives a contraction of the  $\fifJ$-tensor \eqref{fifJtensor} with five boundary coherent $\fourJ$-vectors  \eqref{cohIntwT}, represented by the tensor network notation, 

\begin{equation}\label{cohvertexA}
A_v( j  ,  {\bf n} ) \equiv    \,\, \includegraphics[scale=0.32,valign=c]{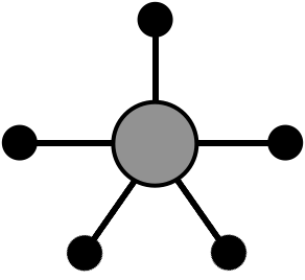}     \q .
\end{equation}

\begin{figure}[ht!]
\centering 
\includegraphics[scale=0.395,valign=c]{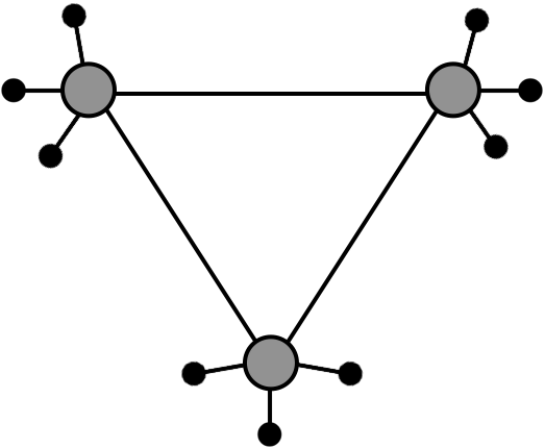}  \hspace{1.7cm}
\includegraphics[scale=0.395,valign=c]{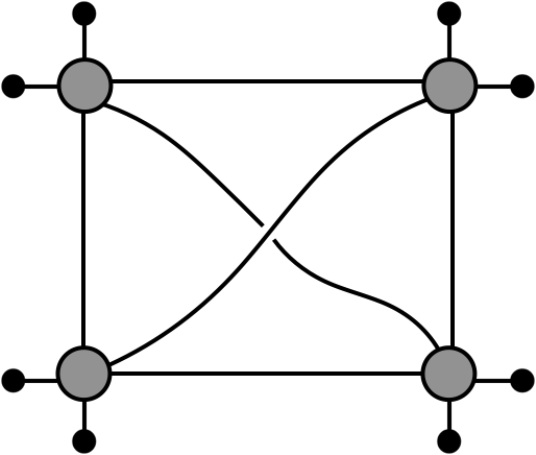}  \hspace{1.7cm}
\includegraphics[scale=0.375,valign=c]{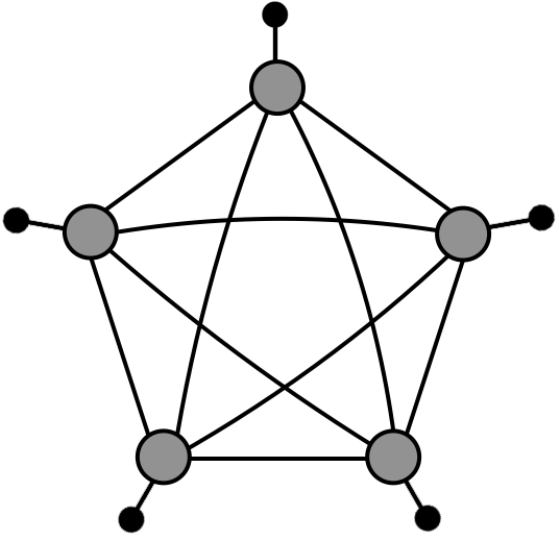}  
\caption{Tensor network diagrams for the contractions of intertwiners in coherent amplitudes for various 2-complexes. From left to right: The 2-complexes are dual to the $T3$ triangulation, $T4$ triangulation and $T5$ triangulation in the 3-3, 4-2 and 5-1 Pachner move configurations \cite{pachner1991pl} respectively. The $T3$ triangulation is often referred to as the Delta-3 ($\Delta_3$) triangulation.    }
\label{fig:complex}
\end{figure}

For a generic 2-complex composed of multiple vertices, the contraction of the intertwiners among vertex amplitudes and boundary coherent intertwiners can be diagrammatically represented by following the connectivity of the components of the corresponding 2-complex.  Figure \ref{fig:complex} illustrates examples of tensor network diagrams\footnote{ Details of the explicit terms or tensors involved in the contractions of the bulk intertwiners are omitted in the tensor diagrams of the 2-complexes  represented in Figure \ref{fig:complex}.} of coherent amplitudes for 2-complexes with multiple vertices. The vertex amplitudes are represented by the grey circles, while the coherent boundary edges are represented by the lines with small black circles at their tips. Each vertex is 5-valent in the intertwiner indices and is dual to a 4-simplex triangulation. 

~ 

\subsection{Spin Foam Numerical Computations: Existing Packages and Libraries } \label{sec:SFnumerics}

Numerical computations of spin foam models primarily focus on evaluating the coherent amplitudes, such as those represented by Equation \eqref{Coh_amplitude} in the spin-network representation. These amplitudes are expressed as sums over bulk spins and intertwiner labels. To compute the coherent amplitudes, coherent boundary data are assigned to the faces associated with each boundary edge of the 2-complex. The state-of-the-art libraries \texttt{sl2cfoam} \cite{Dona:2019dkf} and its improved version \texttt{sl2cfoam-next} \cite{Gozzini:2021kbt} are designed to perform explicit computations for Lorentzian EPRL coherent amplitudes. These libraries can also be used compute SU(2) BF coherent amplitudes. Furthermore, work in \cite{Asante:2022lnp} utilized a numerical code in the \texttt{Julia} programming language to compute SU(2) BF coherent vertex amplitudes. The Lorentzian EPRL and SU(2) BF models are closely related; the boundary data in both spin foam models are given by SU(2) coherent states. Additionally, the EPRL vertex amplitudes can be expressed as an infinite, but convergent, summation over the SU(2) $\fifJ$-symbols for auxiliary labels.

Various well-optimized libraries across different programming languages facilitate efficient computations of SU(2) invariants, such as Clebsch-Gordan coefficients, Wigner \(\threeJ\)-symbols, and \(\sixJ\)-symbols. For example, \texttt{sl2cfoam} and \texttt{sl2cfoam-next} make use of  \texttt{WIGXJPF} and \texttt{FASTWIGXJ} packages within the \texttt{C} programming language. Similarly, the Julia programming language offers efficient packages like \texttt{WignerSymbols.jl} for computing SU(2) invariants, including \texttt{wigner3j} and \texttt{wigner6j} functions. These packages enable efficient computation of the \(\fifJ\)-symbol \eqref{15jsymbol}, expressed as a sum of products of \(\sixJ\)-symbols. Currently, in existing numerical implementations, each vertex amplitude represented by the \(\fifJ\)-symbol of the first kind is computed as a 5-valent tensor in its intertwiner indices, as depicted in \eqref{fifJtensor}. The sum over the intertwiner indices is then performed as tensor contractions between the vertex amplitudes and the coherent vectors associated with the components of a  fixed 2-complex.

However, a significant drawback of computing the $\fifJ$-symbol as a tensor lies in the sheer computational complexity involved.  For larger representation labels, the size of the intertwiner labels grows, and with it, the size of the $\fifJ$-tensor increases exponentially, leading to severe scalability issues. For instance, given $\fifJ$-tensor with uniform spins $j_{ab}=j$, each intertwiner index is of dimension $d_j = (2j+1)$. Therefore, the $\fifJ$-tensor with equal spins has a total size of $ d_j^5$ elements. For a large spin $j \gg 1$, this exponential growth in tensor size highlights the memory-intensive nature for initializing and storing them.

Moreover, the process of contracting these high-valence tensors becomes increasingly computationally expensive for larger spins due to the exponential growth in the number of operations. As an example, consider contracting a \(\fifJ\)-tensor of uniforms spins with a vectors representing an intertwiner index, where the vector is also of dimension $d_j$. Such a contraction requires a computation cost of order $\blue {\cal O}(d_j^5)$. When many such contractions are required, the overall computation can become prohibitively time-consuming. This generally limits the practical use of direct numerical computation methods for many vertices or for large representation labels, which are often needed to explore properties of the amplitudes and the semi-classical regime of spin foam models.

These practical limitations necessitate the exploration of more efficient numerical techniques. In the next section, we shall explore new techniques inspired by tensor network methods to address these challenges.

\section{Efficient Tensor Network Methods for Spin Foam Amplitudes} \label{sec:newvertex}

In this section, we introduce a tensor network algorithm to enhance the efficiency of computing SU(2) invariants and SU(2) BF coherent amplitudes. This method involves rearranging the terms involved in the coherent vertex amplitudes to enable contractions between smaller tensors, particularly matrices, which have lower rank compared to the $\fifJ$-tensor. This strategy can also be applied to derive a tensor network algorithm for evaluating EPRL coherent vertex amplitudes. By decomposing high-valence tensors into networks of lower-valence tensors, the approach not only accelerates computations but also significantly reduces memory usage, as matrices require less memory allocations compared to the 5-valent $\fifJ$-tensor. Additionally, this method  can be employed to compute amplitudes associated with 2-complexes with multiple vertices. We will utilize tensor network notations to clarify and streamline the discussion. These notations will also help illustrate the decomposition process and the resulting computational benefits.

\subsection{Coherent Vertex Amplitudes as Matrix Contractions}\label{sec:matrix_contraction}

To illustrate this tensor network method, we first consider the coherent vertex amplitude defined in Equation \eqref{eqn:cohvertex}. In the expression for the $\fifJ$-symbol, each $\sixJ$-symbol depends on a pair of intertwiner labels. By treating the spins $j_{ab}$ and $x$ as fixed parameters, each $\sixJ$-symbol can be represented as a matrix with the pair of intertwiners as its open indices. We refer to such a matrix as the {\blue $\sixJ$-intertwiner matrix}, where its components are defined by:
\begin{equation}\label{sixJmatrix}  
w^{(j_{a},j_{b},j_{c},x)} _{ \red i_1 \,  i_2} :=  \begin{Bmatrix}  \red  i_1 & j_{a} & x \\   \red i_2 & j_{b} &  j_{c}   \end{Bmatrix} \equiv {\red i_1} \, \includegraphics[scale=0.3,valign=c]{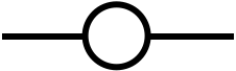}  \, {\red i_2} \,\, . \q \q   
\end{equation}
The graphical notation in Equation \eqref{sixJmatrix} represents the tensor notation for the $\sixJ$-intertwiner matrix $[w_{i_1 i_2}]$, where the two open legs signify the intertwiner indices. Given the range of values for $i_1$ and $i_2$, the $\sixJ$-intertwiner matrix can be efficiently computed for fixed values of spins $j_a, j_b, j_c$, and $x$ using existing libraries. For instance, the package \texttt{WignerSymbols.jl} within \texttt{Julia} language,  has an optimized function, \texttt{wigner6j}, to compute the components \eqref{sixJmatrix} of the $\sixJ$-intertwiner matrix. Additionally, known symmetries of the Wigner $\sixJ$-symbol \cite{Varshalovich:1988ifq} can be utilized to simplify the direct implementation of these matrices. As an example, $w_{i_1 i_2}$ is symmetric in its indices when either $j_a = j_b$ or $j_c = x$.

The expression of the $\fifJ$-symbol \eqref{15jsymbol} involves multiplication of $\sixJ$-symbols, as components of the $\sixJ$-intertwiner matrices, with no summation over the intertwiner labels shared by a pair of $\sixJ$-symbols. To handle such multiplications, we introduce the following notations to represent tensors whose components are expressed as products of smaller tensors without summations over their shared indices.  For example, consider a 3-valent tensor whose components are defined in terms of that of two $\sixJ$-intertwiner matrices by ${\blue T_{i_1 i_2 i_3} }:= w_{i_1 i_2} w_{i_2 i_3} $, with no summation over the index $i_2$. This tensor can be represented using the following notation:
\begin{equation} \label{tensornotation}
T_{i_1 i_2 i_3} = (i_1 \,\includegraphics[scale=0.2,valign=c]{intw_6j} \, i_2 )\,  \, (i_2 \includegraphics[scale=0.2,valign=c]{intw_6j} \, i_3)   \, \equiv \, 
\begin{picture}(62,16) 
\put(10,3) {\includegraphics[scale=0.26,valign=c]{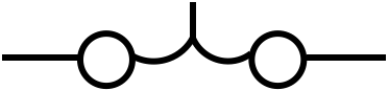} }
\put(0,3){$  i_1$}
\put(30,17){$  i_2$}
\put(62,3){$ i_3$,}
\end{picture}  
\end{equation}
where the open index $i_2$ connects the tensor notations of the two matrices, indicating it is a shared index. Also, a vector defined by $T_{i_1} := w_{i_1 i_1}$ (with $i_1$ not summed over), which is the diagonal vector of the $\sixJ$-intertwiner matrix can be represented by the notation $\includegraphics[scale=0.3,valign=c]{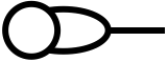}\,i_1$, where $i_1$ is its open index. These notations provide a more detailed structure of the tensor, showing the individual tensors involved.

Using the notation in \eqref{tensornotation}, the $\fifJ$-tensor expressed in terms of the products of components of $\sixJ$-matrices\footnote{The representation of the $\fifJ$-tensor using the $\sixJ$-matrices is similar to a \emph{Matrix Product States} (MPS) \cite{Orus:2013kga} tensor network representation.} can be represented as  
\[  \fifJ^{(J_{ab})}_{ i_1 i_2 i_3 i_4 i_5}  =  \sum_x d_x\, (-1)^{ i_1 + \cdots + i_5}\,     
\begin{picture}(22,33)
\put(12,7){\includegraphics[scale=0.24,rotate=180,valign=c]{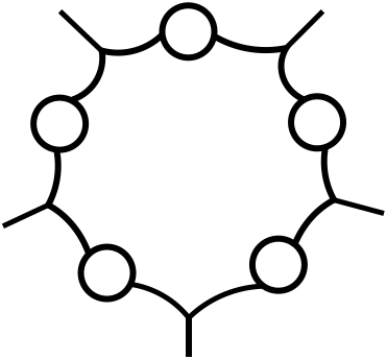}}     
\put(31,34){$ \scriptstyle i_1$}  \put(59,15){$  \scriptstyle i_2$}  \put(50,-15){$ \scriptstyle i_3$}  \put(12,-15){$ \scriptstyle i_4$}  \put(5,16){$ \scriptstyle i_5$}
 \end{picture} \q \q  \]
Again, note that there is no contraction or sum over all the intertwiner indices involved in the above expression. This form of the vertex amplitude is still a 5-valent tensor and is therefore, not yet computationally efficient. As discussed in section \ref{sec:SFnumerics}, computing and storing these 5-valent tensors for evaluating coherent vertex amplitudes demands significant memory allocations, particularly for large values of the spins.  To address this, we present a strategy which avoids the use these 5-valent $\fifJ$-tensors, and instead rely on matrix contractions.

The strategy is based on combining coherent vectors with the $\sixJ$-intertwiner matrices and reorganizing the sums involved in the coherent vertex amplitude \eqref{eqn:cohvertex}. Before performing either the sum over the intertwiner labels or the virtual spin $x$ in the definition of the $\fifJ$-symbol, consider the following combination

\begin{eqnarray}\label{coh6jmatrix}
f^{(J_{12},{\bf n}_1,\,x)}_{ \red i_1 \,  i_2} &:=& (-1)^{\red i_1} d_{\red i_1}  {\cohj }_{\red i_1}(j,{\bf n}_1)   \, w^{(j_{a},j_{b},j_{c},x)} _{ \red i_1 i_2}  \nn \\ 
&=& (-1)^{\red i_1} \left(  \includegraphics[scale=0.36,valign=c]{coh4j} \, {\red i_1 } \right)  \left(  {\red i_1 } \, \includegraphics[scale=0.25,valign=c]{intw_6j}  \, {\red i_2} \right) \\
&\equiv&  {\red i_1}\, \includegraphics[scale=0.34,valign=c]{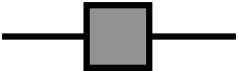}  \, {\red i_2}   . \nn
\end{eqnarray}
This combination defines a matrix $[f_{i_1i_2}]$, termed the {\blue coherent $\sixJ$-intertwiner matrix}, with the intertwiner labels $i_1,i_2$ as its open indices. The components of the coherent $\sixJ$-intertwiner matrix are given by products (without summation) of the $\sixJ$-matrix $w_{i_1 i_2}$ together with the coherent  $\fourJ$-vector ${\cohj }_{i_1}$, a phase factor $(-1)^{i_1}$, and the dimension for the intertwiner $i_1$.  All the spins $j_{ab}$ involved in the $\sixJ$-symbol and the coherent $\fourJ$-symbol  are collected into the parameter $J_{12}$.  Additionally, the normal vectors ${\bf n}_i$ and the virtual spin $x$ are considered as fixed parameters. The last line of Equation \eqref{coh6jmatrix} denotes the tensor notation for the coherent $\sixJ$-intertwiner matrix, where the two legs representing the intertwiner labels are its indices. 

The sums over the intertwiners in the coherent vertex amplitude \eqref{eqn:cohvertex} can thus be expressed as a contraction of the coherent $\sixJ$-intertwiner matrices \eqref{coh6jmatrix}. By first performing the sums over the intertwiner labels, the coherent vertex amplitude \eqref{eqn:cohvertex} can be re-expressed as a sum over the virtual spin $x$ of contractions of coherent $\sixJ$-intertwiner matrices as follows:

\begin{eqnarray}\label{cohvertex_new}
A_v( j  ,  {\bf n} )  &= & (-1)^\chi \sum_{x} d_x  \sum_{ i_1, \ldots ,i_5}   f^{(J_{12},{\bf n}_1,\,x)}_{  i_1 \,  i_2} \,  f^{(J_{23},{\bf n}_2,\,x)}_{  i_2 \,  i_3} \,  f^{(J_{34},{\bf n}_3,\,x)}_{  i_3 \, i_4} \,  f^{(J_{45},{\bf n}_4,\,x)}_{  i_4 \,  i_5} \,  f^{(J_{51},{\bf n}_5,\,x)}_{  i_5 \,  i_1}    \\
&\equiv&  (-1)^\chi \sum_{x} d_x \,\,  \includegraphics[scale=0.3,valign=c]{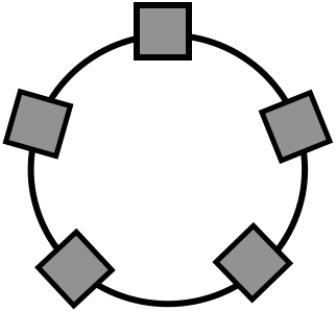}  \nn =: (-1)^\chi \sum_{x} d_x \,\, { F}(j,{\bf n},x) . \nn
\end{eqnarray}
Here, ${ F}(j,{\bf n},x) \in \mathbb C$ is defined as the trace of products/contraction of the coherent $\sixJ$-intertwiner matrices for a fixed set of the spins $j_{ab},x$ and normal vectors $\bf n$ associated with the components of the vertex $v$.  Its tensor network notation is depicted by the diagram in Equation \eqref{cohvertex_new}.

Thus, Equation \eqref{cohvertex_new} provides an alternate and more efficient method to evaluate the SU(2) BF coherent vertex amplitude \eqref{eqn:cohvertex} by summing sequences of the matrix trace ${ F}(j,{\bf n},x)$. This approach is computationally advantageous compared to the contraction involving the $\fifJ$-tensor in Equation \eqref{cohvertexA}. For instance, considering a vertex with uniform spins $j_{ab} = j$ for all $a,b$, where all intertwiners have dimension $d_j = (2j+1)$, the computational cost of evaluating a matrix trace ${ F}(j,{\bf n},x)$ is of order $\blue \mathcal{O}(4 \, d_j^3) $. This is significantly more efficient than the contraction with a $\fifJ$-tensor and a coherent vector (as discussed in Section \ref{sec:SFnumerics}). Consequently, this method not only reduces computational complexity but also enhances scalability, making it feasible to handle larger spin  representation labels.

\begin{algorithm}[htb!]
\small
\caption{{\bf : }SU(2) BF Coherent Vertex Amplitude}\label{alg:cohvtx}
\begin{flushleft}
\textbf{Input: }\\ \quad { \texttt{jays}: a 10-tuple of spin assignments $j_{12},j_{13}, \dots, j_{45}$. }  \\
\quad \,{ \texttt{noms}: sets of unit normal vectors ${\bf n}_1, {\bf n}_2, \dots, {\bf n}_5$ associated with the boundary edges. }  \\
\textbf{Output: }{value of the coherent vertex amplitude $A_v \in \mathbb C$. } \\
\end{flushleft}
\begin{algorithmic}[1]
\For{\emph{each edge label}  $k \in 1,\dots, 5$ } 
\State compute the $\sixJ$-intertwiner matrix $w^{(J_{ab},x)}_{i_k i_{k+1}} $ for fixed parameters $J_{ab},x$. \Comment{  $ w_{i_5 i_{5+1}}= w_{i_5 i_1}$}
\State compute the coherent $\fourJ$-vector $ \cohj_{ i_k}(j,{\bf n}_k).$ 
\State construct the  coherent $\sixJ$-intertwiner matrix $f_{ i_k i_{k+1}}$, as defined in \eqref{coh6jmatrix} .   \Comment{  $ f_{i_5 i_{5+1}}= f_{i_5 i_1}$} 
\EndFor
\State Compute  the trace of the products of the $\sixJ$-intertwiner matrices,  $F(j,{\bf n},x).$ 
\State Sum the product of the dimension factor $d_x$ and  the matrix trace $ F(j,{\bf n},x) $ over the range of values of the virtual spin $x$.
\State \Return{ the coherent amplitude $A_v \in \mathbb C $.}
\end{algorithmic}
\end{algorithm}

Algorithm \ref{alg:cohvtx} summarizes the necessary steps to compute SU(2) BF coherent vertex amplitudes $A_v$ as a sum over trace of matrices. It takes as input a set of spin assignments and unit normal vectors, and it returns the coherent vertex amplitude as a complex number. This tensor network algorithm offers an efficient method, significantly reducing memory usage and computational resources.   
Additionally, steps 2 through 4 within the \texttt{for loop} in Algorithm \ref{alg:cohvtx} can be parallelized to further optimize the computational efficiency.  Algorithm \ref{alg:cohvtx}  has been implemented in \texttt{Julia} programming language and is available on the \href{https://github.com/Seth-Kurankyi/su2bf-TNAlgo}{GitHub repository} \cite{AsanteTN2024}.

\subsection{ EPRL Coherent Vertex Amplitudes as Matrix Contractions }

Here, we adapt the tensor network algorithm to the case of Lorentzian EPRL (Engle-Pereira-Rovelli-Livine) coherent vertex amplitude. The EPRL  spin foam model  provides a framework for implementing a quantized version of gravity, rooted in the canonical loop quantum gravity (LQG) approach. The model is defined on a 2-complex $\Delta^{\!*}$, with each face assigned a unitary irreducible representation of the Lorentz group $\rm SL(2,\mathbb C)$. The key idea is to implement the simplicity constraints of Plebanski formulation of gravity at the quantum level using $\gamma$-simple unitary representations of $\rm SL(2,\mathbb C)$ \cite{Dona:2020xzv}. Here $\gamma$ is the Barbero-Immirzi parameter.

On a 2-complex $\Delta^{\!*}$ dual to a triangulated manifold, the EPRL amplitude takes a similar form as Equation \eqref{transitionAmp}.  The transition amplitudes are defined in terms of the $\gamma$-simple representations of $\rm SL(2,\mathbb C)$. We refer the reader to \cite{Engle:2007wy,Dona:2022yyn,Dona:2019dkf} for details on the definition of the transition amplitudes of the EPRL model.  The coherent representation of EPRL amplitudes make use SU(2) coherent states, labeled by $ \{ j_\ell, {\bf n}_\ell \}$, as boundary data. The coherent vertex amplitude is given by 
\begin{equation}\label{cohamp_EPRL}
 A_v^{\rm EPRL} (j_f, {\bf n})  = \sum_{l_f = j_f}^\infty \sum_{ \{ k_{e'} \} }    \sum_{ \{ i_e \} } \, \fifJ (i_1,j_f,l_f,k_{e'}) \, \prod_{e=1}^5 d_{i_e} \cohj_{i_e} (j,{\bf n})   \prod_{e'=2}^5 B_4^\gamma (j_f,l_f; i_{e'},k_{e'})    
\end{equation}
The infinite summation range over the `internal' spin labels $l_f$ is due to the non-compactness of the gauge group. 
The form of the vertex amplitude arises from using the Cartan decomposition \cite{Speziale:2016axj} of a $\rm SL(2,\mathbb C)$ group element $g$  and an integration measure given by  
\[  g  = u \cdot \exp\left( {\tfrac{r}{2} \sigma_3} \right) \cdot v^{-1} , \q \q \d g = \frac{1}{4\pi} \sinh^2r \,\, \d r \, \d u \, \d v , \]
where $u,v $ are $\rm SU(2)$ group elements, $\sigma_3$ is a Pauli matrix and $r \in [0, \infty) $ is the rapidity. The booster function $B_4^\gamma$ is defined as a one dimensional integral over the rapidity $r$ (again see \cite{Dona:2022yyn,Dona:2019dkf,Speziale:2016axj} for details and definition of the Booster function). $B_4^\gamma$ encodes how the quantum simplicity constraints are implemented on a 2-complex. Furthermore, the $\fifJ$-symbol is explicitly given in terms of the spin and interwiner labels by 
\begin{eqnarray}\label{15jsymbol2} 
\fifJ &=& \begin{Bmatrix}  \red i_1 & j_{13} & \red k_3 & l_{35} &  \red k_5 \\ j_{12} & l_{23} & l_{34} & l_{45} & j_{15}\\  l_{25} &  \red k_2 & l_{24} &  \red k_4 & j_{14}   \end{Bmatrix} . 
\end{eqnarray}

Each booster function depends on two intertwiner indices $i_e$ and $ k_e$, and hence can be represented as a matrix  whose components are given by 
\begin{equation}\label{booster}
b^{(J_2,\gamma)}_{\red i_2 \,k_2} :=  B_4^\gamma(j_{12},j_{23},j_{24},j_{25} ;\,  j_{12},l_{23},l_{24},l_{25} ; {\red i_2, k_2} )  \equiv   {\red i_2}\, \includegraphics[scale=0.38,valign=c]{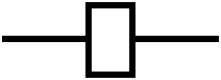}  \, {\red k_2}  
\end{equation}
The rectangular shaped diagram in Equation \eqref{booster} denotes the tensor notation for the booster functions or matrices. In the coherent amplitude \eqref{cohamp_EPRL}, the sum over the intertwiner variables $i_e, e \in \{ 2,\dots,5\}$ can easily be performed by contracting the booster functions with the coherent $\fourJ$-vectors as
\begin{equation} \label{boostervector}
\sum_{ \{ i_e\}  } d_{i_e} \,  \cohj_{i_e} (j,{\bf n})  \, b^{(J_e,\gamma)}_{i_e \, k_e}   =     \includegraphics[scale=0.38,valign=c]{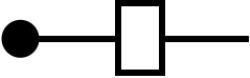}  \, { k_e}  
\end{equation}
We refer to this contraction as the {\blue coherent booster-vector}, where $k_e$ denotes its index. Thus, the EPRL coherent vertex amplitude  written as a contraction of the $\fifJ$-symbol with the coherent $\fourJ$ vector and booster coherent vector can be represented by the notation 

\begin{equation}\label{cohvertexEPRLA}
 A_v^{\rm EPRL} (j_f, {\bf n})  =  \sum_{l_f = j_f}^\infty  \,\, \includegraphics[scale=0.34,valign=c]{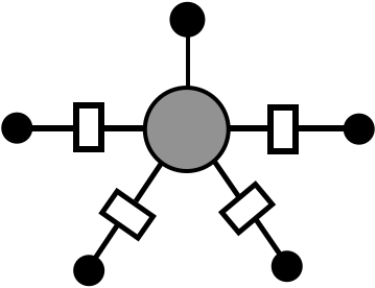}    .
\end{equation}
Notice that the contraction of the EPRL coherent vertex amplitude involves four coherent booster-vectors and one coherent $\fourJ$ vector. This is due to a gauge-fixing of one of the five non-compact $\rm SL(2,\mathbb C)$ group integrals associated to the edges in its integral representation, resulting in a finite vertex amplitude. The EPRL vertex amplitude is therefore similar in structure to that of SU(2) BF model. However, due to the summation over the internal spins $l_f$, the tensor contractions has to be repeated many times for each possible configuration of the spins. The summation over of internal spins can be performed using approximate schemes such as truncated sums using `shells' \cite{Dona:2018nev,Dona:2019dkf} or acceleration operators \cite{Dona:2022dxs,Dona:2023myv,Dittrich:2023rcr} for faster convergence.

To optimize the computation of the coherent vertex amplitude in the EPRL model, we adopt a similar strategy akin to that used in SU(2) BF case. The coherent vertex amplitude can similarly be rewritten as a contraction of matrices by reorganizing the $\rm SL(2,\mathbb C)$ invariant functions and the sums over the spins and intertwiners. To achieve this, we combine components of the $\sixJ$-intertwiner matrix and the coherent booster-vector to give:

\begin{eqnarray}\label{boscoh6jmatrix}
h^{(J_{23},\gamma,{\bf n}_2,\,x)}_{ \red k_2 \,  k_3} &:=& (-1)^{\red k_2} \sum_{i_2} d_{i_2}  \,  \cohj_{i_2} (j,{\bf n}_2)  \, b^{(J_2,\gamma)}_{i_2 \,{\red k_2} } \, w^{(j_{13},l_{24},l_{23},x)} _{ \red k_2 k_3}  \nn \\ 
&=&  (-1)^{\red k_2}  ( \includegraphics[scale=0.37,valign=c]{booster_vec} {\, \red k_2}) ( {\red k_2 \, \includegraphics[scale=0.25,valign=c]{intw_6j} \, k_3} )  \\
&\equiv&  {\red k_2}\, \includegraphics[scale=0.35,valign=c]{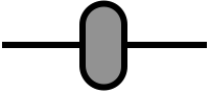}  \, {\red k_3}  \nn
\end{eqnarray}
This resulting matrix is referred to as the {\blue coherent booster-matrix}. The last line represents its tensor notation.  Using these coherent booster-matrices, the sum over the intertwiner variables $i_1, k_e$ in the coherent amplitude can be represented as a (trace of) matrix contractions. The coherent amplitude is therefore given by 

\begin{eqnarray}\label{EPRLtn}
\hspace{-9pt} A_v^{\rm EPRL} (j_f, {\bf n})   &= & (-1)^\chi \!\!\!  \sum_{l_{ab} = j_{ab} }^\infty \!\! \sum_{ x }  d_x \!\!\! \sum_{ i_1, k_2, \ldots ,k_5}   f^{(J_{12},{\bf n}_1,\,x)}_{  i_1 \,  k_2} \,  h^{(J_{23},\gamma,{\bf n}_2,\,x)}_{  k_2 \,  k_3} \,  h^{(J_{34},\gamma,{\bf n}_3,\,x)}_{  k_3 \,  k_4} \,  h^{(J_{45},\gamma,{\bf n}_4,\,x)}_{  k_4 \,  k_5} \,  h^{(J_{45},\gamma,{\bf n}_3,\,x)}_{  k_5 \,  i_1}    \nn \\
 &=&  (-1)^\chi \sum_{l_{ab} = j_{ab} }^\infty \sum_{ x }   d_x  \,  \includegraphics[scale=0.34,valign=c]{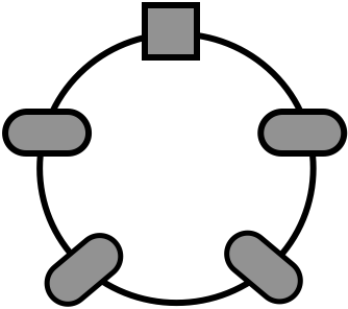}   
\end{eqnarray}
This formulation of the coherent vertex amplitude for the EPRL spin foam model \eqref{EPRLtn} offers a more efficient alternative, and enables a faster and more scalable computations.

\begin{algorithm}[htb!]
\small
\caption{{ EPRL Coherent Vertex Amplitude}}\label{alg:EPRLcohvtx}
\algnotext{EndIf}
\algnotext{EndFor}
\algnotext{EndWhile}
\begin{flushleft}
\textbf{Input: }\\ \quad { \texttt{jays}: a 10-tuple of spin assignments $j_{12},j_{13}, \dots, j_{45}$ }  \\
\quad \,{ \texttt{noms}: sets of normal vectors ${\bf n}_1, {\bf n}_2, \dots, {\bf n}_5$ associated with the boundary edges }  \\
\textbf{Output: }{Value of the coherent vertex amplitude $A_v^{\rm EPRL} \in \mathbb C$. } \\
\end{flushleft}
\begin{algorithmic}[1]
\State Compute the $\sixJ$-matrices $w^{(J_{12},x)}_{i_1 k_2}, \dots, w^{(J_{51},x)}_{k_5 i_1} $ for fixed values of spins $j_{ab}, l_{ab},x$
\State Compute the booster-coherent matrices  $b_{i_2 k_2}, \dots, b_{i_5 k_5}  $ for the four boundary edges 
\State Compute the coherent $\fourJ$-vectors $ {\cohj }_{ i_a}(j,{\bf n}_a)  $ for all five boundary edges $a = 1,\dots,5$
\State Contract the booster coherent matrices with the coherent vectors into the booster coherent vectors as given in Equation \eqref{boostervector}
\State Construct the four booster coherent matrices $ h_{ k_2 k_3} , \dots, h_{k_5 i_1} $ and the coherent matrix $f_{ i_1 k_2} $ for fixed spins $j_{ab},l_{ab},x$
\State Contract the coherent matrix $f_{ i_1 k_2} $ and four booster matrices as trace of matrix contractions 
\State Sum over all values of the virtual spin $x$ of product of the dimension factor $d_x$ and the {matrix trace}
\State Perform the unbounded sums $l_{ab} = j_{ab} $ to $\infty$ using  any method of choice. 
\State \Return{ The coherent amplitude $A_v^{\rm EPRL} \in \mathbb C $}
\end{algorithmic}
\end{algorithm}

Algorithm \ref{alg:EPRLcohvtx} gives a summary of the procedure for computing the EPRL coherent vertex amplitude as a sum of sequences of matrix contractions. This algorithm leverages the matrix formulation and optimized computation techniques to efficiently evaluate the amplitude for a given set of boundary spins and normal vectors. The explicit numerical implementation of Algorithm \ref{alg:EPRLcohvtx} in \texttt{Julia} language is left to future work. This tensor network algorithm can also be applied within the \texttt{sl2cfoam-next} library where the computation of the booster functions have already been implemented.

\subsection{Partial-Coherent Vertex Amplitudes}\label{sec:partial_vertex}

The coherent amplitude $A_v$ \eqref{cohvertexA} is associated with a 2-complex $v$, characterized by a single vertex with five boundary edges which is dual to a 4-simplex triangulation. To study dynamics of quantum geometries, it is essential to analyze spin foam amplitudes on a 2-complex with more than one vertex. In a generic 2-complex, with multiple vertices (see examples in Figure \ref{fig:complex}), a pair of neighbouring vertices are connected by one or more (bulk) edges. In the dual triangulation of a generic 2-complex, any 4-simplex that contains a boundary tetrahedron is dual to a vertex exhibiting a combination of bulk and boundary edges. Such a vertex is referred to as  a \emph{boundary vertex}, for conciseness. A \emph{bulk vertex}, on the other hand, refers to a vertex with each of its five edges dual to bulk tetrahedron.

\begin{figure}[ht!]
\centering 
\includegraphics[scale=0.385,valign=c]{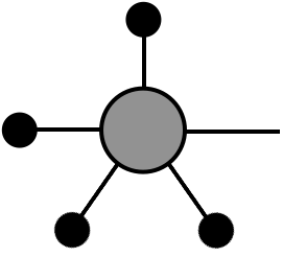}  \hspace{1.7cm}
\includegraphics[scale=0.385,valign=c]{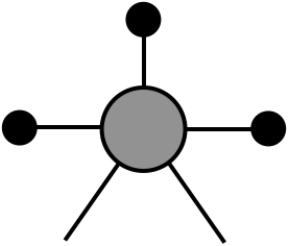}  \hspace{1.7cm}
\includegraphics[scale=0.365,valign=c]{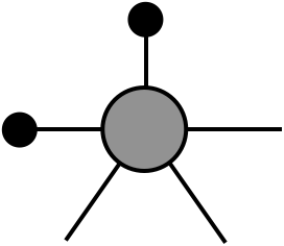}  \hspace{1.7cm}
\includegraphics[scale=0.365,valign=c]{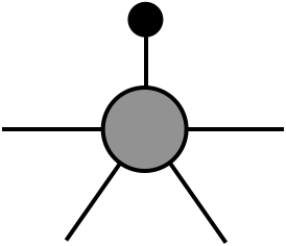}  
\caption{Tensor network notations for partial coherent vertex amplitudes. The open legs represent bulk edges with intertwiner indices while the legs with black circular dots are boundary edges with coherent data.  }
\label{fig:partialvertex}
\end{figure}

In the coherent amplitude \eqref{Coh_amplitude}, coherent data $\{ j_{\ell}, {\bf n}_\ell \}$ are assigned to the faces of boundary edges, while bulk edges are assigned spin and intertwiner data $\{ j_{f} , i_{e} \}$ in the spin-network basis. A boundary vertex endowed with both coherent data for the boundary edges and intertwiner data for the bulk edges  is termed as a {\blue partial-coherent vertex}. We will describe the amplitudes associated with these partial coherent vertices, focusing on the SU(2) BF spin foam model. Figure \ref{fig:partialvertex} illustrates the tensor network notations for all partial coherent vertex amplitudes, each associated with a boundary vertex.  

\noindent The tensor network notations in \eqref{cohvertexA} and \eqref{fifJtensor}  represent a coherent vertex with only boundary edges and a vertex with only bulk edges respectively. The amplitude associated with a bulk vertex is simply given by the $\fifJ$-symbol.

Each of the four partial-coherent vertices in Figure \ref{fig:partialvertex} can be computed as a contraction of the $\fifJ$-tensor with a number of coherent-$\fourJ$ vectors \eqref{cohIntwT}, resulting in a tensor with its open legs indexed by the intertwiner labels for the the bulk edge(s) attached to the vertex. To avoid creating the $\fifJ$-tensor, we shall make use of components of the $\sixJ$-intertwiner matrices and the coherent $\sixJ$-intertwiner matrices to rewrite these partial coherent vertices.  The sums over the virtual spin and the intertwiner labels involved in the partial coherent vertices can also be reorganized such that the resulting tensors are expressed as contractions between the smaller tensors using the matrices defined in Equations \eqref{sixJmatrix} and \eqref{coh6jmatrix}.  We shall also make use of the tensor notations described in Equation \eqref{tensornotation} for the expressions.

Consider the first partial-coherent vertex with one bulk edge and four boundary edges associated with coherent data. It has one open index corresponding to the bulk intertwiner label, and hence represents a vector. Using the expressions and notations for the matrices in \eqref{sixJmatrix} and \eqref{coh6jmatrix}, the 1-valent partial-coherent vertex amplitude can be expressed as 
\begin{eqnarray}\label{partcohvertex1}
\includegraphics[scale=0.32,valign=c]{pat_vertex41} \,\, i_1   &= & (-1)^\chi 
 \sum_{i_2, \cdots ,i_5} \, \, \fifJ ( j_{12}, \ldots ,j_{45}; i_1,\ldots,i_5)    \prod_{k=2}^5  d_{i_k} {\cohj}_{i_k}(j,{\bf n}_k ) \nn \\
&= &  (-1)^\chi \sum_{x} d_x  \sum_{ i_2, \ldots ,i_5}  (-1)^{i_1}\,  w^{(J_{12},\,x)}_{  i_1 \,  i_2} \,  f^{(J_{23},{\bf n}_2,\,x)}_{  i_2 \,  i_3} \,  f^{(J_{34},{\bf n}_3,\,x)}_{  i_3 \, i_4} \,  f^{(J_{45},{\bf n}_4,\,x)}_{  i_4 \,  i_5} \,  f^{(J_{51},{\bf n}_5,\,x)}_{  i_5 \,  i_1} \q \nn \\
& = &  (-1)^\chi  \sum_{x}  d_x \, (-1)^{ i_1} \, \includegraphics[scale=0.28,valign=c,rotate=10]{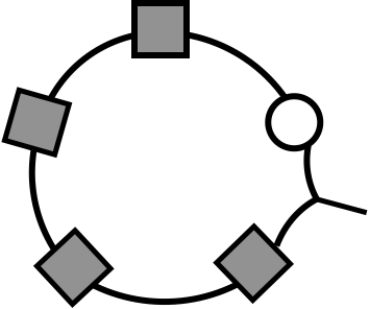} i_1 \, \, .
\end{eqnarray}
In the second line of  Equation \eqref{partcohvertex1}, the four coherent $\fourJ$-vectors associated with the boundary edges are combined with the $\sixJ$-intertwiner matrices into coherent $\sixJ$-intertwiner matrices $f_{ii'}$.  Just as in the case of the coherent vertex amplitude \eqref{cohvertex_new}, the sum over the intertwiner labels of the boundary edges is first performed, followed by the summation over the virtual spin $x$. Since the intertwiner label $i_1$ associated with the bulk edge is not summed over, the contraction results in a vector indexed by $i_1$. The diagram in the last line of Equation \eqref{partcohvertex1} represents the tensor notation of the resulting vector for fixed spins $j_{ab}$ and $x$.

The second partial-coherent vertex amplitude as shown in Figure \ref{fig:partialvertex} involves two bulk edges and three boundary edges with coherent data. It, therefore, represents a matrix with its indices corresponding to the intertwiner labels of the two bulk edges. The initial expression of this vertex amplitude involves a contraction of a $\fifJ$-tensor with three coherent $\fourJ$-vectors associated with the boundary edges. This can also be rewritten as contraction of the $\sixJ$-intertwiner matrices and coherent $\sixJ$-intertwiner matrices after rearranging the sums over the intertwiner labels and virtual spin as follows:

\begin{eqnarray}\label{partcohvertex2}
\hspace{-12pt}
\begin{picture}(50,20)
\put(0,0){\includegraphics[scale=0.32,valign=c]{pat_vertex32}  }
\put(1,-23) {$ i_1$}
\put(35,-23) {$ i_2$}
\end{picture}   &= & (-1)^\chi 
 \sum_{i_3, i_4 ,i_5} \, \, \fifJ ( j_{12}, \ldots ,j_{45}; i_1,\ldots,i_5)    \prod_{k=3}^5  d_{i_k} {\cohj}_{i_k}(j,{\bf n}_k ) \nn \\
 \nn \\
 &= & (-1)^\chi \sum_{x} d_x  \, (-1)^{i_1 + i_2}\, w^{(J_{12},\,x)}_{  i_1 \,  i_2}  \sum_{ i_3, i_4 ,i_5}      w^{(J_{23},\,x)}_{  i_2 \,  i_3} \,  f^{(J_{34},{\bf n}_3,\,x)}_{  i_3 \, i_4} \,  f^{(J_{45},{\bf n}_4,\,x)}_{  i_4 \,  i_5} \,  f^{(J_{51},{\bf n}_5,\,x)}_{  i_5 \,  i_1} \q \nn \\
&= & (-1)^\chi \sum_{x} d_x  \,  (-1)^{i_1 + i_2}\,
\begin{picture}(50,35)
\put(-5,15){ \includegraphics[scale=0.28,valign=c,rotate=-36]{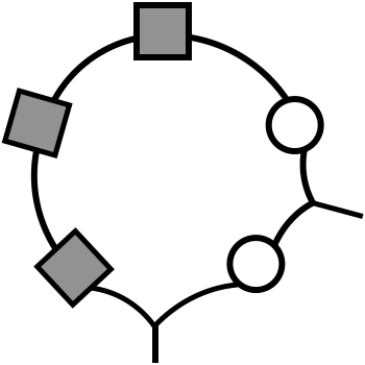} }
\put(9,-25) {$ i_1$}
\put(48,-25) {$ i_2$}
\end{picture}   \q  \q  \\ \nn
\end{eqnarray}
The diagram in the last line of Equation \eqref{partcohvertex2} represent the product of the components of the matrix $w_{i_1 i_2}$ with the contracted matrices involved in the sums over the boundary intertwiner labels $i_3,i_4,i_5$. These partial-coherent vertex amplitudes are required, for example, in computing the amplitude for a 2-complex with one bulk face dual to a $\Delta_3$ triangulation (see Figure \ref{fig:complex}).  Also, a generalization to a 2-complex with one bulk face and $n$ vertices (dual to a triangulation denoted $\Delta_n$, for $n$ finite) implements these 2-valent parital-coherent vertex amplitudes.  
Permuting the boundary and bulk intertwiner labels changes the order of the contracted matrices. However, the resulting partial-coherent amplitude can still be expressed in a similar form.

The third partial-coherent vertex amplitude features three bulk edges and two boundary edges assigned with coherent data. It is a 3-valent tensor with its indices corresponding to the intertwiner labels of the bulk edges. This vertex amplitude is initially expressed as a contraction of a $\fifJ$-tensor with two coherent $\fourJ$-vectors associated with the boundary edges. The coherent $\fourJ$-vectors and the $\sixJ$-intertwiner matrices can again be combined into coherent $\sixJ$-intertwiner matrices $f_{ii'}$. The sum over the intertwiner labels gives contraction of matrices and the sum over virtual spin result in the following expression

\begin{eqnarray}\label{patcoh3}
\\
\begin{picture}(230,23)
\put(0,25){\includegraphics[scale=0.33,valign=c]{pat_vertex23}  }
\put(1,0) {$ i_1$}
\put(37,0) {$ i_2$}
\put(49,25) {$ i_3$}
\put(62,25){$ \displaystyle =  (-1)^\chi  \sum_{x}  d_x \, (-1)^{ i_1+i_2+i_3} \, $ }
\put(183,11){\includegraphics[scale=0.28,valign=c,rotate=33]{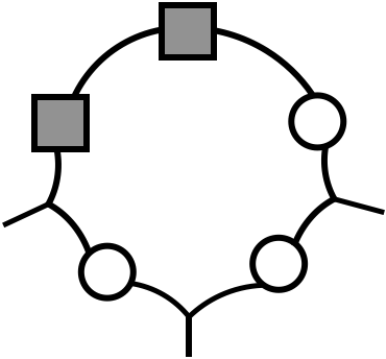} }
\put(193,0) {$ i_1$}
\put(232,0) {$ i_2$}
\put(244,37) {$ i_3$}
\end{picture}   \q \q  .\nn
\end{eqnarray}
The amplitude associated with each boundary vertex of the 2-complex dual to the $T4$ triangulation, as shown in Figure \ref{fig:complex}, is given by such a 3-valent partial-coherent vertex amplitude.

Lastly, the fourth partial-coherent vertex amplitude comprises four bulk edges and one boundary edge with coherent data. It represents a 4-valent tensor with its indices corresponding to the intertwiner labels of the bulk edges. After contracting the matrices involved  in the sum over the boundary intertwiner, the components of this 4-valent vertex can be represented as 

\begin{eqnarray}
\\
\begin{picture}(230,23)
\put(0,25){\includegraphics[scale=0.33,valign=c]{pat_vertex14}  }
\put(-10,25) {$ i_1$}
\put(1,0) {$ i_2$}
\put(37,0) {$ i_3$}
\put(49,25) {$ i_4$}
\put(62,25){$ \displaystyle =  (-1)^\chi  \sum_{x}  d_x \, (-1)^{ i_1+i_2+i_3+i_4} \, $ }
\put(210,11){\includegraphics[scale=0.28,valign=c,rotate=33]{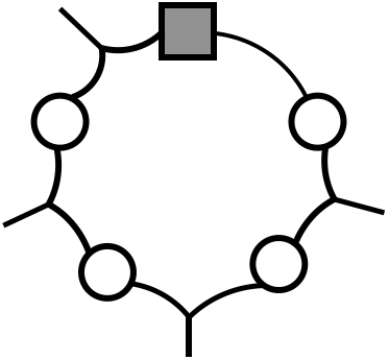} }
\put(209,36) {$ i_1$}
\put(221,0) {$ i_2$}
\put(258,0) {$ i_3$}
\put(272,37) {$ i_4$}
\end{picture}   \q \q \nn
\end{eqnarray}

In summary, each partial-coherent vertex amplitude in Figure \ref{fig:partialvertex} is characterized by the number of $\sixJ$-intertwiner matrices and coherent $\sixJ$-intertwiner matrices, corresponding to the number of bulk and boundary edges with coherent data, respectively. By reorganizing the sums over the virtual spin and intertwiner labels, the expressions for these vertices can be efficiently computed through matrix contractions, thereby enhancing memory efficiency in the computation of coherent SU(2) BF spin foam amplitudes. This new method of evaluating the partial-coherent amplitudes avoids the need to compute the $\fifJ$-symbol as a 5-valent tensor, further optimizing the computation of coherent amplitudes.  Although the tensor network representation of the 3-valent and 4-valent partial-coherent vertex amplitudes theoretically scales better, their performance is comparable to the computation using the $\fifJ$-tensor for small representation labels. These partial-coherent vertices, together with edge amplitudes \cite{Dona:2022dxs} are relevant for computing coherent amplitudes for generic 2-complexes with multiple boundary vertices.

\section{Results and Benchmarks}\label{sec:results}

Now, we proceed to present results of numerical experiments and benchmarks focusing on examples of SU(2) BF coherent vertex amplitudes. These vertex amplitudes are computed as sum of traces of coherent $\sixJ$-matrices according to the formula \eqref{cohvertex_new} and implemented in Algorithm \ref{alg:cohvtx}. To compute coherent vertex amplitudes, specific boundary data must be specified. The boundary data are given by a set of ten spins and twenty unit vectors, denoted as $\{ j_{ab}, {\bf n}_{ab} \}_{ 1\leq a , b \leq 5}$, satisfying $j_{ab} = j_{ba}$  and $a \neq b$ for $a,b \in \{1,\dots,5\}$. The spins correspond to the area of triangles of the 4-simplex, and the unit normal vectors correspond to the collection of face normal vectors of triangles in each tetrahedron. The geometric characterization of a vertex configuration depends on the conditions satisfied by its boundary data. Coherent amplitudes associated with vertices of similar geometric characterizations exhibit similar behavior, particularly in their asymptotic properties. In Appendix  \ref{sec:geoms}, we describe the conditions satisfied by certain subsets of \emph{twisted geometries}, referred to as vector geometries and Regge geometries. Furthermore,  Appendix \ref{sec:Asymptotics} provides the asymptotic formula of the coherent amplitudes for these geometries.
 
Here, we consider examples of vertices endowed with boundary data describing \emph{Regge} and \emph{vector geometries} and compute their coherent amplitudes using Algorithm \ref{alg:cohvtx}. By exploring these examples, we aim to illustrate the versatility and effectiveness of the tensor network methods, providing insights into the behavior of vertex amplitudes across different vertex configurations. Our benchmarks demonstrate the computational efficiency and advantages of using tensor network methods based on matrix contractions over traditional approaches that rely on contractions of the $\fifJ$-tensor.

\subsection{Coherent Amplitude for Equilateral Vertex Configuration}\label{sec:equi_vertex}

As our first example, we examine a vertex $v$ with Regge boundary data, corresponding to an equilateral 4-simplex triangulation. This configuration represents the simplest and most symmetric instance of a vertex. The \emph{equilateral vertex} has boundary data characterized by equal spins $j_{ab} = j$ for all  faces corresponding to the areas of the dual triangles.  Thus, $j$ characterizes the boundary scale. Each tetrahedron dual to an edge is also equilateral, thus for a fixed edge $a$, the unit normal vectors are explicitly given by 

\begin{equation}\label{equinormals}
\{ {\bf n}_{ab} \}_{1\leq b\leq 5, \,   \,b\neq a } =  \{ ( 1, 0, 0 ), \,\,\, (-\frac13,\frac{\sqrt{8}}{3} ,0), \,\,\, (-\frac13, -\frac{\sqrt{2}}{3}, \frac{\sqrt{6}}{3}), \,\,\, (-\frac13, -\frac{\sqrt{2}}{3}, -\frac{\sqrt{6}}{3}) \}  .
\end{equation}
These set of vectors correspond to the unit normals associated with the triangular faces of each equilateral tetrahedron. They can be rotated to form a \emph{twisted spike} configuration. Additionally, this set of boundary spins and normal vectors provides a consistent length geometry for an equilateral 4-simplex, with equal edge lengths given by $\ell=  j (\sqrt{4}/3)$. The external dihedral angles associated with the triangles are also all equal, given by $\theta =  \arccos(-\frac14)$. Appendix \ref{sec:geoms} provide further details for the equilateral vertex configuration. The boundary data are thus completely determined by the single boundary spin value $j$ along with the unit normal vectors \eqref{equinormals} associated with each tetrahedron. This highly symmetric configuration significantly simplifies the computation of its coherent amplitude.

Previous numerical studies for spin foam models \cite{Asante:2022lnp,Dona:2019dkf,Dona:2017dvf} have also considered the coherent amplitude for the equilateral vertex example. We will compare our results and benchmarks to those established in the earlier works, which mostly relied on contracting the $\fifJ$-tensor and coherent boundary vectors. These computations  require significant memory allocations and computational time, particularly for relatively large spins $(j\geq 50)$. For instance, the computation for the equilateral vertex amplitude at spin $\blue j = 50$, using the $\fifJ$-tensor and floating-point double precision, requires a computer with more than $\blue\rm 78.3GB$ of random-access-memory\footnote{ Moreover, it takes more memory usage to perform the contractions of the $\fifJ$-tensor with boundary coherent vectors.} (RAM). Such a computation is expected to take several days even on a high-performance computing cluster with enough memory resources. There may be extra memory costs from performing contractions. These computational limits restrict the practical exploration of spin foam coherent vertex amplitudes for large spins.

In contrast, the tensor network algorithm employed here, leverages the matrix contractions to substantially reduces both computational complexity and memory usage. For instance, it now takes approximately {\blue$ t \approx 1.05$ seconds} (see Table \ref{tab:benchmarks_equi} for more details) to compute the equilateral vertex amplitude for uniform spin $ j = 50$ using the improved algorithm on a consumer laptop\footnote{All the computations and benchmarks in this article were performed using the \texttt{Julia} programming language on a laptop equipped with Apple M2 Pro Chip and 16GB of RAM.}. This significant reduction in computational time underscores the efficiency gained by reorganizing the computations as contractions between smaller tensors (matrices), compared to the direct computation involving the $\fifJ$-tensor. Moreover, the memory-efficient nature of the new algorithm enables the exploration of coherent amplitudes at higher spin values with significantly reduced resource requirements.

\begin{figure}[ht!]
\centering 
\includegraphics[scale=0.47,valign=c]{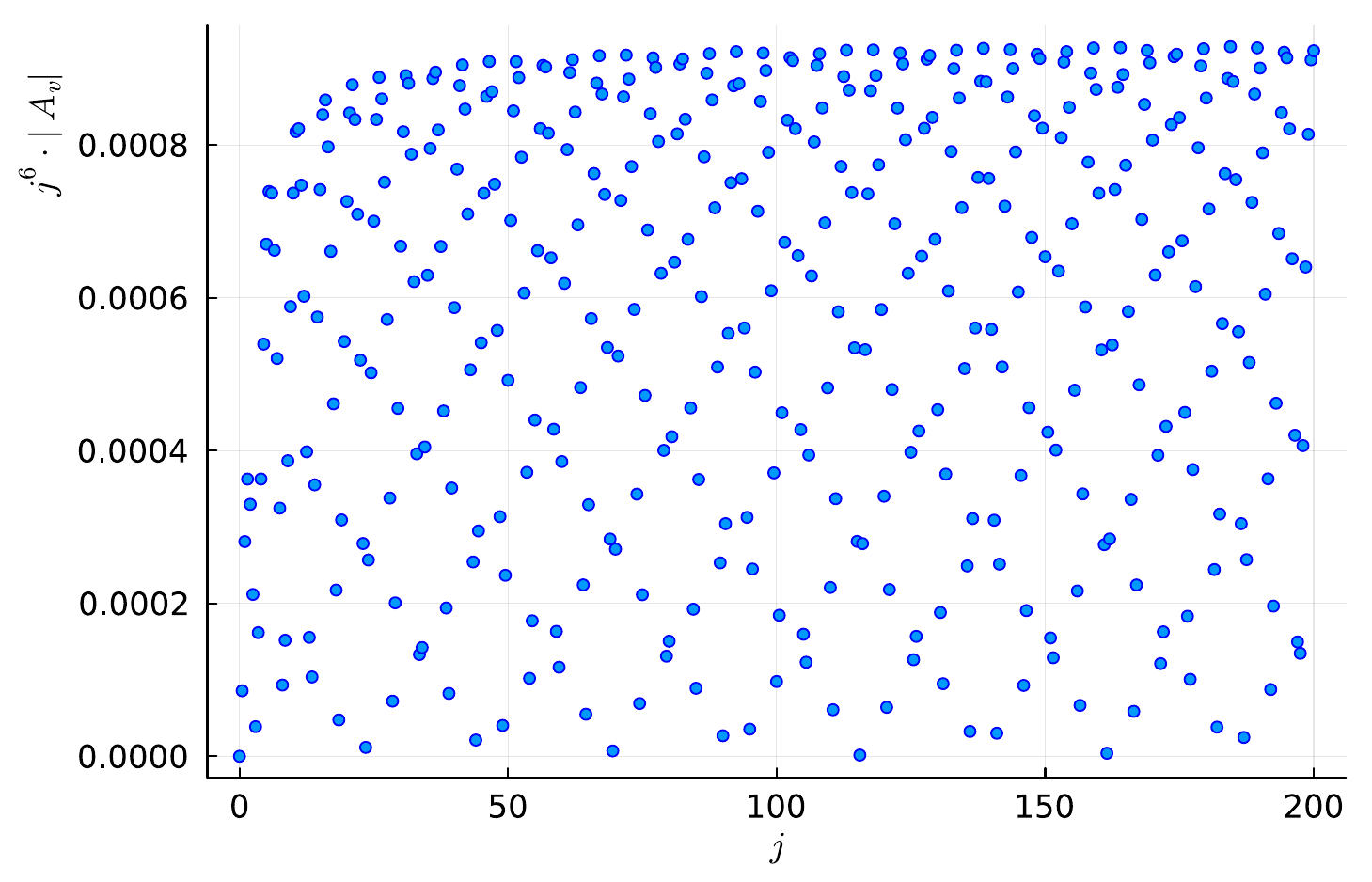} 
\caption{The coherent amplitude for equilateral vertex characterized by equal spins $j$. The plot displays the absolute value of the rescaled amplitude for the spins $0 \leq j \leq 200$ in steps of 0.5. }
\label{fig:cohvertex_equi}
\end{figure}

Figure \ref{fig:cohvertex_equi} displays the absolute value of the coherent amplitude  for the equilateral vertex rescaled by a factor $j^6$ for the boundary spins ranging from $j =0$ to $j=200$ in steps of half integers. The rescaling is due to the power law decay $j^{-6}$ of the vertex amplitude.  The plot demonstrates the capability of the new algorithms to efficiently compute coherent amplitudes across large values of the spins, which was previously impractical due to resource constraints. As shown, the coherent amplitude for the equilateral vertex configuration oscillates as a function of the spins. 

The vertex amplitude for the equilateral configuration has been compared to its asymptotic formula in Appendix \ref{sec:Asymptotics}, as displayed\footnote{The comparison shows a good agreement between the equilateral coherent vertex amplitude and its asymptotic formula with less than $\sim 1\%$ relative error for boundary spins $j\geq 100$. See Figure \ref{fig:cohvertexasy_equi} for more details.} in Figure \ref{fig:cohvertexasy_equi}. In general, the asymptotic analysis of the coherent amplitude for a vertex with Regge boundary data yields two solutions to the critical point equations of its action \cite{Barrett:2009gg,Barrett:2009as}. The frequency of the oscillations is determined by the Regge action associated with the dual 4-simplex triangulation, explaining the persistence of the oscillations for large spins observed in Figure  \ref{fig:cohvertex_equi}.

The coherent vertex amplitude for other examples of Regge geometries, such as the isosceles 4-simplex and non-regular configurations, has been studied in \cite{Steinhaus:2024qov,Dona:2017dvf}. The coherent amplitudes for the non-equilateral examples can also be efficiently computed using Algorithm \ref{alg:cohvtx}. Computational results for examples of non-equilateral configurations are not presented here but are available on the \href{https://github.com/Seth-Kurankyi/su2bf-TNAlgo}{GitHub repository} \cite{AsanteTN2024}.

\subsubsection{Benchmarks for the Equilateral Vertex Amplitude}

\begin{figure}[ht!]
\centering 
\includegraphics[scale=0.386,valign=c]{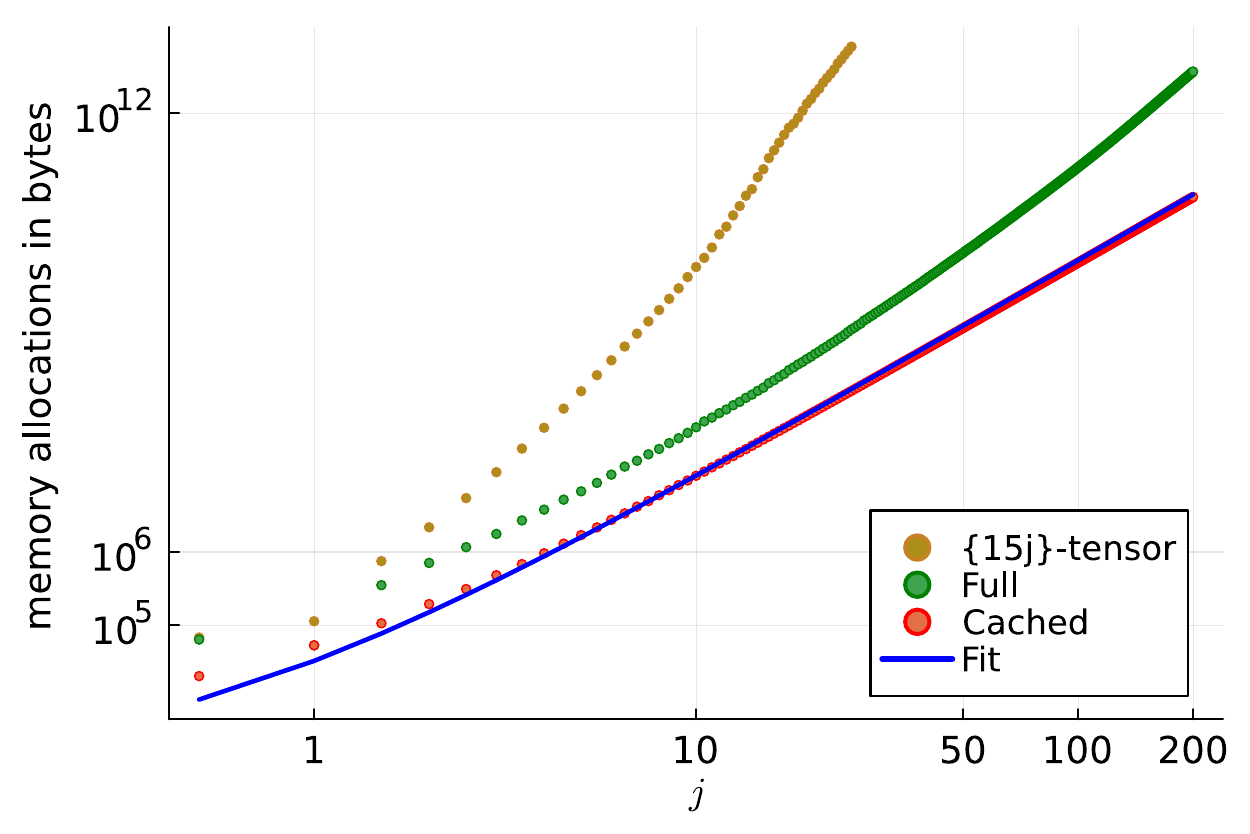} 
\includegraphics[scale=0.386,valign=c]{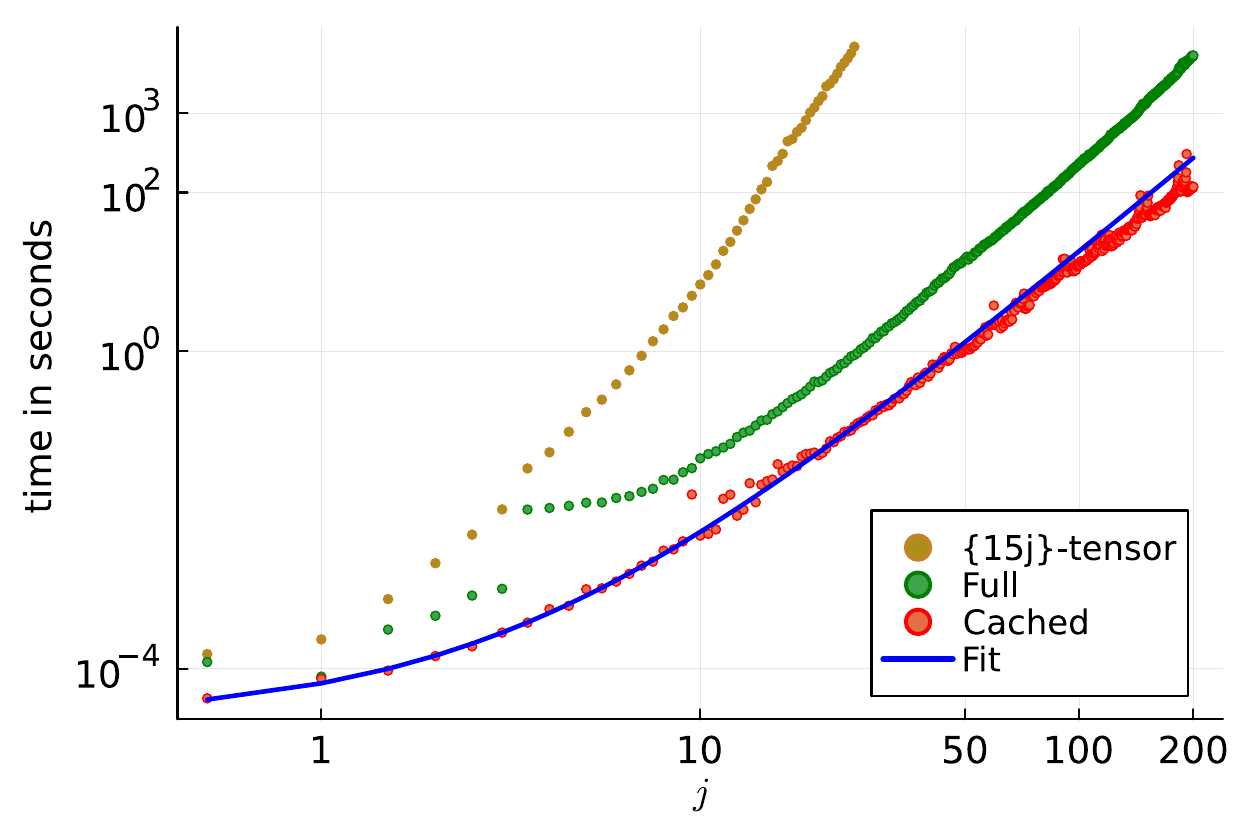} 
\caption{Log-log plots illustrating the benchmarks for computing the coherent amplitude for the equilateral vertex configuration compared to previous computations using the $\fifJ$-tensor. The left plot shows memory allocations, while the right plot shows computational time as a function of spins. In both plots, `\{15j\}-tensor' represent data based on the use of the 5-valent $\fifJ$-tensor, `Full' represents the data involving the storage of certain SU(2) invariants, and `Cached' indicates the computations after the SU(2) invariants are loaded into memory. The `\{15j\}-tensor' performs far worse in both memory usage and computational time.  The Fit (in blue) for the `Cached' data show their scaling behavior. }
\label{fig:benchmarks_equi}
\end{figure}

This section presents the benchmarks for the computations of the coherent amplitude for the equilateral vertex configuration. The benchmarks are illustrated in Figure \ref{fig:benchmarks_equi}, which consists of two plots: one showing memory allocations and the other showing computational time, both as functions of the spins $j$. The plots compare the computational resources used for the tensor network Algorithm \ref{alg:cohvtx} to those used for the previous algorithms based on computing the $\fifJ$-tensor \eqref{fifJtensor}. The data generated from computing the amplitudes by contracting the 5-valent tensor are simply referred to as {\blue `\{15j\}-tensor'}  in Figure \ref{fig:benchmarks_equi}. 

The left plot in Figure \ref{fig:benchmarks_equi} is a log-log plot of memory allocations. The term, {\blue `Full'} in this plot refers to the computations including the preload stage, where the SU(2) invariant functions, such as \texttt{wigner3j}, \texttt{wigner6j}, and \texttt{wignerDjm} functions, are stored in memory. This stage is resource-intensive, as it requires substantial memory to store these functions. In contrast, {\blue `Cached'} refers to data for the computations after the SU(2) invariant functions have been loaded into memory using the \texttt{memoize} package in \texttt{Julia}. The fit for the `Cached' data in Figure \ref{fig:benchmarks_equi} is given by the function $\blue 1200 \times (2j+1)^3$, indicating how the memory allocations scales with spin $j$.

The right plot in Figure \ref{fig:benchmarks_equi} shows a log-log plot of computational time for the equilateral vertex amplitude. Similar to the memory allocations plot, `Full' data includes the overhead of loading and initializing the necessary SU(2) invariants, resulting in higher computational time. The `Cached' data\footnote{The data for the Cached computationally time could be further refined to smooth out with additional benchmark samples. The data presented here is based on only one sample size. } represents the computational time after these functions are cached into memory using the \texttt{memoize} package. The fit\footnote{The fit for the $\fifJ$-tensor data and the Full data are not shown in the plots in Figure \ref{fig:benchmarks_equi}. However, the computational time for the $\fifJ$-tensor data scales approximately as $\blue 10^{-8} \cdot (j+3)^8$ seconds while that for the Full data scales approximately as $\blue 10^{-8} \cdot (2j+7)^{9/2}$ seconds. } for the `Cached' data is given by the function $ \blue 10^{-8} \times (2j+7)^4$, demonstrating the scaling behavior of the computational time with the boundary spin $j$. This fit function aligns with the expectations discussed in Section \ref{sec:matrix_contraction}.

~
\begin{table}[ht!]
\centering
\renewcommand*{\arraystretch}{1.1}
\begin{tabular}{|C{4.5em}|C{6.3em}|C{6.3em}|C{6.3em}||C{6.3em}|C{6.3em}|C{6.3em}|}\hline
 \multirow{2}{5em}{\hspace{0.5cm}  Spin } & \multicolumn{3}{L{16em}||}{Memory allocation / Byte} &  \multicolumn{3}{C{18em}|}{Computational time / second}  \\ \cline{2-7}
& \blue $\fifJ$-tensor & \blue Full  & \blue Cached & \blue $\fifJ$-tensor & \blue Full & \blue Cached\\ \hline 
10   & 7.88211 E9 & 5.0697 E7 & 1.1022 E7 & 6.9577 &0.04486 & 0.00477\\ \hline
25   & 7.12174 E12 & 1.0235 E9 & 1.5058 E8 &5692.7788 &0.86416 & 0.1012 \\ \hline
50   & - & 1.2247 E10 & 1.1526 E9 & - &14.6661 & 1.04734 \\ \hline
100 & - & 1.8121 E11 & 9.0192 E9 & - &234.0320 & 12.3222 \\ \hline
200 & - &  3.7012 E12 & 7.1358 E10 & - &5348.3676 & 118.5069 \\ \hline
\end{tabular}
\caption{Sample of benchmarks for computations of coherent amplitudes for equilateral vertex using the tensor network algorithm based on matrix contractions.}
\label{tab:benchmarks_equi}
\end{table}

The comparison between Full and Cached benchmarks clearly demonstrates the efficiency gains achieved through caching. Memoization of the SU(2) invariant functions significantly reduces the computational overhead and memory requirements, making the coherent amplitude computations faster and more feasible for large spins. Table \ref{tab:benchmarks_equi} displays a sample of the benchmarks in support of the efficiency gained in using Algorithm \ref{alg:cohvtx} based on matrix contractions, compared to computations using the $\fifJ$-tensor.  For spin $\blue j=200$ (a relatively large value), the Full computation takes approximately {\blue 5348 seconds}, while the Cached computation takes approximately {\blue 119 seconds}, showing a dramatic improvement achieved through this optimization. The methods based on the $\fifJ$-tensor become computationally unfeasible for such large spins. 

In summary, the benchmarks for the equilateral vertex amplitude emphasize the importance of optimizing computations using the low-valence tensors and caching. The significant improvements observed in the benchmarks are partly due to the highly symmetric equilateral vertex configuration.  Although non-equilateral vertex amplitudes might show less pronounced memory usage and computational time reductions, the tensor network algorithms are still efficient computational methods. These results and benchmarks provide a clear indication of the performance enhancements and strategies necessary for the effective computation of coherent amplitudes in spin foam models.

\subsection{Coherent Vertex Amplitudes: Vector  Geometries}\label{sec:vec_geoms}

In this section, we present results for the coherent vertex amplitudes with boundary data corresponding to vector geometries. A detailed parametrization for these vector geometries is discussed in Appendix \ref{sec:geoms}. For vector geometries, the set of spins and normal vectors for each edge dual to a tetrahedron satisfies the closure conditions, ensuring that each configuration corresponds to a well-defined Euclidean tetrahedron in $\mathbb{R}^3$. Additionally, the (rotated) unit normal vectors satisfy the anti-parallel condition \({\bf n}_{ab} = -{\bf n}_{ba}\) for all faces \(ab\). Here, we focus on two specific vertices each with a vector geometry configuration, labeled by $v_1$ and $v_2$. The boundary data for both vertices are chosen to have equal spins, i.e., $j_{ab} = j$ for all $a,b$. 

These vertex configurations considered here are selected to be `close' to the equilateral vertex; that is, all the spins associated with the vertices $v_1$ and $v_2$ are equal and in addition, the normal vectors of any two tetrahedra are chosen to correspond to an equilateral configuration. This results in vector geometries with boundary data characterized by two parameters $\blue \{j , \varphi_5\}$, where $j \in \mathbb N/2$ and $-1 \leq \varphi_5 \leq 1$ is an `inner product variable'. More details about the parametrization of vector geometries is discussed in Appendix \ref{sec:geoms}.  For a fixed spin $j$, the value of $\varphi_5$ determines how `close' the vector geometry is to the equilateral vertex.  The boundary data corresponds to the equilateral vertex configuration when $\blue \varphi_5 = \sqrt{5}/3 \approx 0.745$. For any other value of $\varphi_5$, the boundary data do \emph{not} correspond to a Regge geometry since the overall configuration cannot be embedded into a Euclidean 4-simplex triangulation with well-defined lengths. As an example, we chose the inner product variables for vertices $v_1$ and $v_2$ to be  $\blue \varphi_5 = 1/2$ and  $\blue \varphi_5 = 3/5$, respectively. Thus, $v_2$ is `closer' to the equilateral vertex than $v_1$. The corresponding unit normal vectors (\emph{twisted spike configuration}) associated with $v_1$ and $v_2$ are provided in Table \ref{tab:Normals_vg}.

\begin{table}[ht!]
\centering
\renewcommand*{\arraystretch}{1.44}
\begin{tabular}{| l || l|}\hline
Normal vectors for vertex $v_1$ ($\varphi_5 = 1/2$)   & Normal vectors for vertex $v_2$ ($\varphi_5 = 3/5$)  \\ \hline
${\bf n}_{12} = (1.0,0.0,0.0) $ & $ {\bf n}_{12} = (1.0,0.0,0.0) $ \\
${\bf n}_{13} = (-0.33333333, 0.89977173, 0.28160206) $ &  ${\bf n}_{13} = (-0.33333333, 0.8869709, 0.3196428) $ \\
$ {\bf n}_{14} =  (-0.33333333, -0.20601133, -0.92002621)  $ \,\,\, & $ {\bf n}_{14} =  (-0.33333333, -0.16666667, -0.92796073) \,\, $ \\ 
$ {\bf n}_{15} = (-0.33333333, -0.6937604, 0.63842415) $ & $ {\bf n}_{15} = (-0.33333333, -0.72030423, 0.60831793)  $ \\ 
${\bf n}_{23} =  (0.08235735, 0.05089964, -0.99530221) $ \,\,\,\, & $ {\bf n}_{23} =  (0.20111989, 0.10055995, -0.97439134) $ \\ 
$ {\bf n}_{24} = (0.6557091, -0.70053255, 0.28160206) $ & $ {\bf n}_{24} =  (0.50957672, -0.7988492, 0.3196428)  $ \\ 
$ {\bf n}_{25} =  (0.26193354, 0.6496329, 0.71370015) $ & $ {\bf n}_{25} =  (0.28930339, 0.69828926, 0.65474855) $ \\ 
$ {\bf n}_{34} =  (0.4472136, 0.89442719, 0.0) $ & $ {\bf n}_{34} =  (0.6, 0.8, 0.0) $ \\ 
$ {\bf n}_{35} =   (-0.69818958, 0.05624418, -0.71370015)  $ & $ {\bf n}_{35} =   (-0.73221344, 0.18753084, -0.65474855) $ \\ 
$ {\bf n}_{45} =  (0.76958937, -0.01211669, -0.63842415)  $ & $ {\bf n}_{45} = (0.77624338, -0.16551587, -0.60831793)  $ \\ \hline
\end{tabular}
\caption{Boundary unit  normal vectors associated with faces of two vertices $v_1$ and $v_2$ whose boundary data describe vector geometries.  The vectors are given up to 8 decimal places. These vectors satisfy the anti-parallel conditions ${\bf n}_{ab} = - {\bf n}_{ba}$, for all $a,b$.  }
\label{tab:Normals_vg}
\end{table}

\begin{figure}[ht!]
\centering 
\includegraphics[scale=0.38,valign=c]{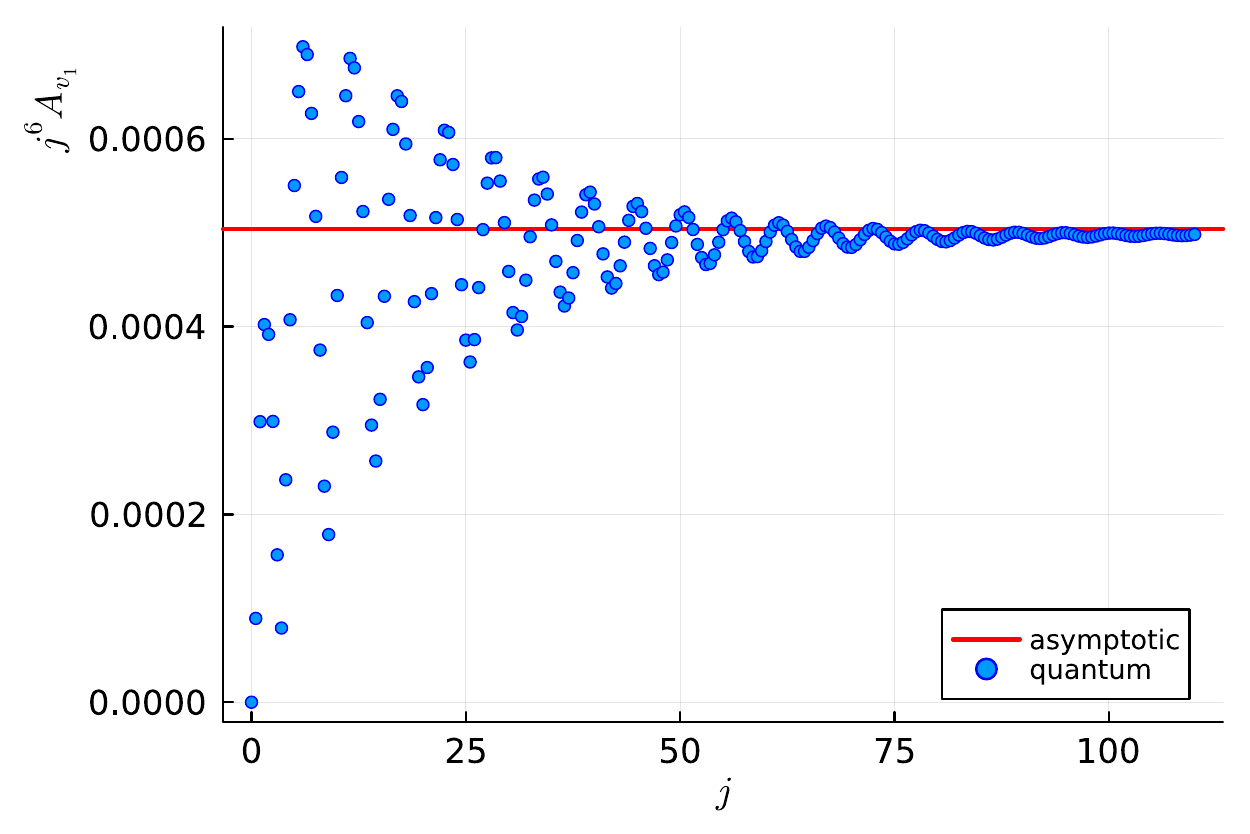}  
\includegraphics[scale=0.38,valign=c]{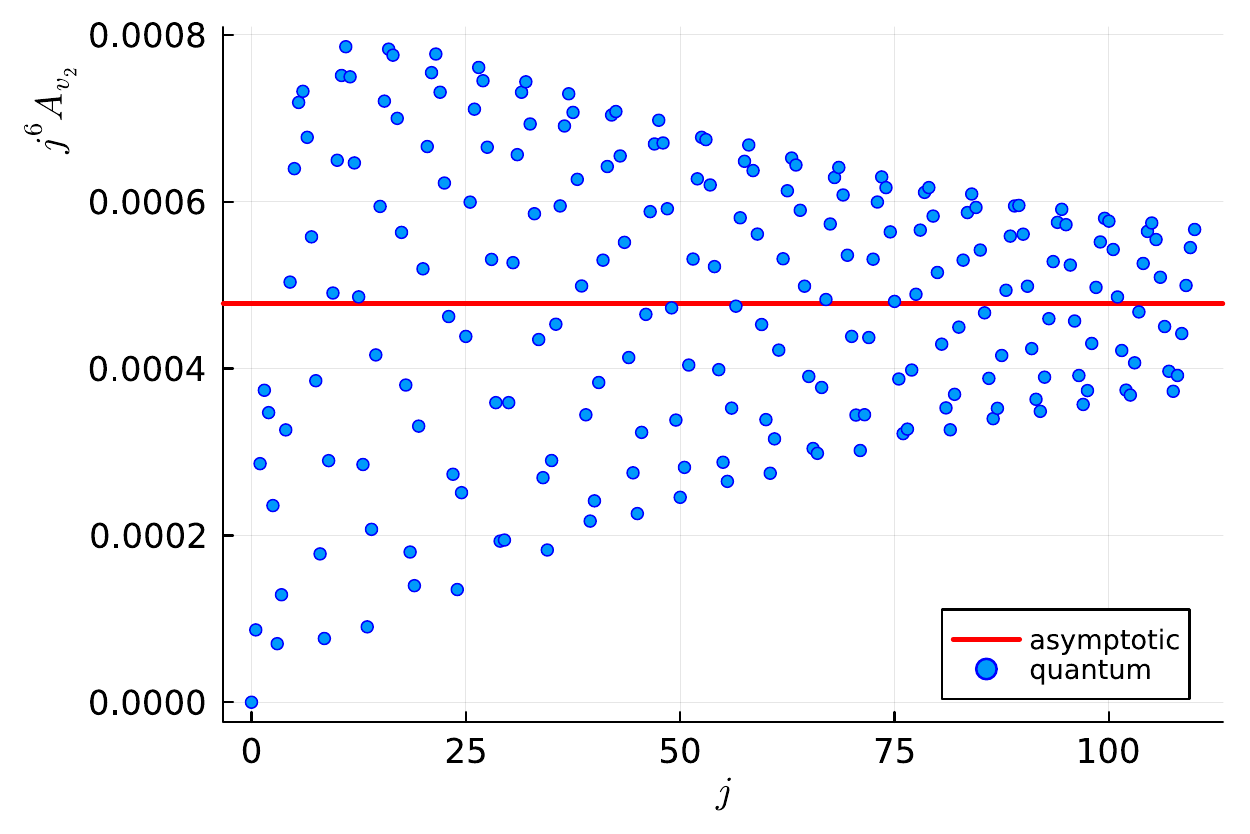}  
\caption{The coherent amplitude for vertices $v_1$ and $v_2$ with boundary data corresponding to vector geometries. The coherent amplitudes (quantum) are compared to their asymptotic formula.}
\label{fig:cohvertex_vgeom}
\end{figure}

Using Algorithm \ref{alg:cohvtx}, we computed the coherent amplitudes for vertices $v_1$ and $v_2$ for boundary spins ranging between $0 \leq j \leq 110$. The resulting rescaled coherent amplitudes are displayed in Figure \ref{fig:cohvertex_vgeom}, together with their asymptotic formulae. The power-law scaling behavior $j^{-6}$ also holds for these vector geometry vertex amplitudes. The asymptotic formula for a vector geometry is described in Equation \eqref{eqn:vg_asmp} in Appendix \ref{sec:Asymptotics}. The vector geometries which do not correspond to a geometric 4-simplex yield one solution to the critical point equations of the vertex amplitude.  The asymptotic formula, thus, after rescaling by a factor $j^6$ gives a constant term.  This explains the behavior of the vertex amplitudes shown in  Figure \ref{fig:cohvertex_vgeom}, which exhibit damped oscillations around their asymptotic values.  The oscillations decreases with large $j$ as the amplitude approaches its asymptotic value.

The plots in Figure \ref{fig:cohvertex_vgeom} illustrate different oscillating behaviors for the vector geometry configurations $v_1$ and $v_2$. The vertex amplitude for vector geometries that are `close' to a Regge geometry exhibit oscillations that persist for large spins, and the asymptotic limit can only be reached for very large spin. Conversely, the vertex amplitude for vector geometries `not close' to a Regge geometry exhibit fewer oscillations, and the asymptotic limit is reached for relatively small spin values. Specifically, in the examples chosen here, the asymptotic formula for vertex $v_1$ matches its quantum amplitude $A_{v_1}$ within approximately  $ \sim 10\% $ relative error for spins $j \geq 25$, while for $v_2$, its asymptotic formula matches the amplitude $A_{v_2}$ within $\sim 10\%$ relative error for larger spins $j \geq 100$. This analysis also highlights the optimization achieved with the tensor algorithms in exploring such examples comprehensively, providing insights into the quantum geometric behavior for different types of vertex configurations across varying spin regimes.

\begin{figure}[ht!]
\centering 
\includegraphics[scale=0.41,valign=c]{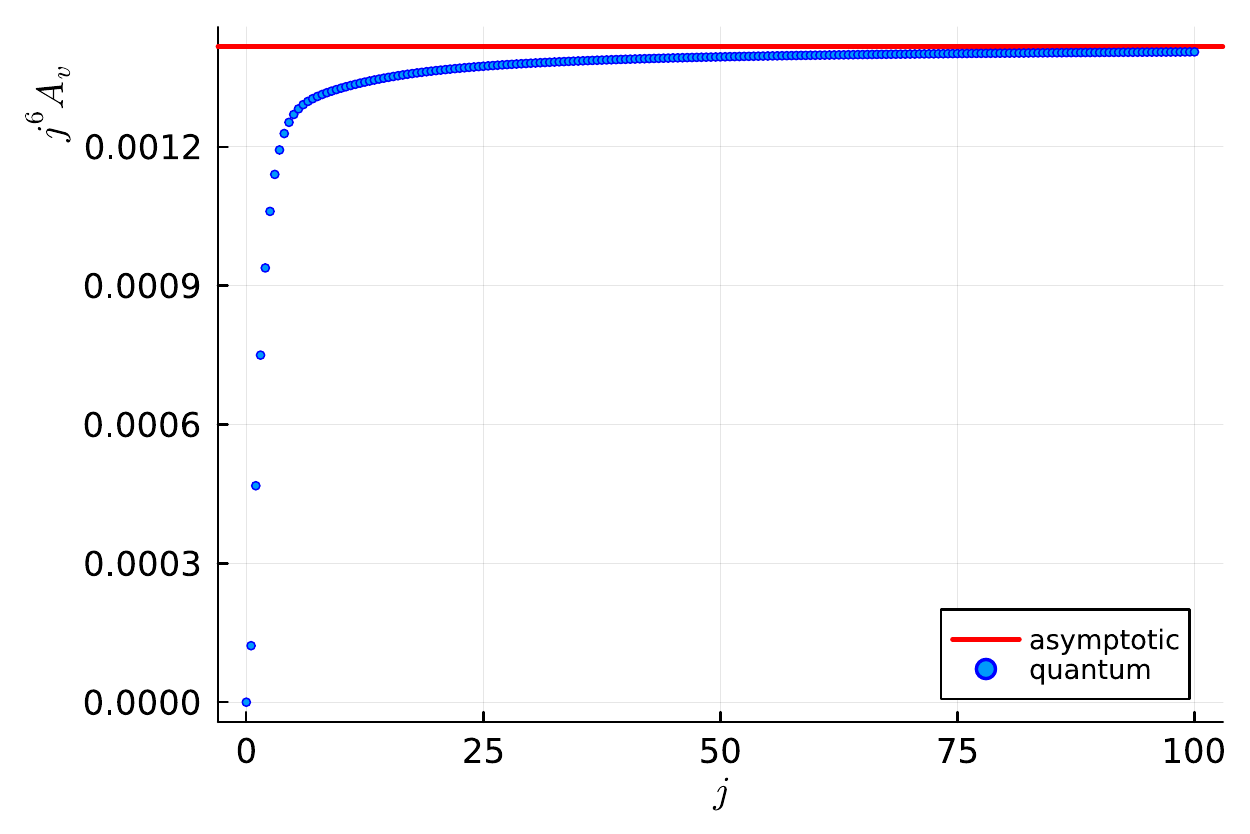} 
\includegraphics[scale=0.36,valign=c]{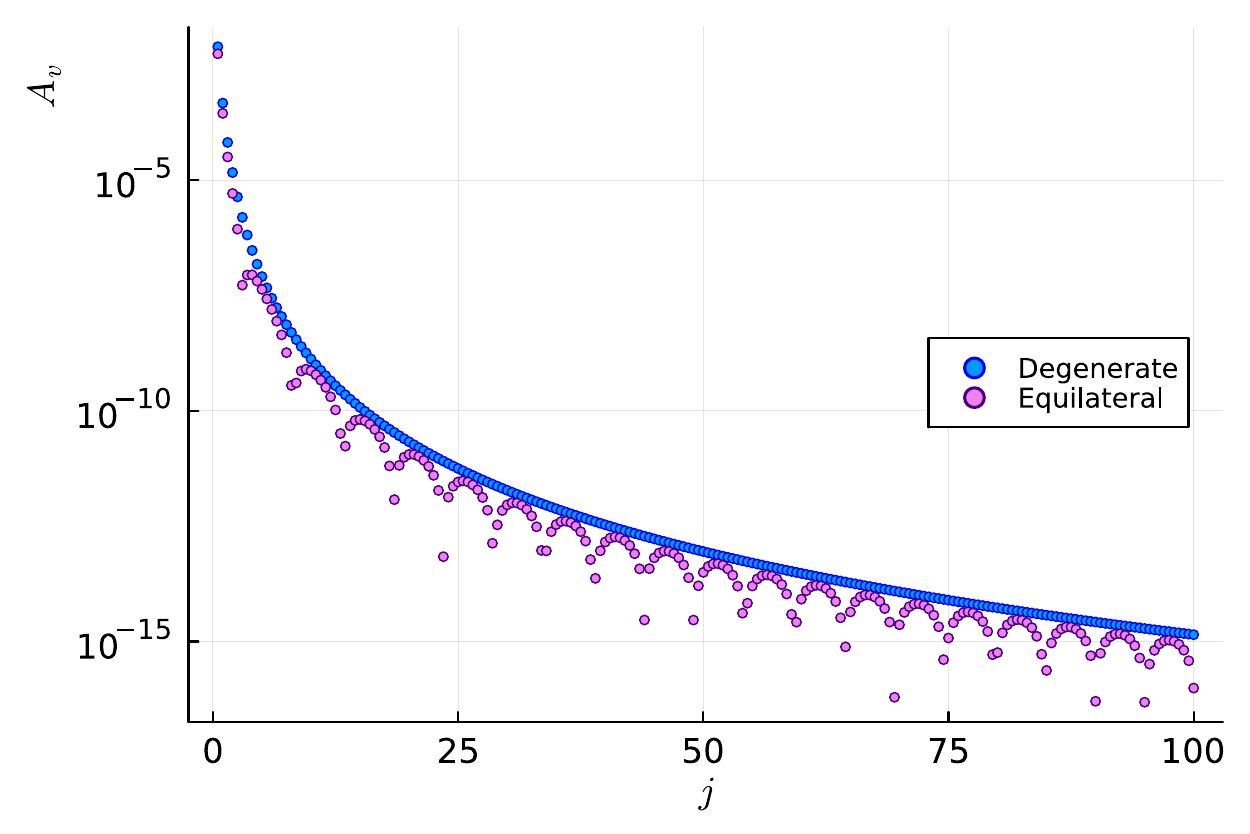} 
\caption{{\it Left panel:} The rescaled coherent vertex amplitude for a degenerate vector geometry compared to its asymptotic formula. {\it Right panel:} A comparison of coherent amplitudes of a degenerate vector geometry to that of a Regge geometry (the equilateral vertex). Both coherent vertex amplitudes have same power-law scaling but differ in their oscillating characteristics.}
\label{fig:vecgeom_equi}
\end{figure}

Lastly, we consider an example of a \emph{degenerate vector geometry}, also with equal boundary spins $j_{ab} = j$ and its (rotated or twisted spike) boundary unit normal vectors, $ {\bf n} =  \{ {\bf n}_{ab}\}_{1\leq a<b\leq5}$ chosen to be
\[  {\bf n} =    \big\{ (1, 0, 0), (0, 0, 1),  (0, 0, -1),
 (-1, 0, 0),
 (0, 0, -1),
 (0, 0, 1),
 (1, 0, 0),
 (0, 1, 0),
 (0, -1, 0),
 (0, 1, 0) \big\} . \]
Geometrically, these unit normal vectors form the edge vectors of a (one-sided) open unit cube\footnote{Vector geometries can generically be parametrized by the deformations of a three dimensional cube \cite{Barrett:2002ur}.} in $\mathbb R^3$. Each face of the cube corresponds to the normal vectors of a tetrahedron. This example is degenerate since, for each of the tetrahedron, the face normal vectors form a two-dimensional subspace, resulting in a vanishing volume. 

The coherent amplitude for this degenerate vector geometry example is shown in the left plot in Figure \ref{fig:vecgeom_equi}. The vertex amplitude shows a power-law scaling behavior $ j^{-6}$ similar to the previous equilateral vertex and vector geometry configurations. However, unlike those examples, this amplitude shows no oscillations with respect to the spins. The right panel of Figure \ref{fig:vecgeom_equi} compares a logarithmic plot of the coherent amplitude of degenerate vertex configuration with that of the equilateral configuration, showing the different oscillation patterns between the amplitudes. For this degenerate vertex example, the asymptotic formula closely matches its coherent amplitude even at low spin values, with approximately $\sim 3\%$ relative error observed at spin $j=25$.

\section{Ideas to Go Beyond Boundary Vertices} \label{sec:ideas}

There are two key reasons underpinning the optimization obtained in the tensor network algorithms. First, defining the 5-valent tensor of $\fifJ$-symbols is computationally costly and requires substantial storage, as its size grows exponentially with the range of its index labels (assuming for simplicity that all labels have the same range).  Second, the contraction of the 5-valent tensor with coherent vectors of boundary states, while also computationally expensive, is relatively sub-leading compared to the initial evaluation of the $\fifJ$-tensor. For a single vertex with coherent boundary data, we circumvent these issues by rewriting the contractions as a matrix trace, thereby avoiding the definition of the $\fifJ$-tensor entirely. Consequently, when expressing calculations as tensor network contractions, it is advantageous to use low-valent tensors whenever possible (at best at all times), as low-valent tensors are less costly to define and store and their contractions are computationally faster.

Spin foam amplitudes associated with a 2-complex, whether consisting of a single vertex or multiple vertices, can be expressed as a sum over tensor networks, where each vertex is represented by a 5-valent tensor of $\fifJ$-symbols. This raises the question of whether our algorithm can be extended to handle multiple vertices, particularly those entirely within the bulk, where all intertwiners are contracted with other vertices. As discussed in Section \ref{sec:partial_vertex}, the intertwiner labels that will be contracted later must be explicitly retained; if all five labels must be retained, we essentially revert to defining a 5-valent tensor. Thus, the tensor network method does not appear to be readily applicable to generic bulk vertices. Moreover, if two bulk vertices are contracted along a common edge, it results in an 8-valent tensor, as explicit dependence on the remaining indices to be contracted later must be kept. Such a contraction is costly and might thus limit, both in time and memory, the ability to run these simulations. In the following, we explore different scenarios involving bulk vertices, analyzing where the tensor network algorithms can or cannot be effectively applied to compute spin foam amplitudes.

\begin{figure}[ht!]
\centering 
\includegraphics[scale=0.4,valign=c]{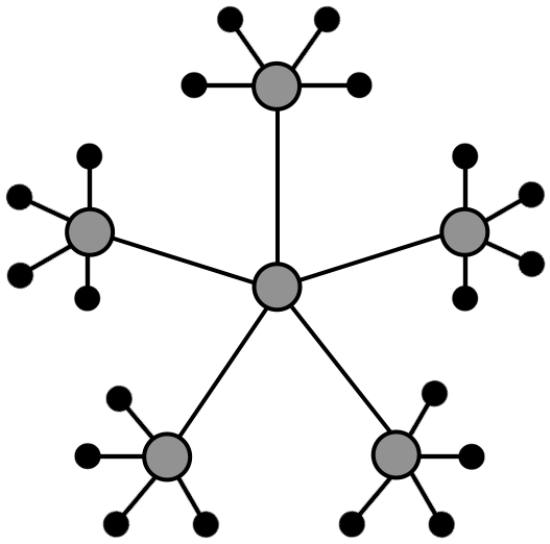}  \hspace{3cm}
\includegraphics[scale=0.4,valign=c]{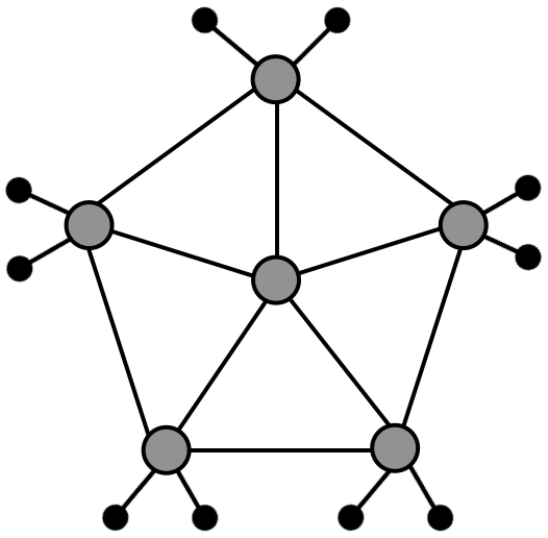}  
\caption{2-complexes with a bulk vertex. \emph{On the left:} The boundary vertices linked to the bulk vertex are disconnected.  \emph{On the right:} The bulk vertex is linked to connected boundary vertices.   }
\label{fig:blkvertex}
\end{figure}

Importantly, when evaluating spin foam amplitude associated with a 2-complex with multiple vertices, it is more beneficial to perform tensor contractions `inwards': that is, first contract the tensors associated with boundary vertices\footnote{For example, the partial-coherent vertex amplitudes} and then contract with the bulk vertex amplitudes. This procedure can effectively reduce the number of overall contractions to be performed in the full amplitude. 
If a bulk vertex is connected with boundary vertices which are themselves disconnected, then by contracting `inwards', one can avoid computing the $\fifJ$-tensor. Consider, for example, a bulk vertex connected with five 1-valent partial-coherent vertex (see Figure \ref{fig:blkvertex}). In this case, each partial-coherent boundary vertex amplitude can be contracted separately, resulting in a 1-valent tensor or a vector. The resulting vectors can be contracted with the bulk vertex using the same techniques for the coherent amplitude \eqref{cohvertex_new}. Hence, in this scenario, the 5-valent $\fifJ$-tensor can be avoided in the computation of the full amplitude using the methods in Algorithm \ref{alg:cohvtx}. On the other hand, if the boundary vertices linked to a bulk vertex are connected amongst each other (see Figure \ref{fig:blkvertex}), then the Algorithm  \ref{alg:cohvtx} cannot be directly employed to evaluate the full amplitude using matrix contractions.

Still, low-valent tensors can help optimize the contractions between $\fifJ$-tensors. The $\fifJ$-symbol is expressed as a sum of products of several $\sixJ$-symbols, as shown in Equation \eqref{15jsymbol}. We can leverage this structure by reorganizing the summations to contract the intertwiner labels first, before summing over the virtual spin label. For example, consider an intertwiner label $i_2$ shared by two $\sixJ$-symbols; these can be combined into a 3-valent tensor $T_{i_1 i_2 i_3}$ (same as the notation in Equation \eqref{tensornotation}), with additional indices $i_1$ and $i_3$. Thus, the contraction of the intertwiner index $i_2$ can be performed locally using this 3-valent tensor rather than the full 5-valent $\fifJ$-tensor. For instance, when contracting the intertwiner label $i_2$ between two bulk vertices, we can use two such 3-valent tensors, resulting in a 4-valent tensor:
\[  \sum_{i_2}  T_{i_1 i_2 i_3} \, T_{i_1' i_2 i_3'} =   {\tilde T}_{i_1 i_1' i_3 i_3'}.  \]
The remaining open indices are then contracted with other tensors involved in the amplitude computation. Additionally, the sum over bulk and virtual spins are performed later. In essence, by reorganizing the summations, the contractions of the intertwiner labels can be performed locally using lower-valent tensors rather than the $\fifJ$-tensors. This method is therefore, expected to scale better and be more computationally efficient as the dimensions of the intertwiner labels grow. However, the additional sums over the bulk and virtual spins could potentially increase the computational costs. Therefore, further analysis of these aspects is necessary, which we defer to future research.

To sum up, applying these tensor network ideas to generic bulk vertices is possible but not straightforward and will most likely be numerically costly for a large 2-complex. The methods depend a lot on the structure of the connected components of the spin foam amplitude of the associated 2-complex in order to optimize the computation of the corresponding amplitude.  For large complexes, it may be necessary to complement our methods with Monte Carlo methods, such as importance sampling of coherent boundary intertwiners \cite{Steinhaus:2024qov} and random sampling of the bulk spins \cite{Dona:2023myv}, to achieve faster convergent results. 

~

\section{Discussion and Conclusion}\label{sec:conclude}

In this paper, we have conducted a detailed analysis of numerical computations of coherent amplitudes for the topological SU(2) BF and the Lorentzian EPRL spin foam models. Given a 2-complex $\Delta^{\!*}$, the coherent amplitude associated with $\Delta^{\!*}$ can be expressed as a contraction of a tensor network involving invariant functions derived from the specifics of the corresponding spin foam model. By introducing tensor network algorithms, we demonstrated significant improvements in the efficiency of computing these amplitudes. Our approach reorganizes the computations into sums of contractions of low-valence tensors, particularly matrices, which substantially reduces both computational complexity and memory usage. This reorganization enables scaling up the computations to include large representation labels, which were previously computationally infeasible.

The primary goal was to address the computational inefficiencies associated with traditional methods that rely on evaluating high-valence tensors, such as the 5-valent $\fifJ$-tensor, for computing SU(2) BF or Lorentzian EPRL spin foam amplitudes. Evaluating a 5-valent $\fifJ$-tensor is resource-intensive in terms of both time and memory, especially for large representation labels, due to the exponential growth in its size. For this reason, traditional numerical methods are limited to relatively small values for the spin representation labels. The concept of tensor network methods, which evaluate large tensors as sequences of contractions between smaller tensors, provides a valuable alternative for optimizing spin foam computations. For computing coherent vertex amplitudes, we have introduced tensor network algorithms that employ matrix contractions. These are less costly to define and store, and their contractions are computationally faster. Thus, these algorithms significantly improve the computation of spin foam amplitudes, allowing us to explore higher spin regimes.

The results and benchmarks, focusing on SU(2) BF vertex amplitudes, highlight significant improvements achieved through tensor network algorithms. By restructuring the amplitudes into matrix contractions, the tensor network approach not only scales more efficiently but also renders the computation of spin foam coherent amplitudes feasible on standard consumer hardware. Using these algorithms, we conducted experiments across a wide range of spins for SU(2) BF coherent amplitudes of various vertex configurations. Vertex configurations characterized by boundary data in Regge geometries and vector geometries exhibit similar scaling behavior but differ in their oscillation patterns. The distinct oscillation patterns influence the spin regime where the asymptotic formula remains applicable: vector geometries `close' to a Regge boundary show rapid oscillations, extending into higher spin regimes before matching their asymptotic formula, whereas vector geometries `not close' to a Regge boundary exhibit fewer oscillations, reaching their asymptotic limit at smaller spin values.

The tensor network methods are also applicable to spin foam computations beyond vertex amplitudes. For a generic 2-complex with multiple vertices, the corresponding amplitude is computed via tensor contractions between amplitudes assigned to the vertices and edges of the 2-complex.  Boundary vertices are comprised of a mixture of bulk edges and boundary edges, therefore, they are assigned `partial-coherent vertex' amplitudes. The tensor network methods simplify computations of these partial-coherent vertex amplitudes using matrix contractions. In certain cases, computations involving bulk vertices can also be simplified with these methods.

The results presented in this paper open several avenues for future research and development.  A direct implementation of these tensor network methods for other spin foam models such as the EPRL model in \texttt{Julia} language is a promising next step.  Additionally, applying these methods to more complex scenarios, such as amplitudes for many-vertex configurations, will expand the applicability and robustness of these computational techniques. Future efforts will focus on further optimization, including the implementation of parallelization and GPU computing into the existing algorithms.  As part of the ongoing efforts, the tensor network algorithms for the SU(2) BF model have been made available on the \href{https://github.com/Seth-Kurankyi/su2bf-TNAlgo}{GitHub repository} \cite{AsanteTN2024} for continued exploration and verification. Future work includes the potential development of these algorithms into a dedicated package for easy access. 

In conclusion, the efficient tensor network algorithms developed here represent a significant step forward in the numerical computation of spin foam models.  The techniques enhance the feasibility of practical evaluations and establish a solid foundation for future advancements in the numerical study of quantum gravity.

\begin{acknowledgments}
This project is supported by  the Deutsche Forschungsgemeinschaft (DFG, German Research Foundation) project number 422809950. SKA acknowledges the support of the Alexander von Humboldt Foundation in the early stages of this project. 
\end{acknowledgments}

~

\appendix 

\section{ Geometries Associated with Boundary Data for a Vertex}\label{sec:geoms}

This appendix explores the geometries related to the boundary data of a vertex within a 2-complex, combinatorially dual to a 4-simplex. Each vertex is connected by edges labeled \( e_a \) where \( a \in \{1, 2, \dots, 5\} \). The coherent boundary data for a vertex is described by a set of spins and unit normal vectors, denoted as $\{ j_{ab}, \mathbf{n}_{ab} \}_{1 \leq a, b \leq 5}$, with $ j_{ab} = j_{ba} $ and $a \neq b $. This set includes 10 spins corresponding to the triangle areas of the 4-simplex and 20 unit normal vectors in $S^2$ representing the face normals of these triangles within each tetrahedron. Generically, such a boundary data is described by a $10+ 2(20) =50$ dimensional  parameter space. {\blue Twisted geometries} represent a sub-class of the boundary data described by gluing together classical tetrahedra along their faces \cite{Dittrich:2008ar,Freidel:2010aq}.  However, not every twisted geometry correspond to a geometric 4-simplex that can be embedded in $\mathbb{R}^4$.

Below, we outline specific subsets of twisted geometries---{ vector geometries and Regge geometries }---that satisfy the critical point equations of the coherent vertex amplitude (refer to Appendix \ref{sec:Asymptotics}). Examples of vertices with these boundary data have been considered in the results Section \ref{sec:results} for computing their coherent vertex amplitudes as matrix contractions.

~

{ \bf  Vector geometries:} 
Vector geometries are a subset of twisted geometries characterized by the following properties:
\begin{enumerate}
\item  Each pair of unit normal vectors \(\mathbf{n}_{ab}\) and \(\mathbf{n}_{ba}\) can be rotated into each other by an SO(3) group element such that the rotated vectors are anti-parallel: \(\mathbf{n}'_{ab} = -\mathbf{n}'_{ba}\) for all \( a < b \).
\item  The data associated with each edge satisfies the closure conditions:
\begin{equation}\label{eqn:closure}
\sum_{b \neq a} j_{ab} \mathbf{n}_{ab} = \mathbf{0}, \quad \forall a \in \{1, \dots, 5\},
\end{equation}
such that via Minkowki's theorem that each edge is dual to a well-defined tetrahedron in \(\mathbb{R}^3\). 
\end{enumerate}
The anti-parallel  conditions reduce the normal vectors to $10$ unit vectors while the closure conditions introduce a total of 15 constraints. Hence, a vector geometry is described by a 15 dimensional parameter space. Various geometric interpretations of vector geometries have been discussed in the references \cite{Dona:2017dvf,Barrett:2002ur}.

Vector geometries can be parameterized by the ten spins $\{ j_{ab} \}_{1 \leq a < b \leq 5}$, representing areas of the dual faces, and five additional variables $\blue \{ \varphi_\alpha \}_{\alpha=1}^5$. These extra five variables are defined through specific `inner products' variables, chosen to be:
\begin{equation}
\varphi_1 = \mathbf{n}_{12} \cdot \mathbf{n}_{13}, \quad \varphi_2 = \mathbf{n}_{12} \cdot \mathbf{n}_{14}, \quad \varphi_3 = \mathbf{n}_{12} \cdot \mathbf{n}_{23}, \quad \varphi_4 = \mathbf{n}_{12} \cdot \mathbf{n}_{24}, \quad \varphi_5 = \mathbf{n}_{12} \cdot \mathbf{n}_{34}
\end{equation}
where $-1 \leq \varphi_i \leq 1$ for all $i$.
The pairs $\{ \varphi_1, \varphi_2 \}$ and $\{ \varphi_3, \varphi_4 \}$ represent the (cosine of) dihedral angles for two adjacent edges of the tetrahedra dual to edges \( e_1 \) and \( e_2 \), respectively. Any two tetrahedra and their adjacent dihedral angles can be chosen for parametrization. The inner product \(\varphi_5\) does not correspond to a dihedral angle of a tetrahedron, since the normal vectors $\mathbf{n}_{12}$ and \(\mathbf{n}_{34}\) are associated with faces from different tetrahedra. Additionally, the spins $j_{ab}$ can be expressed as inner products, satisfying $ j_{ab}^2 = {\bf u}_{ab} \cdot {\bf u}_{ab}$, where ${\bf u}_{ab}  = j_{ab} {\bf n}_{ab}$ is a non-normalized vector. In summary, the set of fifteen inner product variables, including the areas or spins $j_{ab} $, uniquely determine boundary data satisfying the anti-parallel and closure conditions, and hence defines a vector geometry.

Consider, as an example, a vector geometry with data corresponding to equal spins, i.e.,  $j_{ab} = j$ for all $a,b$. Furthermore, the two tetrahedra dual to the edges $e_1,e_2$ are chosen to be equilateral, thus their inner products satisfy  $\{ \varphi_1, \varphi_2 \} = \{-\tfrac13, -\tfrac13\}$ and $\{ \varphi_3, \varphi_4 \} = \{-\tfrac13, -\tfrac13\}$.  The last variable $\varphi_5$ remains arbitrary.  Hence, for this example, the boundary data is parametrized by the two parameters $\{j , \varphi_5 \}$. In Section \ref{sec:results}, we have considered coherent amplitude for vertices $v_1$ and $v_2$, having this vector geometries example with equal spins chosen for their boundary data, where $\varphi_5 = 1/2$ for vertex $v_1$ and $\varphi_5 = 3/5$ for vertex $v_2$.

~

{\bf Regge geometries:}
Regge geometries are a subset of vector geometries, and hence satisfy the anti-parallel and closure conditions described above. In addition, the boundary data are such that they correspond to a Euclidean 4-simplex in $\mathbb R^4$ with a well-defined length geometry with 10 degrees of freedom. However, specifying the 10 spins corresponding to areas does not uniquely determine a geometric 4-simplex: see \cite{Asante:2024rrd} for an extensive study on inverting areas and lengths of a 4-simplex.  The degrees of freedom for a Regge geometry can be parametrized by the edge lengths  $\{ \ell_{vv'} \}_{1\leq v<v'\leq 5},$ of the corresponding 4-simplex, where $v,v'$ label the vertices of the 4-simplex. This set of edge lengths ensures a unique 4-simplex geometric configuration.

From the edge lengths, other geometric quantities such as dihedral angles can be derived. Given the edge lengths of a 4-simplex, the boundary data $\{ j_{ab}, \mathbf{n}_{ab} \}_{1 \leq a, b \leq 5}$ associated with the corresponding 2-complex vertex can be determined. Conversely, a set of spins and unit normal vectors correspond to a boundary data for a geometric 4-simplex if the edge lengths of the 4-simplex can uniquely be determined from the boundary data. 

Consider, for example, an equilateral 4-simplex with its edge lengths given by $\ell = j (\sqrt{4}/3) $. This equilateral configuration corresponds to the vector geometry with equal areas $j$ and inner product variables given by 
$ \{ \varphi_1,\varphi_2,\varphi_3,\varphi_4,\varphi_5  \} = \{-\tfrac13, -\tfrac13,-\tfrac13, -\tfrac13,\tfrac{\sqrt{5}}{3} \}  $. Results and benchmarks for the coherent equilateral vertex amplitude are discussed in Section \ref{sec:equi_vertex}.

\section{Asymptotic Formulae of SU(2) Coherent Vertex Amplitudes}\label{sec:Asymptotics}

The starting point of the asymptotic analysis for the SU(2) vertex amplitude is to rewrite the integral form of the coherent amplitude \eqref{su2cohvertex} as 
\begin{equation}
A_v( j  ,  {\bf n} ) =  (-1)^\chi \int_{\rm SU(2)} \left( \prod_{a=1}^5 \d G_a \right)\, \delta (G_1) \,  \exp {\left(  S( j, {\bf n}) \right)}
\end{equation}
where 
\begin{equation}\label{eqn:action} 
S (j, {\bf n})  = \sum_{a <b}  2 j_{ab} \, \ln   \langle  - {\bf n}_{ba} | G_{a}^\dagger \, G_b \,| {\bf n}_{ba}  \rangle 
\end{equation}
is the associated \emph{action} for the asymptotic problem. By scaling all the spins equally, i.e., $j_{ab} \rightarrow \lambda j_{ab}$, the asymptotic limit can be investigated in the limit $\lambda \rightarrow \infty$,  where $\lambda$ takes values in positive integers. 
The critical point equations obtained by varying the action with respect to the group elements given by 
\begin{equation}\label{eqn:critical}  
G_a {\bf n}_{ab} = - G_{b}  {\bf n}_{ba} \,\,\,\,\, \forall \, ab, \q \text{and} \q     \q  \sum_b j_{ab}\frac{\langle  - {\bf n}_{ba} | \sigma_I \, G_{a}^\dagger \, G_b \,| {\bf n}_{ba}  \rangle }{ \langle  - {\bf n}_{ba} | G_{a}^\dagger \, G_b \,| {\bf n}_{ba}  \rangle  } = {\bf 0} ,\,\,\,\,\, \forall \, a
\end{equation}
where $\sigma_I , I \in  \{1,2,3\}$ are the Pauli matrices. 
These equations imply gluing conditions  for the triangles and closure conditions \eqref{eqn:closure} for each tetrahedron respectively. Therefore, the existence of solutions to the critical point equations depends on the nature of the  coherent boundary data of the vertex. 

The asymptotic analysis for various boundary data has extensively been studied in \cite{Barrett:2009gg,Barrett:2009as,Dona:2017dvf}. Here, we only state their results. If the boundary data associated with the vertex correspond to a vector geometry, then the critical point equations in \eqref{eqn:critical} are satisfied. If the vector geometry data is non-geometric (does not correspond to a geometric 4-simplex), then there is at most one solution (up to equivalence) to the critical point equations, given by $ G^*_a = \pm \mathds 1, \, \forall a $. The asymptotic formula for these non-geometric data is given by 
\begin{equation}\label{eqn:vg_asmp} 
A_v(\lambda j, {\bf n})  \,\, \sim \,\, (-1)^{\chi}  \, \left( \frac{2\pi}{\lambda} \right)^6 \left( \frac{2}{4\pi^2} \right)^4      \frac{1}{\sqrt{\det H} } ,      \q  
\end{equation}
where $H$ is the Hessian of the action \eqref{eqn:action}  evaluated at the critical point. The asymptotic formulae compared to the coherent amplitudes for several vector geometry configurations are discussed in Section \ref{sec:vec_geoms}.

\begin{figure}[ht!]
\centering 
\includegraphics[scale=0.41,valign=c]{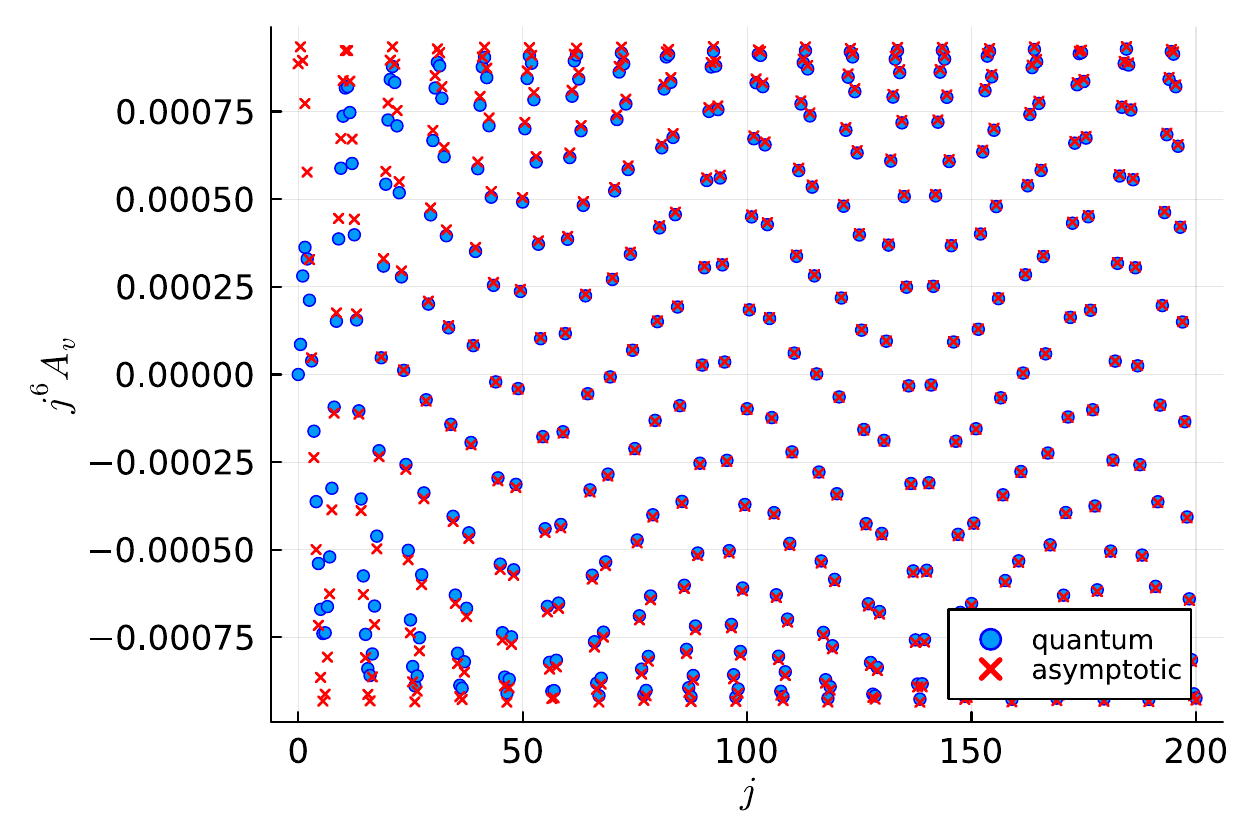} 
\includegraphics[scale=0.36,valign=c]{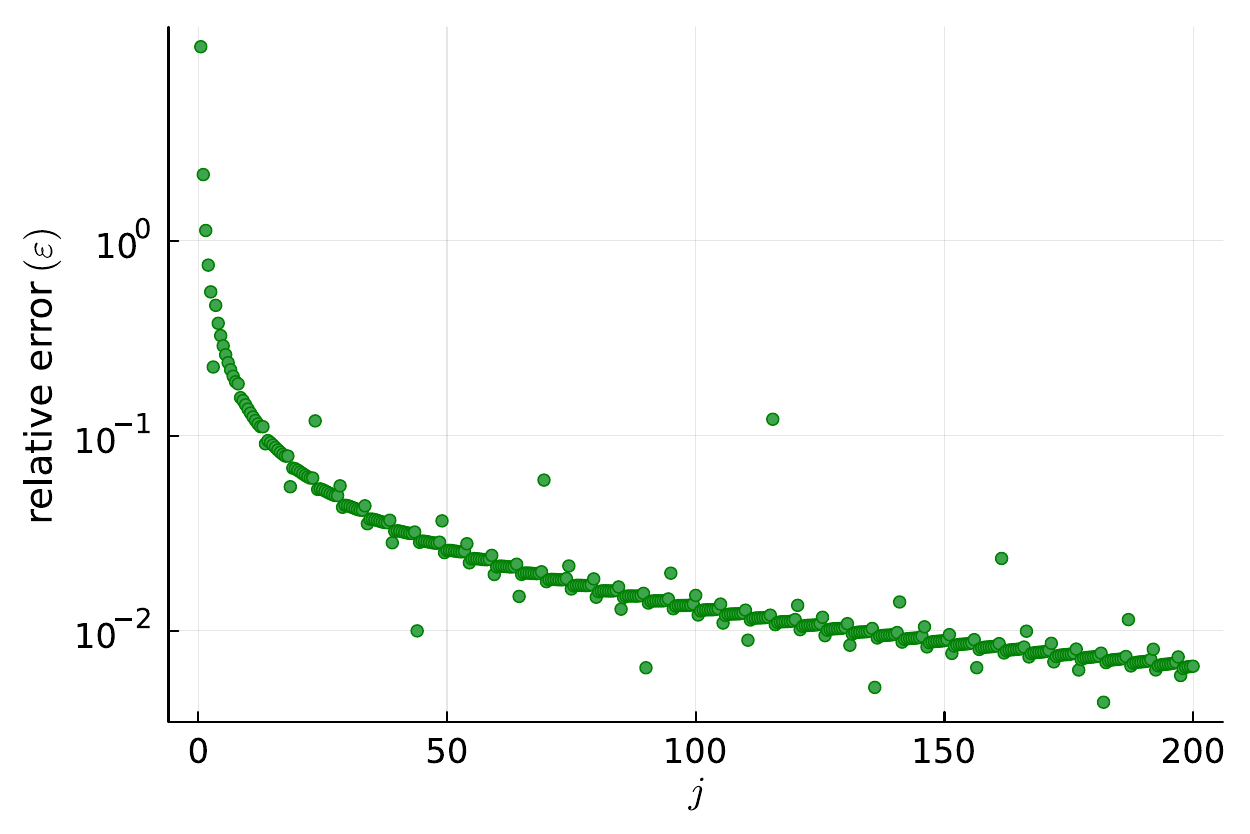} 
\caption{The left panel displays the equilateral coherent vertex amplitude overlapped with its asymptotic formula. The right panel depicts the logarithmic plot of the relative error $\varepsilon$ as a function of spin $j$. }
\label{fig:cohvertexasy_equi}
\end{figure}

On the other hand, if the boundary data for a vertex corresponds to a well-defined 4-simplex in $\mathbb R^4$, i.e., a Regge data, then the critical points equations have two inequivalent solutions.  This leads to an oscillatory asymptotic behavior given by the formula
\begin{equation}\label{Reggeasym}  
A_v(\lambda j, {\bf n})  \,\, \sim \,\, (-1)^{\chi}  \, \left( \frac{2\pi}{\lambda} \right)^6 \left( \frac{2}{4\pi^2} \right)^4   \left(   \frac{e^{+ \i \, S_R(\lambda j,\theta)  }}{\sqrt{\det H_+} } + \frac{e^{-\i \,S_R(\lambda j,\theta)  }}{ \sqrt{ \det   H_-}  }  \right)  , \q  
\end{equation}
where $H_\pm$ are the Hessians of the action evaluated at the two solutions. $S_R (j,\theta):= \sum_f j_f \theta_f$ is the boundary Regge action for the 4-simplex which is dual to the vertex, where $\theta_f$ is the external dihedral angle corresponding to the face $f$ dual to a triangle. For all other boundary data that are neither non-degenerate nor a vector geometry, the amplitude is exponentially suppressed for large boundary scale $\lambda$. 

The coherent amplitude for the equilateral vertex is compared to its asymptotic formula in Figure \ref{fig:cohvertexasy_equi}. The asymptotic formula matches the coherent amplitude well for relatively small spins. The right panel of  Figure \ref{fig:cohvertexasy_equi} shows a log plot for the relative error $\varepsilon = |(A_v - A_v^{\text{asy}} )/A_v| $ between the asymptotic formula and the equilateral vertex amplitude. At spin $j=20$, the relative error is approximately 10\%,  and it further decreases to below 1\% relative error for spins $j \geq 110$.

~

\bibliography{bibsample}

\begin{thebibliography}{59}%
\makeatletter
\providecommand \@ifxundefined [1]{%
 \@ifx{#1\undefined}
}%
\providecommand \@ifnum [1]{%
 \ifnum #1\expandafter \@firstoftwo
 \else \expandafter \@secondoftwo
 \fi
}%
\providecommand \@ifx [1]{%
 \ifx #1\expandafter \@firstoftwo
 \else \expandafter \@secondoftwo
 \fi
}%
\providecommand \natexlab [1]{#1}%
\providecommand \enquote  [1]{``#1''}%
\providecommand \bibnamefont  [1]{#1}%
\providecommand \bibfnamefont [1]{#1}%
\providecommand \citenamefont [1]{#1}%
\providecommand \href@noop [0]{\@secondoftwo}%
\providecommand \href [0]{\begingroup \@sanitize@url \@href}%
\providecommand \@href[1]{\@@startlink{#1}\@@href}%
\providecommand \@@href[1]{\endgroup#1\@@endlink}%
\providecommand \@sanitize@url [0]{\catcode `\\12\catcode `\$12\catcode
  `\&12\catcode `\#12\catcode `\^12\catcode `\_12\catcode `\%12\relax}%
\providecommand \@@startlink[1]{}%
\providecommand \@@endlink[0]{}%
\providecommand \url  [0]{\begingroup\@sanitize@url \@url }%
\providecommand \@url [1]{\endgroup\@href {#1}{\urlprefix }}%
\providecommand \urlprefix  [0]{URL }%
\providecommand \Eprint [0]{\href }%
\providecommand \doibase [0]{https://doi.org/}%
\providecommand \selectlanguage [0]{\@gobble}%
\providecommand \bibinfo  [0]{\@secondoftwo}%
\providecommand \bibfield  [0]{\@secondoftwo}%
\providecommand \translation [1]{[#1]}%
\providecommand \BibitemOpen [0]{}%
\providecommand \bibitemStop [0]{}%
\providecommand \bibitemNoStop [0]{.\EOS\space}%
\providecommand \EOS [0]{\spacefactor3000\relax}%
\providecommand \BibitemShut  [1]{\csname bibitem#1\endcsname}%
\let\auto@bib@innerbib\@empty
%</preamble>
\bibitem [{\citenamefont {Bambi}\ \emph {et~al.}(2023)\citenamefont {Bambi},
  \citenamefont {Modesto},\ and\ \citenamefont {Shapiro}}]{Bambi:2023jiz}%
  \BibitemOpen
  \bibinfo {editor} {\bibfnamefont {C.}~\bibnamefont {Bambi}}, \bibinfo
  {editor} {\bibfnamefont {L.}~\bibnamefont {Modesto}},\ and\ \bibinfo {editor}
  {\bibfnamefont {I.}~\bibnamefont {Shapiro}},\ eds.,\ \href
  {https://doi.org/10.1007/978-981-19-3079-9} {\emph {\bibinfo {title}
  {{Handbook of Quantum Gravity}}}}\ (\bibinfo  {publisher} {Springer},\
  \bibinfo {year} {2023})\BibitemShut {NoStop}%
\bibitem [{\citenamefont {Engle}\ and\ \citenamefont
  {Speziale}(2023)}]{Engle:2023qsu}%
  \BibitemOpen
  \bibfield  {author} {\bibinfo {author} {\bibfnamefont {J.}~\bibnamefont
  {Engle}}\ and\ \bibinfo {author} {\bibfnamefont {S.}~\bibnamefont
  {Speziale}},\ }\bibinfo {title} {{Spin Foams: Foundations}}\ (\bibinfo {year}
  {2023})\ \Eprint {https://arxiv.org/abs/2310.20147} {arXiv:2310.20147
  [gr-qc]} \BibitemShut {NoStop}%
\bibitem [{\citenamefont {Livine}(2024)}]{Livine:2024hhc}%
  \BibitemOpen
  \bibfield  {author} {\bibinfo {author} {\bibfnamefont {E.~R.}\ \bibnamefont
  {Livine}},\ }\bibfield  {title} {\bibinfo {title} {{Spinfoam Models for
  Quantum Gravity: Overview}},\ }\href@noop {} {\  (\bibinfo {year} {2024})},\
  \Eprint {https://arxiv.org/abs/2403.09364} {arXiv:2403.09364 [gr-qc]}
  \BibitemShut {NoStop}%
\bibitem [{\citenamefont {Dona}\ \emph {et~al.}(2023)\citenamefont {Dona},
  \citenamefont {Han},\ and\ \citenamefont {Liu}}]{Dona:2022yyn}%
  \BibitemOpen
  \bibfield  {author} {\bibinfo {author} {\bibfnamefont {P.}~\bibnamefont
  {Dona}}, \bibinfo {author} {\bibfnamefont {M.}~\bibnamefont {Han}},\ and\
  \bibinfo {author} {\bibfnamefont {H.}~\bibnamefont {Liu}},\ }\bibinfo {title}
  {{Spinfoams and High-Performance Computing}},\ in\ \href
  {https://doi.org/10.1007/978-981-19-3079-9_100-1} {\emph {\bibinfo
  {booktitle} {{Handbook of Quantum Gravity}}}},\ \bibinfo {editor} {edited by\
  \bibinfo {editor} {\bibfnamefont {C.}~\bibnamefont {Bambi}}, \bibinfo
  {editor} {\bibfnamefont {L.}~\bibnamefont {Modesto}},\ and\ \bibinfo {editor}
  {\bibfnamefont {I.}~\bibnamefont {Shapiro}}}\ (\bibinfo {year} {2023})\ pp.\
  \bibinfo {pages} {1--38},\ \Eprint {https://arxiv.org/abs/2212.14396}
  {arXiv:2212.14396 [gr-qc]} \BibitemShut {NoStop}%
\bibitem [{\citenamefont {Don\`a}\ \emph {et~al.}(2019)\citenamefont {Don\`a},
  \citenamefont {Fanizza}, \citenamefont {Sarno},\ and\ \citenamefont
  {Speziale}}]{Dona:2019dkf}%
  \BibitemOpen
  \bibfield  {author} {\bibinfo {author} {\bibfnamefont {P.}~\bibnamefont
  {Don\`a}}, \bibinfo {author} {\bibfnamefont {M.}~\bibnamefont {Fanizza}},
  \bibinfo {author} {\bibfnamefont {G.}~\bibnamefont {Sarno}},\ and\ \bibinfo
  {author} {\bibfnamefont {S.}~\bibnamefont {Speziale}},\ }\bibfield  {title}
  {\bibinfo {title} {{Numerical study of the Lorentzian
  Engle-Pereira-Rovelli-Livine spin foam amplitude}},\ }\href
  {https://doi.org/10.1103/PhysRevD.100.106003} {\bibfield  {journal} {\bibinfo
   {journal} {Phys. Rev. D}\ }\textbf {\bibinfo {volume} {100}},\ \bibinfo
  {pages} {106003} (\bibinfo {year} {2019})},\ \Eprint
  {https://arxiv.org/abs/1903.12624} {arXiv:1903.12624 [gr-qc]} \BibitemShut
  {NoStop}%
\bibitem [{\citenamefont {Gozzini}(2021)}]{Gozzini:2021kbt}%
  \BibitemOpen
  \bibfield  {author} {\bibinfo {author} {\bibfnamefont {F.}~\bibnamefont
  {Gozzini}},\ }\bibfield  {title} {\bibinfo {title} {{A high-performance code
  for EPRL spin foam amplitudes}},\ }\href
  {https://doi.org/10.1088/1361-6382/ac2b0b} {\bibfield  {journal} {\bibinfo
  {journal} {Class. Quant. Grav.}\ }\textbf {\bibinfo {volume} {38}},\ \bibinfo
  {pages} {225010} (\bibinfo {year} {2021})},\ \Eprint
  {https://arxiv.org/abs/2107.13952} {arXiv:2107.13952 [gr-qc]} \BibitemShut
  {NoStop}%
\bibitem [{\citenamefont {Dona}\ and\ \citenamefont
  {Frisoni}(2022)}]{Dona:2022dxs}%
  \BibitemOpen
  \bibfield  {author} {\bibinfo {author} {\bibfnamefont {P.}~\bibnamefont
  {Dona}}\ and\ \bibinfo {author} {\bibfnamefont {P.}~\bibnamefont {Frisoni}},\
  }\bibfield  {title} {\bibinfo {title} {{How-to Compute EPRL Spin Foam
  Amplitudes}},\ }\href {https://doi.org/10.3390/universe8040208} {\bibfield
  {journal} {\bibinfo  {journal} {Universe}\ }\textbf {\bibinfo {volume} {8}},\
  \bibinfo {pages} {208} (\bibinfo {year} {2022})},\ \Eprint
  {https://arxiv.org/abs/2202.04360} {arXiv:2202.04360 [gr-qc]} \BibitemShut
  {NoStop}%
\bibitem [{\citenamefont {Engle}\ \emph {et~al.}(2007)\citenamefont {Engle},
  \citenamefont {Pereira},\ and\ \citenamefont {Rovelli}}]{Engle:2007uq}%
  \BibitemOpen
  \bibfield  {author} {\bibinfo {author} {\bibfnamefont {J.}~\bibnamefont
  {Engle}}, \bibinfo {author} {\bibfnamefont {R.}~\bibnamefont {Pereira}},\
  and\ \bibinfo {author} {\bibfnamefont {C.}~\bibnamefont {Rovelli}},\
  }\bibfield  {title} {\bibinfo {title} {{The Loop-quantum-gravity
  vertex-amplitude}},\ }\href {https://doi.org/10.1103/PhysRevLett.99.161301}
  {\bibfield  {journal} {\bibinfo  {journal} {Phys. Rev. Lett.}\ }\textbf
  {\bibinfo {volume} {99}},\ \bibinfo {pages} {161301} (\bibinfo {year}
  {2007})},\ \Eprint {https://arxiv.org/abs/0705.2388} {arXiv:0705.2388
  [gr-qc]} \BibitemShut {NoStop}%
\bibitem [{\citenamefont {Engle}\ \emph {et~al.}(2008)\citenamefont {Engle},
  \citenamefont {Livine}, \citenamefont {Pereira},\ and\ \citenamefont
  {Rovelli}}]{Engle:2007wy}%
  \BibitemOpen
  \bibfield  {author} {\bibinfo {author} {\bibfnamefont {J.}~\bibnamefont
  {Engle}}, \bibinfo {author} {\bibfnamefont {E.}~\bibnamefont {Livine}},
  \bibinfo {author} {\bibfnamefont {R.}~\bibnamefont {Pereira}},\ and\ \bibinfo
  {author} {\bibfnamefont {C.}~\bibnamefont {Rovelli}},\ }\bibfield  {title}
  {\bibinfo {title} {{LQG vertex with finite Immirzi parameter}},\ }\href
  {https://doi.org/10.1016/j.nuclphysb.2008.02.018} {\bibfield  {journal}
  {\bibinfo  {journal} {Nucl. Phys. B}\ }\textbf {\bibinfo {volume} {799}},\
  \bibinfo {pages} {136} (\bibinfo {year} {2008})},\ \Eprint
  {https://arxiv.org/abs/0711.0146} {arXiv:0711.0146 [gr-qc]} \BibitemShut
  {NoStop}%
\bibitem [{\citenamefont {Don\`a}\ \emph {et~al.}(2020)\citenamefont {Don\`a},
  \citenamefont {Gozzini},\ and\ \citenamefont {Sarno}}]{Dona:2020tvv}%
  \BibitemOpen
  \bibfield  {author} {\bibinfo {author} {\bibfnamefont {P.}~\bibnamefont
  {Don\`a}}, \bibinfo {author} {\bibfnamefont {F.}~\bibnamefont {Gozzini}},\
  and\ \bibinfo {author} {\bibfnamefont {G.}~\bibnamefont {Sarno}},\ }\bibfield
   {title} {\bibinfo {title} {{Numerical analysis of spin foam dynamics and the
  flatness problem}},\ }\href {https://doi.org/10.1103/PhysRevD.102.106003}
  {\bibfield  {journal} {\bibinfo  {journal} {Phys. Rev. D}\ }\textbf {\bibinfo
  {volume} {102}},\ \bibinfo {pages} {106003} (\bibinfo {year} {2020})},\
  \Eprint {https://arxiv.org/abs/2004.12911} {arXiv:2004.12911 [gr-qc]}
  \BibitemShut {NoStop}%
\bibitem [{\citenamefont {Don\`a}\ \emph {et~al.}(2022)\citenamefont {Don\`a},
  \citenamefont {Frisoni},\ and\ \citenamefont {Wilson-Ewing}}]{Dona:2022vyh}%
  \BibitemOpen
  \bibfield  {author} {\bibinfo {author} {\bibfnamefont {P.}~\bibnamefont
  {Don\`a}}, \bibinfo {author} {\bibfnamefont {P.}~\bibnamefont {Frisoni}},\
  and\ \bibinfo {author} {\bibfnamefont {E.}~\bibnamefont {Wilson-Ewing}},\
  }\bibfield  {title} {\bibinfo {title} {{Radiative corrections to the
  Lorentzian Engle-Pereira-Rovelli-Livine spin foam propagator}},\ }\href
  {https://doi.org/10.1103/PhysRevD.106.066022} {\bibfield  {journal} {\bibinfo
   {journal} {Phys. Rev. D}\ }\textbf {\bibinfo {volume} {106}},\ \bibinfo
  {pages} {066022} (\bibinfo {year} {2022})},\ \Eprint
  {https://arxiv.org/abs/2206.14755} {arXiv:2206.14755 [gr-qc]} \BibitemShut
  {NoStop}%
\bibitem [{\citenamefont {Frisoni}\ \emph {et~al.}(2023)\citenamefont
  {Frisoni}, \citenamefont {Gozzini},\ and\ \citenamefont
  {Vidotto}}]{Frisoni:2022urv}%
  \BibitemOpen
  \bibfield  {author} {\bibinfo {author} {\bibfnamefont {P.}~\bibnamefont
  {Frisoni}}, \bibinfo {author} {\bibfnamefont {F.}~\bibnamefont {Gozzini}},\
  and\ \bibinfo {author} {\bibfnamefont {F.}~\bibnamefont {Vidotto}},\
  }\bibfield  {title} {\bibinfo {title} {{Markov chain Monte Carlo methods for
  graph refinement in spinfoam cosmology}},\ }\href
  {https://doi.org/10.1088/1361-6382/acc5d6} {\bibfield  {journal} {\bibinfo
  {journal} {Class. Quant. Grav.}\ }\textbf {\bibinfo {volume} {40}},\ \bibinfo
  {pages} {105001} (\bibinfo {year} {2023})},\ \Eprint
  {https://arxiv.org/abs/2207.02881} {arXiv:2207.02881 [gr-qc]} \BibitemShut
  {NoStop}%
\bibitem [{\citenamefont {Han}\ \emph {et~al.}(2021)\citenamefont {Han},
  \citenamefont {Huang}, \citenamefont {Liu}, \citenamefont {Qu},\ and\
  \citenamefont {Wan}}]{Han:2020npv}%
  \BibitemOpen
  \bibfield  {author} {\bibinfo {author} {\bibfnamefont {M.}~\bibnamefont
  {Han}}, \bibinfo {author} {\bibfnamefont {Z.}~\bibnamefont {Huang}}, \bibinfo
  {author} {\bibfnamefont {H.}~\bibnamefont {Liu}}, \bibinfo {author}
  {\bibfnamefont {D.}~\bibnamefont {Qu}},\ and\ \bibinfo {author}
  {\bibfnamefont {Y.}~\bibnamefont {Wan}},\ }\bibfield  {title} {\bibinfo
  {title} {{Spinfoam on a Lefschetz thimble: Markov chain Monte Carlo
  computation of a Lorentzian spinfoam propagator}},\ }\href
  {https://doi.org/10.1103/PhysRevD.103.084026} {\bibfield  {journal} {\bibinfo
   {journal} {Phys. Rev. D}\ }\textbf {\bibinfo {volume} {103}},\ \bibinfo
  {pages} {084026} (\bibinfo {year} {2021})},\ \Eprint
  {https://arxiv.org/abs/2012.11515} {arXiv:2012.11515 [gr-qc]} \BibitemShut
  {NoStop}%
\bibitem [{\citenamefont {Don\`a}\ and\ \citenamefont
  {Frisoni}(2023)}]{Dona:2023myv}%
  \BibitemOpen
  \bibfield  {author} {\bibinfo {author} {\bibfnamefont {P.}~\bibnamefont
  {Don\`a}}\ and\ \bibinfo {author} {\bibfnamefont {P.}~\bibnamefont
  {Frisoni}},\ }\bibfield  {title} {\bibinfo {title} {{Summing bulk quantum
  numbers with Monte~Carlo in spin foam theories}},\ }\href
  {https://doi.org/10.1103/PhysRevD.107.106008} {\bibfield  {journal} {\bibinfo
   {journal} {Phys. Rev. D}\ }\textbf {\bibinfo {volume} {107}},\ \bibinfo
  {pages} {106008} (\bibinfo {year} {2023})},\ \Eprint
  {https://arxiv.org/abs/2302.00072} {arXiv:2302.00072 [gr-qc]} \BibitemShut
  {NoStop}%
\bibitem [{\citenamefont {Steinhaus}(2024)}]{Steinhaus:2024qov}%
  \BibitemOpen
  \bibfield  {author} {\bibinfo {author} {\bibfnamefont {S.}~\bibnamefont
  {Steinhaus}},\ }\bibfield  {title} {\bibinfo {title} {{A Monte Carlo
  algorithm for spin foam intertwiners}},\ }\href@noop {} {\  (\bibinfo {year}
  {2024})},\ \Eprint {https://arxiv.org/abs/2403.04836} {arXiv:2403.04836
  [gr-qc]} \BibitemShut {NoStop}%
\bibitem [{\citenamefont {Han}\ \emph {et~al.}(2022)\citenamefont {Han},
  \citenamefont {Huang}, \citenamefont {Liu},\ and\ \citenamefont
  {Qu}}]{Han:2021kll}%
  \BibitemOpen
  \bibfield  {author} {\bibinfo {author} {\bibfnamefont {M.}~\bibnamefont
  {Han}}, \bibinfo {author} {\bibfnamefont {Z.}~\bibnamefont {Huang}}, \bibinfo
  {author} {\bibfnamefont {H.}~\bibnamefont {Liu}},\ and\ \bibinfo {author}
  {\bibfnamefont {D.}~\bibnamefont {Qu}},\ }\bibfield  {title} {\bibinfo
  {title} {{Complex critical points and curved geometries in four-dimensional
  Lorentzian spinfoam quantum gravity}},\ }\href
  {https://doi.org/10.1103/PhysRevD.106.044005} {\bibfield  {journal} {\bibinfo
   {journal} {Phys. Rev. D}\ }\textbf {\bibinfo {volume} {106}},\ \bibinfo
  {pages} {044005} (\bibinfo {year} {2022})},\ \Eprint
  {https://arxiv.org/abs/2110.10670} {arXiv:2110.10670 [gr-qc]} \BibitemShut
  {NoStop}%
\bibitem [{\citenamefont {Han}\ \emph {et~al.}(2023)\citenamefont {Han},
  \citenamefont {Liu},\ and\ \citenamefont {Qu}}]{Han:2023cen}%
  \BibitemOpen
  \bibfield  {author} {\bibinfo {author} {\bibfnamefont {M.}~\bibnamefont
  {Han}}, \bibinfo {author} {\bibfnamefont {H.}~\bibnamefont {Liu}},\ and\
  \bibinfo {author} {\bibfnamefont {D.}~\bibnamefont {Qu}},\ }\bibfield
  {title} {\bibinfo {title} {{Complex critical points in Lorentzian spinfoam
  quantum gravity: Four-simplex amplitude and effective dynamics on a
  double-\ensuremath{\Delta}3 complex}},\ }\href
  {https://doi.org/10.1103/PhysRevD.108.026010} {\bibfield  {journal} {\bibinfo
   {journal} {Phys. Rev. D}\ }\textbf {\bibinfo {volume} {108}},\ \bibinfo
  {pages} {026010} (\bibinfo {year} {2023})},\ \Eprint
  {https://arxiv.org/abs/2301.02930} {arXiv:2301.02930 [gr-qc]} \BibitemShut
  {NoStop}%
\bibitem [{\citenamefont {Han}\ \emph {et~al.}(2024)\citenamefont {Han},
  \citenamefont {Liu},\ and\ \citenamefont {Qu}}]{Han:2024lti}%
  \BibitemOpen
  \bibfield  {author} {\bibinfo {author} {\bibfnamefont {M.}~\bibnamefont
  {Han}}, \bibinfo {author} {\bibfnamefont {H.}~\bibnamefont {Liu}},\ and\
  \bibinfo {author} {\bibfnamefont {D.}~\bibnamefont {Qu}},\ }\bibfield
  {title} {\bibinfo {title} {{A Mathematica program for numerically computing
  real and complex critical points in 4-dimensional Lorentzian spinfoam
  amplitude}},\ }\href@noop {} {\  (\bibinfo {year} {2024})},\ \Eprint
  {https://arxiv.org/abs/2404.10563} {arXiv:2404.10563 [gr-qc]} \BibitemShut
  {NoStop}%
\bibitem [{\citenamefont {Asante}\ \emph
  {et~al.}(2023{\natexlab{a}})\citenamefont {Asante}, \citenamefont {Sim\~ao},\
  and\ \citenamefont {Steinhaus}}]{Asante:2022lnp}%
  \BibitemOpen
  \bibfield  {author} {\bibinfo {author} {\bibfnamefont {S.~K.}\ \bibnamefont
  {Asante}}, \bibinfo {author} {\bibfnamefont {J.~D.}\ \bibnamefont
  {Sim\~ao}},\ and\ \bibinfo {author} {\bibfnamefont {S.}~\bibnamefont
  {Steinhaus}},\ }\bibfield  {title} {\bibinfo {title} {{Spin-foams as
  semiclassical vertices: Gluing constraints and a hybrid algorithm}},\ }\href
  {https://doi.org/10.1103/PhysRevD.107.046002} {\bibfield  {journal} {\bibinfo
   {journal} {Phys. Rev. D}\ }\textbf {\bibinfo {volume} {107}},\ \bibinfo
  {pages} {046002} (\bibinfo {year} {2023}{\natexlab{a}})},\ \Eprint
  {https://arxiv.org/abs/2206.13540} {arXiv:2206.13540 [gr-qc]} \BibitemShut
  {NoStop}%
\bibitem [{\citenamefont {Bahr}\ and\ \citenamefont
  {Steinhaus}(2016{\natexlab{a}})}]{Bahr:2015gxa}%
  \BibitemOpen
  \bibfield  {author} {\bibinfo {author} {\bibfnamefont {B.}~\bibnamefont
  {Bahr}}\ and\ \bibinfo {author} {\bibfnamefont {S.}~\bibnamefont
  {Steinhaus}},\ }\bibfield  {title} {\bibinfo {title} {{Investigation of the
  Spinfoam Path integral with Quantum Cuboid Intertwiners}},\ }\href
  {https://doi.org/10.1103/PhysRevD.93.104029} {\bibfield  {journal} {\bibinfo
  {journal} {Phys. Rev. D}\ }\textbf {\bibinfo {volume} {93}},\ \bibinfo
  {pages} {104029} (\bibinfo {year} {2016}{\natexlab{a}})},\ \Eprint
  {https://arxiv.org/abs/1508.07961} {arXiv:1508.07961 [gr-qc]} \BibitemShut
  {NoStop}%
\bibitem [{\citenamefont {Bahr}\ and\ \citenamefont
  {Steinhaus}(2016{\natexlab{b}})}]{Bahr:2016hwc}%
  \BibitemOpen
  \bibfield  {author} {\bibinfo {author} {\bibfnamefont {B.}~\bibnamefont
  {Bahr}}\ and\ \bibinfo {author} {\bibfnamefont {S.}~\bibnamefont
  {Steinhaus}},\ }\bibfield  {title} {\bibinfo {title} {{Numerical evidence for
  a phase transition in 4d spin foam quantum gravity}},\ }\href
  {https://doi.org/10.1103/PhysRevLett.117.141302} {\bibfield  {journal}
  {\bibinfo  {journal} {Phys. Rev. Lett.}\ }\textbf {\bibinfo {volume} {117}},\
  \bibinfo {pages} {141302} (\bibinfo {year} {2016}{\natexlab{b}})},\ \Eprint
  {https://arxiv.org/abs/1605.07649} {arXiv:1605.07649 [gr-qc]} \BibitemShut
  {NoStop}%
\bibitem [{\citenamefont {Asante}\ \emph {et~al.}(2020)\citenamefont {Asante},
  \citenamefont {Dittrich},\ and\ \citenamefont {Haggard}}]{Asante:2020qpa}%
  \BibitemOpen
  \bibfield  {author} {\bibinfo {author} {\bibfnamefont {S.~K.}\ \bibnamefont
  {Asante}}, \bibinfo {author} {\bibfnamefont {B.}~\bibnamefont {Dittrich}},\
  and\ \bibinfo {author} {\bibfnamefont {H.~M.}\ \bibnamefont {Haggard}},\
  }\bibfield  {title} {\bibinfo {title} {{Effective Spin Foam Models for
  Four-Dimensional Quantum Gravity}},\ }\href
  {https://doi.org/10.1103/PhysRevLett.125.231301} {\bibfield  {journal}
  {\bibinfo  {journal} {Phys. Rev. Lett.}\ }\textbf {\bibinfo {volume} {125}},\
  \bibinfo {pages} {231301} (\bibinfo {year} {2020})},\ \Eprint
  {https://arxiv.org/abs/2004.07013} {arXiv:2004.07013 [gr-qc]} \BibitemShut
  {NoStop}%
\bibitem [{\citenamefont {Asante}\ \emph
  {et~al.}(2021{\natexlab{a}})\citenamefont {Asante}, \citenamefont
  {Dittrich},\ and\ \citenamefont {Padua-Arguelles}}]{Asante:2021zzh}%
  \BibitemOpen
  \bibfield  {author} {\bibinfo {author} {\bibfnamefont {S.~K.}\ \bibnamefont
  {Asante}}, \bibinfo {author} {\bibfnamefont {B.}~\bibnamefont {Dittrich}},\
  and\ \bibinfo {author} {\bibfnamefont {J.}~\bibnamefont {Padua-Arguelles}},\
  }\bibfield  {title} {\bibinfo {title} {{Effective spin foam models for
  Lorentzian quantum gravity}},\ }\href
  {https://doi.org/10.1088/1361-6382/ac1b44} {\bibfield  {journal} {\bibinfo
  {journal} {Class. Quant. Grav.}\ }\textbf {\bibinfo {volume} {38}},\ \bibinfo
  {pages} {195002} (\bibinfo {year} {2021}{\natexlab{a}})},\ \Eprint
  {https://arxiv.org/abs/2104.00485} {arXiv:2104.00485 [gr-qc]} \BibitemShut
  {NoStop}%
\bibitem [{\citenamefont {Asante}\ \emph
  {et~al.}(2021{\natexlab{b}})\citenamefont {Asante}, \citenamefont
  {Dittrich},\ and\ \citenamefont {Haggard}}]{Asante:2020iwm}%
  \BibitemOpen
  \bibfield  {author} {\bibinfo {author} {\bibfnamefont {S.~K.}\ \bibnamefont
  {Asante}}, \bibinfo {author} {\bibfnamefont {B.}~\bibnamefont {Dittrich}},\
  and\ \bibinfo {author} {\bibfnamefont {H.~M.}\ \bibnamefont {Haggard}},\
  }\bibfield  {title} {\bibinfo {title} {{Discrete gravity dynamics from
  effective spin foams}},\ }\href {https://doi.org/10.1088/1361-6382/ac011b}
  {\bibfield  {journal} {\bibinfo  {journal} {Class. Quant. Grav.}\ }\textbf
  {\bibinfo {volume} {38}},\ \bibinfo {pages} {145023} (\bibinfo {year}
  {2021}{\natexlab{b}})},\ \Eprint {https://arxiv.org/abs/2011.14468}
  {arXiv:2011.14468 [gr-qc]} \BibitemShut {NoStop}%
\bibitem [{\citenamefont {Dittrich}(2021)}]{Dittrich:2021kzs}%
  \BibitemOpen
  \bibfield  {author} {\bibinfo {author} {\bibfnamefont {B.}~\bibnamefont
  {Dittrich}},\ }\bibfield  {title} {\bibinfo {title} {{Modified Graviton
  Dynamics From Spin Foams: The Area Regge Action}},\ }\href@noop {} {\
  (\bibinfo {year} {2021})},\ \Eprint {https://arxiv.org/abs/2105.10808}
  {arXiv:2105.10808 [gr-qc]} \BibitemShut {NoStop}%
\bibitem [{\citenamefont {Dittrich}\ and\ \citenamefont
  {Kogios}(2023)}]{Dittrich:2022yoo}%
  \BibitemOpen
  \bibfield  {author} {\bibinfo {author} {\bibfnamefont {B.}~\bibnamefont
  {Dittrich}}\ and\ \bibinfo {author} {\bibfnamefont {A.}~\bibnamefont
  {Kogios}},\ }\bibfield  {title} {\bibinfo {title} {{From spin foams to area
  metric dynamics to gravitons}},\ }\href
  {https://doi.org/10.1088/1361-6382/acc5d9} {\bibfield  {journal} {\bibinfo
  {journal} {Class. Quant. Grav.}\ }\textbf {\bibinfo {volume} {40}},\ \bibinfo
  {pages} {095011} (\bibinfo {year} {2023})},\ \Eprint
  {https://arxiv.org/abs/2203.02409} {arXiv:2203.02409 [gr-qc]} \BibitemShut
  {NoStop}%
\bibitem [{\citenamefont {Borissova}\ and\ \citenamefont
  {Dittrich}(2023)}]{Borissova:2022clg}%
  \BibitemOpen
  \bibfield  {author} {\bibinfo {author} {\bibfnamefont {J.~N.}\ \bibnamefont
  {Borissova}}\ and\ \bibinfo {author} {\bibfnamefont {B.}~\bibnamefont
  {Dittrich}},\ }\bibfield  {title} {\bibinfo {title} {{Towards effective
  actions for the continuum limit of spin foams}},\ }\href
  {https://doi.org/10.1088/1361-6382/accbfb} {\bibfield  {journal} {\bibinfo
  {journal} {Class. Quant. Grav.}\ }\textbf {\bibinfo {volume} {40}},\ \bibinfo
  {pages} {105006} (\bibinfo {year} {2023})},\ \Eprint
  {https://arxiv.org/abs/2207.03307} {arXiv:2207.03307 [gr-qc]} \BibitemShut
  {NoStop}%
\bibitem [{\citenamefont {Speziale}(2017)}]{Speziale:2016axj}%
  \BibitemOpen
  \bibfield  {author} {\bibinfo {author} {\bibfnamefont {S.}~\bibnamefont
  {Speziale}},\ }\bibfield  {title} {\bibinfo {title} {{Boosting
  Wigner\textquoteright{}s nj-symbols}},\ }\href
  {https://doi.org/10.1063/1.4977752} {\bibfield  {journal} {\bibinfo
  {journal} {J. Math. Phys.}\ }\textbf {\bibinfo {volume} {58}},\ \bibinfo
  {pages} {032501} (\bibinfo {year} {2017})},\ \Eprint
  {https://arxiv.org/abs/1609.01632} {arXiv:1609.01632 [gr-qc]} \BibitemShut
  {NoStop}%
\bibitem [{\citenamefont {Orus}(2014)}]{Orus:2013kga}%
  \BibitemOpen
  \bibfield  {author} {\bibinfo {author} {\bibfnamefont {R.}~\bibnamefont
  {Orus}},\ }\bibfield  {title} {\bibinfo {title} {{A Practical Introduction to
  Tensor Networks: Matrix Product States and Projected Entangled Pair
  States}},\ }\href {https://doi.org/10.1016/j.aop.2014.06.013} {\bibfield
  {journal} {\bibinfo  {journal} {Annals Phys.}\ }\textbf {\bibinfo {volume}
  {349}},\ \bibinfo {pages} {117} (\bibinfo {year} {2014})},\ \Eprint
  {https://arxiv.org/abs/1306.2164} {arXiv:1306.2164 [cond-mat.str-el]}
  \BibitemShut {NoStop}%
\bibitem [{\citenamefont {Biamonte}\ and\ \citenamefont
  {Bergholm}(2017)}]{Biamonte:2017dgr}%
  \BibitemOpen
  \bibfield  {author} {\bibinfo {author} {\bibfnamefont {J.}~\bibnamefont
  {Biamonte}}\ and\ \bibinfo {author} {\bibfnamefont {V.}~\bibnamefont
  {Bergholm}},\ }\bibfield  {title} {\bibinfo {title} {{Tensor Networks in a
  Nutshell}},\ }\href@noop {} {\  (\bibinfo {year} {2017})},\ \Eprint
  {https://arxiv.org/abs/1708.00006} {arXiv:1708.00006 [quant-ph]} \BibitemShut
  {NoStop}%
\bibitem [{\citenamefont {Dittrich}\ \emph {et~al.}(2012)\citenamefont
  {Dittrich}, \citenamefont {Eckert},\ and\ \citenamefont
  {Martin-Benito}}]{Dittrich:2011zh}%
  \BibitemOpen
  \bibfield  {author} {\bibinfo {author} {\bibfnamefont {B.}~\bibnamefont
  {Dittrich}}, \bibinfo {author} {\bibfnamefont {F.~C.}\ \bibnamefont
  {Eckert}},\ and\ \bibinfo {author} {\bibfnamefont {M.}~\bibnamefont
  {Martin-Benito}},\ }\bibfield  {title} {\bibinfo {title} {{Coarse graining
  methods for spin net and spin foam models}},\ }\href
  {https://doi.org/10.1088/1367-2630/14/3/035008} {\bibfield  {journal}
  {\bibinfo  {journal} {New J. Phys.}\ }\textbf {\bibinfo {volume} {14}},\
  \bibinfo {pages} {035008} (\bibinfo {year} {2012})},\ \Eprint
  {https://arxiv.org/abs/1109.4927} {arXiv:1109.4927 [gr-qc]} \BibitemShut
  {NoStop}%
\bibitem [{\citenamefont {Dittrich}\ \emph {et~al.}(2014)\citenamefont
  {Dittrich}, \citenamefont {Martin-Benito},\ and\ \citenamefont
  {Steinhaus}}]{Dittrich:2013voa}%
  \BibitemOpen
  \bibfield  {author} {\bibinfo {author} {\bibfnamefont {B.}~\bibnamefont
  {Dittrich}}, \bibinfo {author} {\bibfnamefont {M.}~\bibnamefont
  {Martin-Benito}},\ and\ \bibinfo {author} {\bibfnamefont {S.}~\bibnamefont
  {Steinhaus}},\ }\bibfield  {title} {\bibinfo {title} {{Quantum group spin
  nets: refinement limit and relation to spin foams}},\ }\href
  {https://doi.org/10.1103/PhysRevD.90.024058} {\bibfield  {journal} {\bibinfo
  {journal} {Phys. Rev. D}\ }\textbf {\bibinfo {volume} {90}},\ \bibinfo
  {pages} {024058} (\bibinfo {year} {2014})},\ \Eprint
  {https://arxiv.org/abs/1312.0905} {arXiv:1312.0905 [gr-qc]} \BibitemShut
  {NoStop}%
\bibitem [{\citenamefont {Dittrich}\ \emph {et~al.}(2016)\citenamefont
  {Dittrich}, \citenamefont {Mizera},\ and\ \citenamefont
  {Steinhaus}}]{Dittrich:2014mxa}%
  \BibitemOpen
  \bibfield  {author} {\bibinfo {author} {\bibfnamefont {B.}~\bibnamefont
  {Dittrich}}, \bibinfo {author} {\bibfnamefont {S.}~\bibnamefont {Mizera}},\
  and\ \bibinfo {author} {\bibfnamefont {S.}~\bibnamefont {Steinhaus}},\
  }\bibfield  {title} {\bibinfo {title} {{Decorated tensor network
  renormalization for lattice gauge theories and spin foam models}},\ }\href
  {https://doi.org/10.1088/1367-2630/18/5/053009} {\bibfield  {journal}
  {\bibinfo  {journal} {New J. Phys.}\ }\textbf {\bibinfo {volume} {18}},\
  \bibinfo {pages} {053009} (\bibinfo {year} {2016})},\ \Eprint
  {https://arxiv.org/abs/1409.2407} {arXiv:1409.2407 [gr-qc]} \BibitemShut
  {NoStop}%
\bibitem [{\citenamefont {Delcamp}\ and\ \citenamefont
  {Dittrich}(2017)}]{Delcamp:2016dqo}%
  \BibitemOpen
  \bibfield  {author} {\bibinfo {author} {\bibfnamefont {C.}~\bibnamefont
  {Delcamp}}\ and\ \bibinfo {author} {\bibfnamefont {B.}~\bibnamefont
  {Dittrich}},\ }\bibfield  {title} {\bibinfo {title} {{Towards a phase diagram
  for spin foams}},\ }\href {https://doi.org/10.1088/1361-6382/aa8f24}
  {\bibfield  {journal} {\bibinfo  {journal} {Class. Quant. Grav.}\ }\textbf
  {\bibinfo {volume} {34}},\ \bibinfo {pages} {225006} (\bibinfo {year}
  {2017})},\ \Eprint {https://arxiv.org/abs/1612.04506} {arXiv:1612.04506
  [gr-qc]} \BibitemShut {NoStop}%
\bibitem [{\citenamefont {Asante}\ \emph
  {et~al.}(2023{\natexlab{b}})\citenamefont {Asante}, \citenamefont
  {Dittrich},\ and\ \citenamefont {Steinhaus}}]{Asante:2022dnj}%
  \BibitemOpen
  \bibfield  {author} {\bibinfo {author} {\bibfnamefont {S.~K.}\ \bibnamefont
  {Asante}}, \bibinfo {author} {\bibfnamefont {B.}~\bibnamefont {Dittrich}},\
  and\ \bibinfo {author} {\bibfnamefont {S.}~\bibnamefont {Steinhaus}},\
  }\bibinfo {title} {{Spin Foams, Refinement Limit, and Renormalization}}\
  (\bibinfo {year} {2023})\ \Eprint {https://arxiv.org/abs/2211.09578}
  {arXiv:2211.09578 [gr-qc]} \BibitemShut {NoStop}%
\bibitem [{\citenamefont {Asante}(2024)}]{AsanteTN2024}%
  \BibitemOpen
  \bibfield  {author} {\bibinfo {author} {\bibfnamefont {S.~K.}\ \bibnamefont
  {Asante}},\ }\href@noop {} {\bibinfo {title} {{Tensor network algorithms for
  SU(2) BF spin foam model}}},\ \bibinfo {howpublished}
  {\url{https://github.com/Seth-Kurankyi/su2bf-TNAlgo}} (\bibinfo {year}
  {2024})\BibitemShut {NoStop}%
\bibitem [{\citenamefont {Baez}(2000)}]{Baez:1999sr}%
  \BibitemOpen
  \bibfield  {author} {\bibinfo {author} {\bibfnamefont {J.~C.}\ \bibnamefont
  {Baez}},\ }\bibfield  {title} {\bibinfo {title} {{An Introduction to Spin
  Foam Models of $BF$ Theory and Quantum Gravity}},\ }\href
  {https://doi.org/10.1007/3-540-46552-9_2} {\bibfield  {journal} {\bibinfo
  {journal} {Lect. Notes Phys.}\ }\textbf {\bibinfo {volume} {543}},\ \bibinfo
  {pages} {25} (\bibinfo {year} {2000})},\ \Eprint
  {https://arxiv.org/abs/gr-qc/9905087} {arXiv:gr-qc/9905087} \BibitemShut
  {NoStop}%
\bibitem [{\citenamefont {Barrett}\ \emph {et~al.}(2010)\citenamefont
  {Barrett}, \citenamefont {Fairbairn},\ and\ \citenamefont
  {Hellmann}}]{Barrett:2009as}%
  \BibitemOpen
  \bibfield  {author} {\bibinfo {author} {\bibfnamefont {J.~W.}\ \bibnamefont
  {Barrett}}, \bibinfo {author} {\bibfnamefont {W.~J.}\ \bibnamefont
  {Fairbairn}},\ and\ \bibinfo {author} {\bibfnamefont {F.}~\bibnamefont
  {Hellmann}},\ }\bibfield  {title} {\bibinfo {title} {{Quantum gravity
  asymptotics from the SU(2) 15j symbol}},\ }\href
  {https://doi.org/10.1142/S0217751X10049281} {\bibfield  {journal} {\bibinfo
  {journal} {Int. J. Mod. Phys. A}\ }\textbf {\bibinfo {volume} {25}},\
  \bibinfo {pages} {2897} (\bibinfo {year} {2010})},\ \Eprint
  {https://arxiv.org/abs/0912.4907} {arXiv:0912.4907 [gr-qc]} \BibitemShut
  {NoStop}%
\bibitem [{\citenamefont {Freidel}\ and\ \citenamefont
  {Krasnov}(2008)}]{Freidel:2007py}%
  \BibitemOpen
  \bibfield  {author} {\bibinfo {author} {\bibfnamefont {L.}~\bibnamefont
  {Freidel}}\ and\ \bibinfo {author} {\bibfnamefont {K.}~\bibnamefont
  {Krasnov}},\ }\bibfield  {title} {\bibinfo {title} {{A New Spin Foam Model
  for 4d Gravity}},\ }\href {https://doi.org/10.1088/0264-9381/25/12/125018}
  {\bibfield  {journal} {\bibinfo  {journal} {Class. Quant. Grav.}\ }\textbf
  {\bibinfo {volume} {25}},\ \bibinfo {pages} {125018} (\bibinfo {year}
  {2008})},\ \Eprint {https://arxiv.org/abs/0708.1595} {arXiv:0708.1595
  [gr-qc]} \BibitemShut {NoStop}%
\bibitem [{\citenamefont {Barrett}\ and\ \citenamefont
  {Crane}(1998)}]{Barrett:1997gw}%
  \BibitemOpen
  \bibfield  {author} {\bibinfo {author} {\bibfnamefont {J.~W.}\ \bibnamefont
  {Barrett}}\ and\ \bibinfo {author} {\bibfnamefont {L.}~\bibnamefont
  {Crane}},\ }\bibfield  {title} {\bibinfo {title} {{Relativistic spin networks
  and quantum gravity}},\ }\href {https://doi.org/10.1063/1.532254} {\bibfield
  {journal} {\bibinfo  {journal} {J. Math. Phys.}\ }\textbf {\bibinfo {volume}
  {39}},\ \bibinfo {pages} {3296} (\bibinfo {year} {1998})},\ \Eprint
  {https://arxiv.org/abs/gr-qc/9709028} {arXiv:gr-qc/9709028} \BibitemShut
  {NoStop}%
\bibitem [{\citenamefont {Barrett}\ and\ \citenamefont
  {Crane}(2000)}]{Barrett:1999qw}%
  \BibitemOpen
  \bibfield  {author} {\bibinfo {author} {\bibfnamefont {J.~W.}\ \bibnamefont
  {Barrett}}\ and\ \bibinfo {author} {\bibfnamefont {L.}~\bibnamefont
  {Crane}},\ }\bibfield  {title} {\bibinfo {title} {{A Lorentzian signature
  model for quantum general relativity}},\ }\href
  {https://doi.org/10.1088/0264-9381/17/16/302} {\bibfield  {journal} {\bibinfo
   {journal} {Class. Quant. Grav.}\ }\textbf {\bibinfo {volume} {17}},\
  \bibinfo {pages} {3101} (\bibinfo {year} {2000})},\ \Eprint
  {https://arxiv.org/abs/gr-qc/9904025} {arXiv:gr-qc/9904025} \BibitemShut
  {NoStop}%
\bibitem [{\citenamefont {Jercher}\ \emph {et~al.}(2022)\citenamefont
  {Jercher}, \citenamefont {Oriti},\ and\ \citenamefont
  {Pithis}}]{Jercher:2022mky}%
  \BibitemOpen
  \bibfield  {author} {\bibinfo {author} {\bibfnamefont {A.~F.}\ \bibnamefont
  {Jercher}}, \bibinfo {author} {\bibfnamefont {D.}~\bibnamefont {Oriti}},\
  and\ \bibinfo {author} {\bibfnamefont {A.~G.~A.}\ \bibnamefont {Pithis}},\
  }\bibfield  {title} {\bibinfo {title} {{Complete Barrett-Crane model and its
  causal structure}},\ }\href {https://doi.org/10.1103/PhysRevD.106.066019}
  {\bibfield  {journal} {\bibinfo  {journal} {Phys. Rev. D}\ }\textbf {\bibinfo
  {volume} {106}},\ \bibinfo {pages} {066019} (\bibinfo {year} {2022})},\
  \Eprint {https://arxiv.org/abs/2206.15442} {arXiv:2206.15442 [gr-qc]}
  \BibitemShut {NoStop}%
\bibitem [{\citenamefont {Ashtekar}\ and\ \citenamefont
  {Lewandowski}(1997)}]{Ashtekar:1996eg}%
  \BibitemOpen
  \bibfield  {author} {\bibinfo {author} {\bibfnamefont {A.}~\bibnamefont
  {Ashtekar}}\ and\ \bibinfo {author} {\bibfnamefont {J.}~\bibnamefont
  {Lewandowski}},\ }\bibfield  {title} {\bibinfo {title} {{Quantum theory of
  geometry. 1: Area operators}},\ }\href
  {https://doi.org/10.1088/0264-9381/14/1A/006} {\bibfield  {journal} {\bibinfo
   {journal} {Class. Quant. Grav.}\ }\textbf {\bibinfo {volume} {14}},\
  \bibinfo {pages} {A55} (\bibinfo {year} {1997})},\ \Eprint
  {https://arxiv.org/abs/gr-qc/9602046} {arXiv:gr-qc/9602046} \BibitemShut
  {NoStop}%
\bibitem [{\citenamefont {Rovelli}\ and\ \citenamefont
  {Smolin}(1995)}]{Rovelli:1995ac}%
  \BibitemOpen
  \bibfield  {author} {\bibinfo {author} {\bibfnamefont {C.}~\bibnamefont
  {Rovelli}}\ and\ \bibinfo {author} {\bibfnamefont {L.}~\bibnamefont
  {Smolin}},\ }\bibfield  {title} {\bibinfo {title} {{Spin networks and quantum
  gravity}},\ }\href {https://doi.org/10.1103/PhysRevD.52.5743} {\bibfield
  {journal} {\bibinfo  {journal} {Phys. Rev. D}\ }\textbf {\bibinfo {volume}
  {52}},\ \bibinfo {pages} {5743} (\bibinfo {year} {1995})},\ \Eprint
  {https://arxiv.org/abs/gr-qc/9505006} {arXiv:gr-qc/9505006} \BibitemShut
  {NoStop}%
\bibitem [{\citenamefont {Bianchi}\ \emph {et~al.}(2010)\citenamefont
  {Bianchi}, \citenamefont {Regoli},\ and\ \citenamefont
  {Rovelli}}]{Bianchi:2010fj}%
  \BibitemOpen
  \bibfield  {author} {\bibinfo {author} {\bibfnamefont {E.}~\bibnamefont
  {Bianchi}}, \bibinfo {author} {\bibfnamefont {D.}~\bibnamefont {Regoli}},\
  and\ \bibinfo {author} {\bibfnamefont {C.}~\bibnamefont {Rovelli}},\
  }\bibfield  {title} {\bibinfo {title} {{Face amplitude of spinfoam quantum
  gravity}},\ }\href {https://doi.org/10.1088/0264-9381/27/18/185009}
  {\bibfield  {journal} {\bibinfo  {journal} {Class. Quant. Grav.}\ }\textbf
  {\bibinfo {volume} {27}},\ \bibinfo {pages} {185009} (\bibinfo {year}
  {2010})},\ \Eprint {https://arxiv.org/abs/1005.0764} {arXiv:1005.0764
  [gr-qc]} \BibitemShut {NoStop}%
\bibitem [{\citenamefont {Ooguri}(1992)}]{Ooguri:1992eb}%
  \BibitemOpen
  \bibfield  {author} {\bibinfo {author} {\bibfnamefont {H.}~\bibnamefont
  {Ooguri}},\ }\bibfield  {title} {\bibinfo {title} {{Topological lattice
  models in four-dimensions}},\ }\href
  {https://doi.org/10.1142/S0217732392004171} {\bibfield  {journal} {\bibinfo
  {journal} {Mod. Phys. Lett. A}\ }\textbf {\bibinfo {volume} {7}},\ \bibinfo
  {pages} {2799} (\bibinfo {year} {1992})},\ \Eprint
  {https://arxiv.org/abs/hep-th/9205090} {arXiv:hep-th/9205090} \BibitemShut
  {NoStop}%
\bibitem [{\citenamefont {Yutsis}\ \emph {et~al.}(1962)\citenamefont {Yutsis},
  \citenamefont {Levinson},\ and\ \citenamefont
  {Vanagas}}]{yutsis1962mathematical}%
  \BibitemOpen
  \bibfield  {author} {\bibinfo {author} {\bibfnamefont {A.~P.}\ \bibnamefont
  {Yutsis}}, \bibinfo {author} {\bibfnamefont {I.~B.}\ \bibnamefont
  {Levinson}},\ and\ \bibinfo {author} {\bibfnamefont {V.~V.}\ \bibnamefont
  {Vanagas}},\ }\bibfield  {title} {\bibinfo {title} {Mathematical apparatus of
  the theory of angular momentum},\ }\href@noop {} {\bibfield  {journal}
  {\bibinfo  {journal} {Academy of Sciences of the Lithuanian SS R}\ }
  (\bibinfo {year} {1962})}\BibitemShut {NoStop}%
\bibitem [{\citenamefont {Livine}\ and\ \citenamefont
  {Speziale}(2007)}]{Livine:2007vk}%
  \BibitemOpen
  \bibfield  {author} {\bibinfo {author} {\bibfnamefont {E.~R.}\ \bibnamefont
  {Livine}}\ and\ \bibinfo {author} {\bibfnamefont {S.}~\bibnamefont
  {Speziale}},\ }\bibfield  {title} {\bibinfo {title} {{A New spinfoam vertex
  for quantum gravity}},\ }\href {https://doi.org/10.1103/PhysRevD.76.084028}
  {\bibfield  {journal} {\bibinfo  {journal} {Phys. Rev. D}\ }\textbf {\bibinfo
  {volume} {76}},\ \bibinfo {pages} {084028} (\bibinfo {year} {2007})},\
  \Eprint {https://arxiv.org/abs/0705.0674} {arXiv:0705.0674 [gr-qc]}
  \BibitemShut {NoStop}%
\bibitem [{\citenamefont {Don\`a}\ \emph {et~al.}(2018)\citenamefont {Don\`a},
  \citenamefont {Fanizza}, \citenamefont {Sarno},\ and\ \citenamefont
  {Speziale}}]{Dona:2017dvf}%
  \BibitemOpen
  \bibfield  {author} {\bibinfo {author} {\bibfnamefont {P.}~\bibnamefont
  {Don\`a}}, \bibinfo {author} {\bibfnamefont {M.}~\bibnamefont {Fanizza}},
  \bibinfo {author} {\bibfnamefont {G.}~\bibnamefont {Sarno}},\ and\ \bibinfo
  {author} {\bibfnamefont {S.}~\bibnamefont {Speziale}},\ }\bibfield  {title}
  {\bibinfo {title} {{SU(2) graph invariants, Regge actions and polytopes}},\
  }\href {https://doi.org/10.1088/1361-6382/aaa53a} {\bibfield  {journal}
  {\bibinfo  {journal} {Class. Quant. Grav.}\ }\textbf {\bibinfo {volume}
  {35}},\ \bibinfo {pages} {045011} (\bibinfo {year} {2018})},\ \Eprint
  {https://arxiv.org/abs/1708.01727} {arXiv:1708.01727 [gr-qc]} \BibitemShut
  {NoStop}%
\bibitem [{\citenamefont {Pachner}(1991)}]{pachner1991pl}%
  \BibitemOpen
  \bibfield  {author} {\bibinfo {author} {\bibfnamefont {U.}~\bibnamefont
  {Pachner}},\ }\bibfield  {title} {\bibinfo {title} {Pl homeomorphic manifolds
  are equivalent by elementary shellings},\ }\href@noop {} {\bibfield
  {journal} {\bibinfo  {journal} {European journal of Combinatorics}\ }\textbf
  {\bibinfo {volume} {12}},\ \bibinfo {pages} {129} (\bibinfo {year}
  {1991})}\BibitemShut {NoStop}%
\bibitem [{\citenamefont {Varshalovich}\ \emph {et~al.}(1988)\citenamefont
  {Varshalovich}, \citenamefont {Moskalev},\ and\ \citenamefont
  {Khersonskii}}]{Varshalovich:1988ifq}%
  \BibitemOpen
  \bibfield  {author} {\bibinfo {author} {\bibfnamefont {D.~A.}\ \bibnamefont
  {Varshalovich}}, \bibinfo {author} {\bibfnamefont {A.~N.}\ \bibnamefont
  {Moskalev}},\ and\ \bibinfo {author} {\bibfnamefont {V.~K.}\ \bibnamefont
  {Khersonskii}},\ }\href {https://doi.org/10.1142/0270} {\emph {\bibinfo
  {title} {{Quantum Theory of Angular Momentum}: {Irreducible Tensors,
  Spherical Harmonics, Vector Coupling Coefficients, 3nj Symbols}}}}\ (\bibinfo
   {publisher} {World Scientific Publishing Company},\ \bibinfo {year}
  {1988})\BibitemShut {NoStop}%
\bibitem [{\citenamefont {Dona}\ \emph {et~al.}(2022)\citenamefont {Dona},
  \citenamefont {Fanizza}, \citenamefont {Martin-Dussaud},\ and\ \citenamefont
  {Speziale}}]{Dona:2020xzv}%
  \BibitemOpen
  \bibfield  {author} {\bibinfo {author} {\bibfnamefont {P.}~\bibnamefont
  {Dona}}, \bibinfo {author} {\bibfnamefont {M.}~\bibnamefont {Fanizza}},
  \bibinfo {author} {\bibfnamefont {P.}~\bibnamefont {Martin-Dussaud}},\ and\
  \bibinfo {author} {\bibfnamefont {S.}~\bibnamefont {Speziale}},\ }\bibfield
  {title} {\bibinfo {title} {{Asymptotics of $\mathrm {SL}(2,{{\mathbb {C}}})$
  coherent invariant tensors}},\ }\href
  {https://doi.org/10.1007/s00220-021-04154-3} {\bibfield  {journal} {\bibinfo
  {journal} {Commun. Math. Phys.}\ }\textbf {\bibinfo {volume} {389}},\
  \bibinfo {pages} {399} (\bibinfo {year} {2022})},\ \Eprint
  {https://arxiv.org/abs/2011.13909} {arXiv:2011.13909 [gr-qc]} \BibitemShut
  {NoStop}%
\bibitem [{\citenamefont {Dona}\ and\ \citenamefont
  {Sarno}(2018)}]{Dona:2018nev}%
  \BibitemOpen
  \bibfield  {author} {\bibinfo {author} {\bibfnamefont {P.}~\bibnamefont
  {Dona}}\ and\ \bibinfo {author} {\bibfnamefont {G.}~\bibnamefont {Sarno}},\
  }\bibfield  {title} {\bibinfo {title} {{Numerical methods for EPRL spin foam
  transition amplitudes and Lorentzian recoupling theory}},\ }\href
  {https://doi.org/10.1007/s10714-018-2452-7} {\bibfield  {journal} {\bibinfo
  {journal} {Gen. Rel. Grav.}\ }\textbf {\bibinfo {volume} {50}},\ \bibinfo
  {pages} {127} (\bibinfo {year} {2018})},\ \Eprint
  {https://arxiv.org/abs/1807.03066} {arXiv:1807.03066 [gr-qc]} \BibitemShut
  {NoStop}%
\bibitem [{\citenamefont {Dittrich}\ and\ \citenamefont
  {Padua-Arg\"uelles}(2023)}]{Dittrich:2023rcr}%
  \BibitemOpen
  \bibfield  {author} {\bibinfo {author} {\bibfnamefont {B.}~\bibnamefont
  {Dittrich}}\ and\ \bibinfo {author} {\bibfnamefont {J.}~\bibnamefont
  {Padua-Arg\"uelles}},\ }\bibfield  {title} {\bibinfo {title} {{Lorentzian
  quantum cosmology from effective spin foams}},\ }\href@noop {} {\  (\bibinfo
  {year} {2023})},\ \Eprint {https://arxiv.org/abs/2306.06012}
  {arXiv:2306.06012 [gr-qc]} \BibitemShut {NoStop}%
\bibitem [{\citenamefont {Barrett}\ \emph {et~al.}(2009)\citenamefont
  {Barrett}, \citenamefont {Dowdall}, \citenamefont {Fairbairn}, \citenamefont
  {Gomes},\ and\ \citenamefont {Hellmann}}]{Barrett:2009gg}%
  \BibitemOpen
  \bibfield  {author} {\bibinfo {author} {\bibfnamefont {J.~W.}\ \bibnamefont
  {Barrett}}, \bibinfo {author} {\bibfnamefont {R.~J.}\ \bibnamefont
  {Dowdall}}, \bibinfo {author} {\bibfnamefont {W.~J.}\ \bibnamefont
  {Fairbairn}}, \bibinfo {author} {\bibfnamefont {H.}~\bibnamefont {Gomes}},\
  and\ \bibinfo {author} {\bibfnamefont {F.}~\bibnamefont {Hellmann}},\
  }\bibfield  {title} {\bibinfo {title} {{Asymptotic analysis of the EPRL
  four-simplex amplitude}},\ }\href {https://doi.org/10.1063/1.3244218}
  {\bibfield  {journal} {\bibinfo  {journal} {J. Math. Phys.}\ }\textbf
  {\bibinfo {volume} {50}},\ \bibinfo {pages} {112504} (\bibinfo {year}
  {2009})},\ \Eprint {https://arxiv.org/abs/0902.1170} {arXiv:0902.1170
  [gr-qc]} \BibitemShut {NoStop}%
\bibitem [{\citenamefont {Barrett}\ and\ \citenamefont
  {Steele}(2003)}]{Barrett:2002ur}%
  \BibitemOpen
  \bibfield  {author} {\bibinfo {author} {\bibfnamefont {J.~W.}\ \bibnamefont
  {Barrett}}\ and\ \bibinfo {author} {\bibfnamefont {C.~M.}\ \bibnamefont
  {Steele}},\ }\bibfield  {title} {\bibinfo {title} {{Asymptotics of
  relativistic spin networks}},\ }\href
  {https://doi.org/10.1088/0264-9381/20/7/307} {\bibfield  {journal} {\bibinfo
  {journal} {Class. Quant. Grav.}\ }\textbf {\bibinfo {volume} {20}},\ \bibinfo
  {pages} {1341} (\bibinfo {year} {2003})},\ \Eprint
  {https://arxiv.org/abs/gr-qc/0209023} {arXiv:gr-qc/0209023} \BibitemShut
  {NoStop}%
\bibitem [{\citenamefont {Dittrich}\ and\ \citenamefont
  {Ryan}(2011)}]{Dittrich:2008ar}%
  \BibitemOpen
  \bibfield  {author} {\bibinfo {author} {\bibfnamefont {B.}~\bibnamefont
  {Dittrich}}\ and\ \bibinfo {author} {\bibfnamefont {J.~P.}\ \bibnamefont
  {Ryan}},\ }\bibfield  {title} {\bibinfo {title} {{Phase space descriptions
  for simplicial 4d geometries}},\ }\href
  {https://doi.org/10.1088/0264-9381/28/6/065006} {\bibfield  {journal}
  {\bibinfo  {journal} {Class. Quant. Grav.}\ }\textbf {\bibinfo {volume}
  {28}},\ \bibinfo {pages} {065006} (\bibinfo {year} {2011})},\ \Eprint
  {https://arxiv.org/abs/0807.2806} {arXiv:0807.2806 [gr-qc]} \BibitemShut
  {NoStop}%
\bibitem [{\citenamefont {Freidel}\ and\ \citenamefont
  {Speziale}(2010)}]{Freidel:2010aq}%
  \BibitemOpen
  \bibfield  {author} {\bibinfo {author} {\bibfnamefont {L.}~\bibnamefont
  {Freidel}}\ and\ \bibinfo {author} {\bibfnamefont {S.}~\bibnamefont
  {Speziale}},\ }\bibfield  {title} {\bibinfo {title} {{Twisted geometries: A
  geometric parametrisation of SU(2) phase space}},\ }\href
  {https://doi.org/10.1103/PhysRevD.82.084040} {\bibfield  {journal} {\bibinfo
  {journal} {Phys. Rev. D}\ }\textbf {\bibinfo {volume} {82}},\ \bibinfo
  {pages} {084040} (\bibinfo {year} {2010})},\ \Eprint
  {https://arxiv.org/abs/1001.2748} {arXiv:1001.2748 [gr-qc]} \BibitemShut
  {NoStop}%
\bibitem [{\citenamefont {Asante}\ and\ \citenamefont
  {Brysiewicz}(2024)}]{Asante:2024rrd}%
  \BibitemOpen
  \bibfield  {author} {\bibinfo {author} {\bibfnamefont {S.~K.}\ \bibnamefont
  {Asante}}\ and\ \bibinfo {author} {\bibfnamefont {T.}~\bibnamefont
  {Brysiewicz}},\ }\bibfield  {title} {\bibinfo {title} {{Solving the
  area-length systems in discrete gravity using homotopy continuation}},\
  }\href@noop {} {\  (\bibinfo {year} {2024})},\ \Eprint
  {https://arxiv.org/abs/2402.17080} {arXiv:2402.17080 [gr-qc]} \BibitemShut
  {NoStop}%
\end{thebibliography}%

\end{document}